\documentclass[a4paper,fleqn,usenatbib]{mnras}

\usepackage[T1]{fontenc}
\usepackage{ae,aecompl}

\usepackage{graphicx}	% Including figure files
\usepackage{amsmath}	% Advanced maths commands
\usepackage{amssymb}	% Extra maths symbols
\usepackage{multirow}
\usepackage{threeparttable}
\usepackage{color}

\title[A modified method for FRD and length properties]{A modified method for determining the FRD and length properties of optical fibres in astronomy}

\author[Y. Yan et al.]{
Yunxiang Yan$^{1,2}$\thanks{E-mail: yxyan@nao.cas.cn (Y. Yan)},
Gang Wang$^{1}$,
Weimin Sun$^{3}$,
A-Li Luo$^{1}$,
Zhenyu Ma$^{3}$,
\newauthor
Jian Li$^{1}$
and Shuqing Wang$^{1}$
\\
$^{1}$Key Laboratory of Optical Astronomy, National Astronomical Observatories, Chinese Academy of Sciences, Beijing 100012, China \\
$^{2}$University of Chinese Academy of Sciences, Beijing 100049, China \\
$^{3}$Key Lab of In-fiber Integrated Optics, Ministry Education of China, Harbin Engineering University, Harbin 150001, China \\
}

\date{Accepted XXX. Received YYY; in original form ZZZ}

\pubyear{2016}

\begin{document}
\label{firstpage}
\pagerange{\pageref{firstpage}--\pageref{lastpage}}
\maketitle

\begin{abstract}
Focal ratio degradation (FRD) is a major contributor to throughput and light loss in a fibre spectroscopic telescope system. We combine the guided mode theory in geometric optics and a well-known model, power distribution model (PDM), to predict and explain the FRD dependence properties. We present a robust method by modifying the energy distribution method (EDM) with \emph{f-intercept} to control the input condition. This method provides a way to determine the proper position of the fibre end on the focal plane to improve energy utilization and FRD performance, which lifts the relative throughput up to 95\% with variation of output focal ratio less than 2\%. And this method can also help to optimize the arrangement of the position of focal-plane plate to enhance the coupling efficiency in a telescope. To investigate length properties, we modified PDM by introducing a new parameter, focal distance \emph{f}, into the original model to make it available for multi-position measurement system. The results show that the modified model is robust and feasible for measuring the key parameter \emph{d}$_0$ to simulate the transmission characteristics. The output focal ratio in the experiment does not follow the prediction trend but shows an interesting phenomenon that the output focal ratio increases at first to the peak, then decreases and remains stable finally with increasing fibre length longer than 15m, which provides a reference for choosing appropriate length of fibre to improve the FRD performance for the design of the fibre system in a telescope.
\end{abstract}


%%%%%%%%%%%%%%%%%%%%%%%%%%%%%%%%%%%%%%%%%%%%%%%%%%

%%%%%%%%%%%%%%%%% BODY OF PAPER %%%%%%%%%%%%%%%%%%

\section{Introduction}
Optical fibres have been for a long time used and are key technology for highly-multiplexed and precision spectroscopy in astronomy. Multi-objects observation makes it much more effective and efficient for the acquisition of spectrum of celestial objects \citep{Angel1977A}. More and more extremely large telescopes have been equipped with fibres to significantly improve observation efficiency and the number of fibres increases really fast from hundreds (400 fibres in AAT \citep{Lewis2002The} and 640 fibres in SDSS, \citep{york2000sloan}) to thousands (2400 fibres in SUBARU \citep{Sugai2015Prime}, 2400 fibers in 4MOST \citep{Jong20144MOST}, 4000 fibres in LAMOST \citep{cui2012large} and 5000 fibres in DESI \citep{Flaugher2014The}). Despite the imperfections of fibres such as optical internal absorption, the non-conservation of Optical Etendue, recognized as focal ratio degradation, and the mode noise which decreases and limits the SNR especially in high resolution spectroscopy, they still are the best option and play an important role in multiplexed spectroscopy for large fields of view.

However, optical fibres are not entirely stable light guides, and the more fibres in telescope, the more difficult to guarantee the consistency of each fibre. Transmission and FRD are the two important properties of optical fibres for astronomical instruments. To better understand the propagation characteristics of light in fibres, an alternative way is to model the fibre according to the wave theory or geometric optics and then to simulate some of these effects. In Sec. 2, the combination of modal analysis and geometric optics explain the relationship between the guided modes and the input conditions, and it also provides the intuitional power distribution on the output end with eccentric input beam.

Many groups have studied the characteristics comprehensive and systematic way \citep{Ramsey1988Focal,Clayton1989The,Avila1998Results,Schmoll1998FRD,Crause2008Investigation,Bryant2010Hexabundles,Bryant2011Characterization,Poppett2010The,Poppett2010A,Haynes2011Relative}. But the inconsistency in the results of different experimenters has been attributed to the lack of a standard way to ensure the experimental environment unchanged and the calibrating method effective and precise. Moreover, there is not only a single source of FRD, making measurements more complicate related to many effects of polishing, mounting and on-telescope application that brings more various causes of generating a divergent output light. Common methods in experiments for testing FRD are using a collimated beam \citep{Ferwana2004All,Haynes2004New, Haynes2008AAOmicron, Haynes2011Relative} or a cone beam \citep{Lee2002Properties,Oliveira2004Studying,Oliveira2011FRD,Santos2014Studying,Pazder2014The} or an annual spot beam \citep{Murphy2008Focal,Murphy2013The} as the light source. In the cone beam technique, a particular focal ratio is set to inject into the fibre and the FRD is measured from the encircled energy (EE) of a given focal ratio at the output. It is good for estimating the light loss and the energy distribution of near-field or far field can be easily acquired, but it is sensitive to alignment errors and position errors on the input fibre end. There are some similarities in both collimated beam method and annual spot beam method. Both the technologies are to inject the light at a selected angle (represent a particular focal ratio) and the FRD is determined from the radial profile width from output spot. And the Optical Etendue caused by FRD scattering rays into the outer halo and central obscuration can be observed at the same time. The Gaussian fitting method measuring the full width at half-maximum (\emph{FWHM}) sometimes can underestimate the FRD if the output distribution deviates away from a Gaussian profile due to scattering effect \citep{Haynes2011Relative}.

In this paper, firstly we combine the geometric optics and wave theory in Sec. 2 to describe the transmission properties in guided modes and the power distribution, which also lays the foundation of modifying the model in the later. Then we proposed a new method in Sec. 3 to improve the illumination system and to keep the consistency of input condition to provide a reliable experiment environment. In the experiments, we choose the large core multimode fibres in LAMOST to investigate the FRD properties. The collimated beam is convenient for testing the influence of changing incident angle and each time the fibre is measured with light injected at positive and negative angles to account for any misalignment. And the cone light will be used for testing both FRD and throughput to improve comparability for the parameters which can be acquired simultaneously. In Sec. 4, a new factor of focal distance \emph{f} is introduced to optimize the PDM model to investigate the FRD and length properties.

\section{Models of optical fibre system}
Introducing fibres into a large telescope to build the multi-object observe system can significantly improve the survey efficiency with a large field of view. Generally, there are three types of configuration of optical fibre system commonly applied in large telescope. Type I is for traditional fibre multiplexed spectrographs and type II \& III are mostly used in integral field spectrographs (IFS). For IFS the input fibre end is equipped with a lenslet array which can eliminate the dead space between fibre cores and the FRD can be minimized by adapting to the input focal ratio.

Type I: In traditional fibre-fed spectrographs, the optical fibres are placed directly at the focal surface and arranged into a pseudoslit which matches the curvature of the focal surface and orients the fibres to the exit pupil of the telescope. Therefore the nearfield out from the fibre end is projected on the detector while the grating is illuminated by the far field.

Type II: The pseudoslit in cludes a lenslet array to convert back to the native focal ratio of the telescope. Therefore the far field out from the fibre end is projected on the detector and the disperser is illuminated by the reimaged near-field.

Type III: The pseudoslit consists of the bare ends of the fibres equipped in a glass plate to reduce scattering losses. It's kind of like type I, the detector receives the near-field out from the fibre end and the disperser is illuminated by the far field.

\begin{figure}
   \centering
   \includegraphics[width=\hsize]{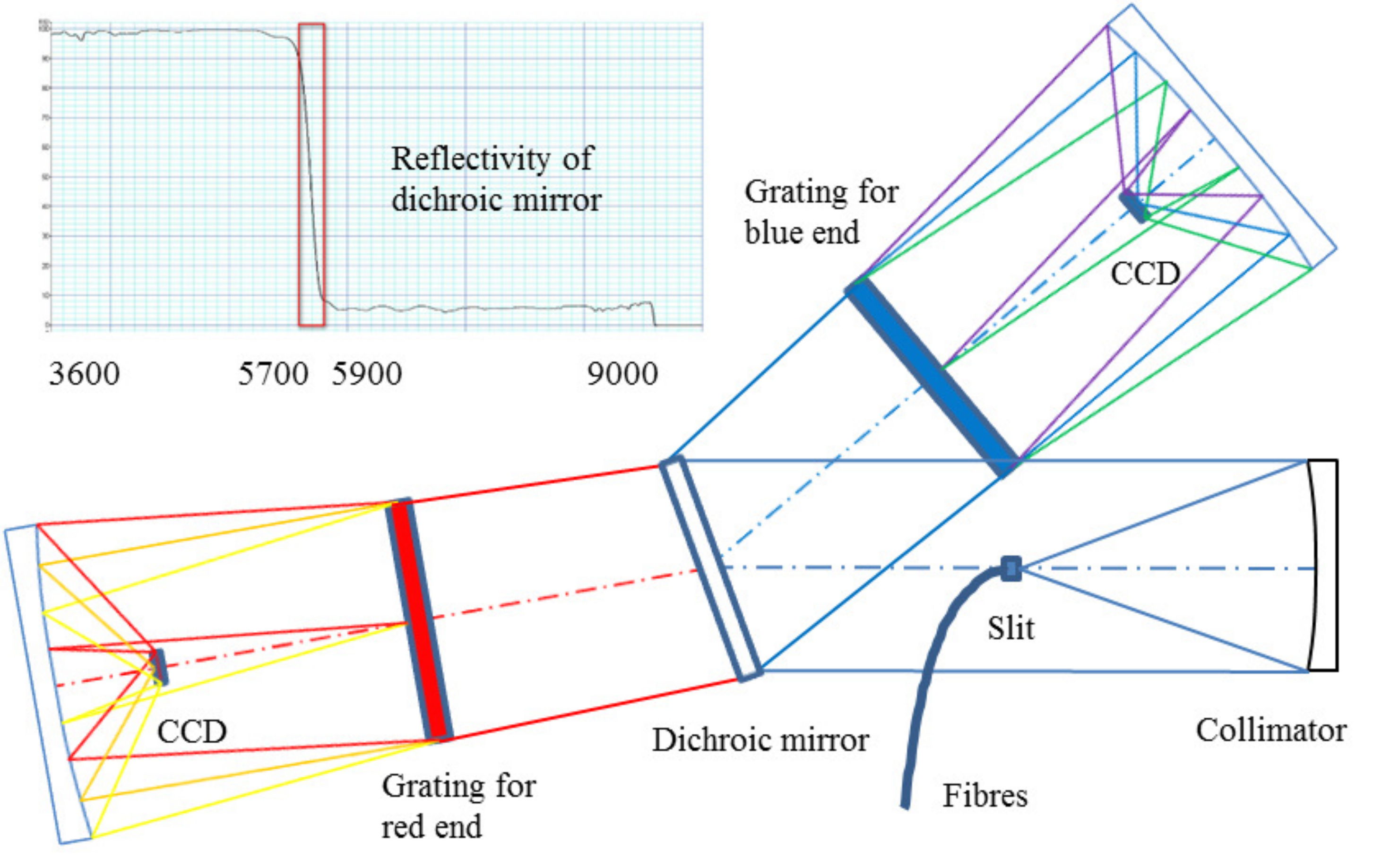}
      \caption{The schematic diagram of spectrograph in LAMOST.}
         \label{fig:1}
   \end{figure}

Practically it is the accuracy required in the near-field or far field and the impact of illumination of the disperser that determine the optimum optical fibre system configuration in the large telescope. Fig.\ref{fig:1} shows the fibre system in the spectrograph of type I applied in LAMOST.

\subsection{Geometry of optical fibres}
\begin{figure}
   \centering
   \includegraphics[width=\hsize]{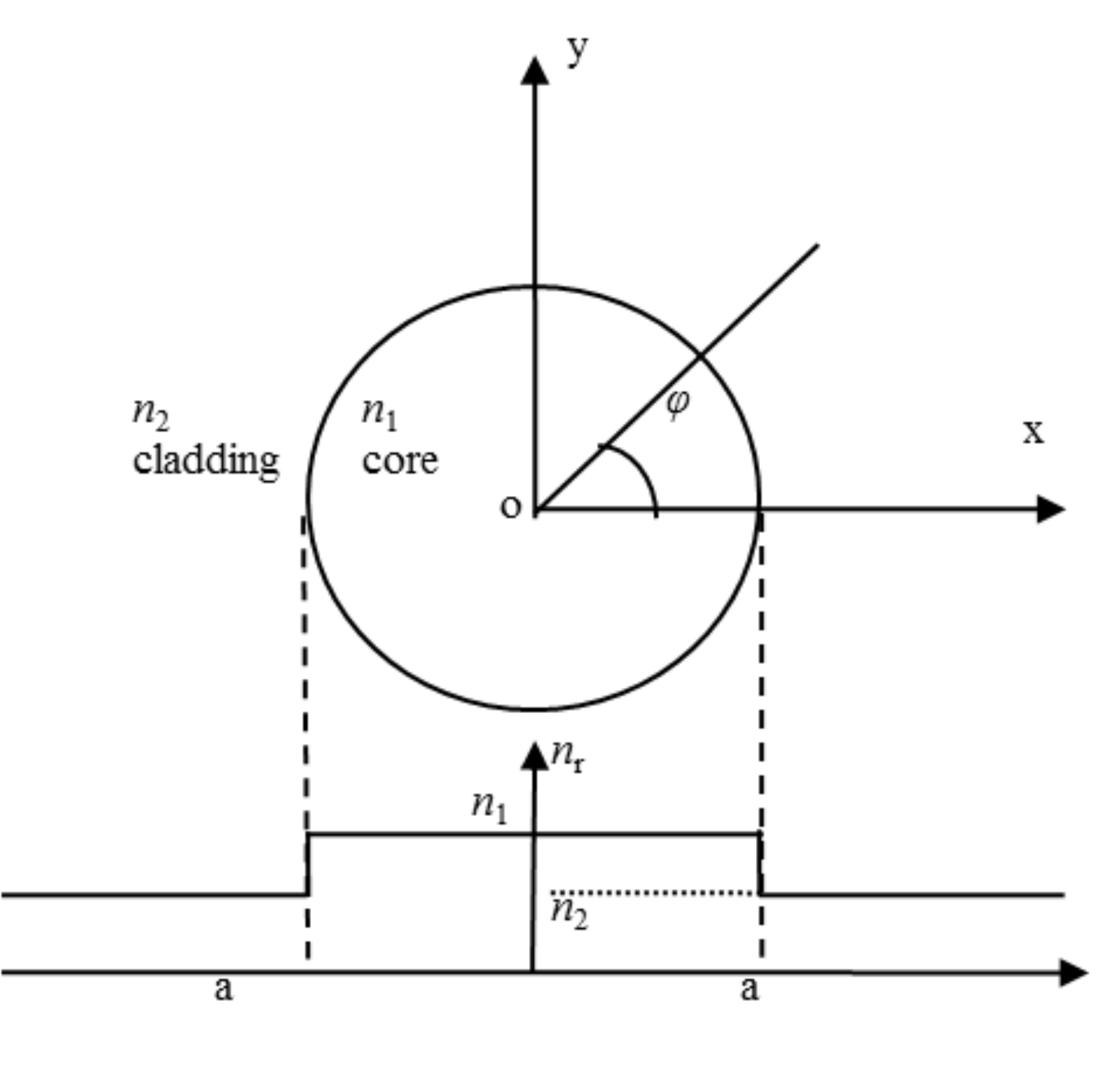}
      \caption{Refraction index distribution of a typical double-layered fibre.}
         \label{fig:2}
   \end{figure}

The optical fibre used in LAMOST manufactured by PolyMicro Tech is low-OH$^-$ multimode fibre with a limiting numerical aperture (sine of the maximum acceptance cone angle, denoted as \emph{N.A.}) of 0.22$\pm$0.02 and the diameters of the core and the cladding are 320$\mu$m and 355$\mu$m, respectively. The typical double-layered fibre structure and the refractive index distribution are shown in Fig.\ref{fig:2}, including the fibre core and cladding.

   \begin{figure}
   \centering
   \includegraphics[width=\hsize]{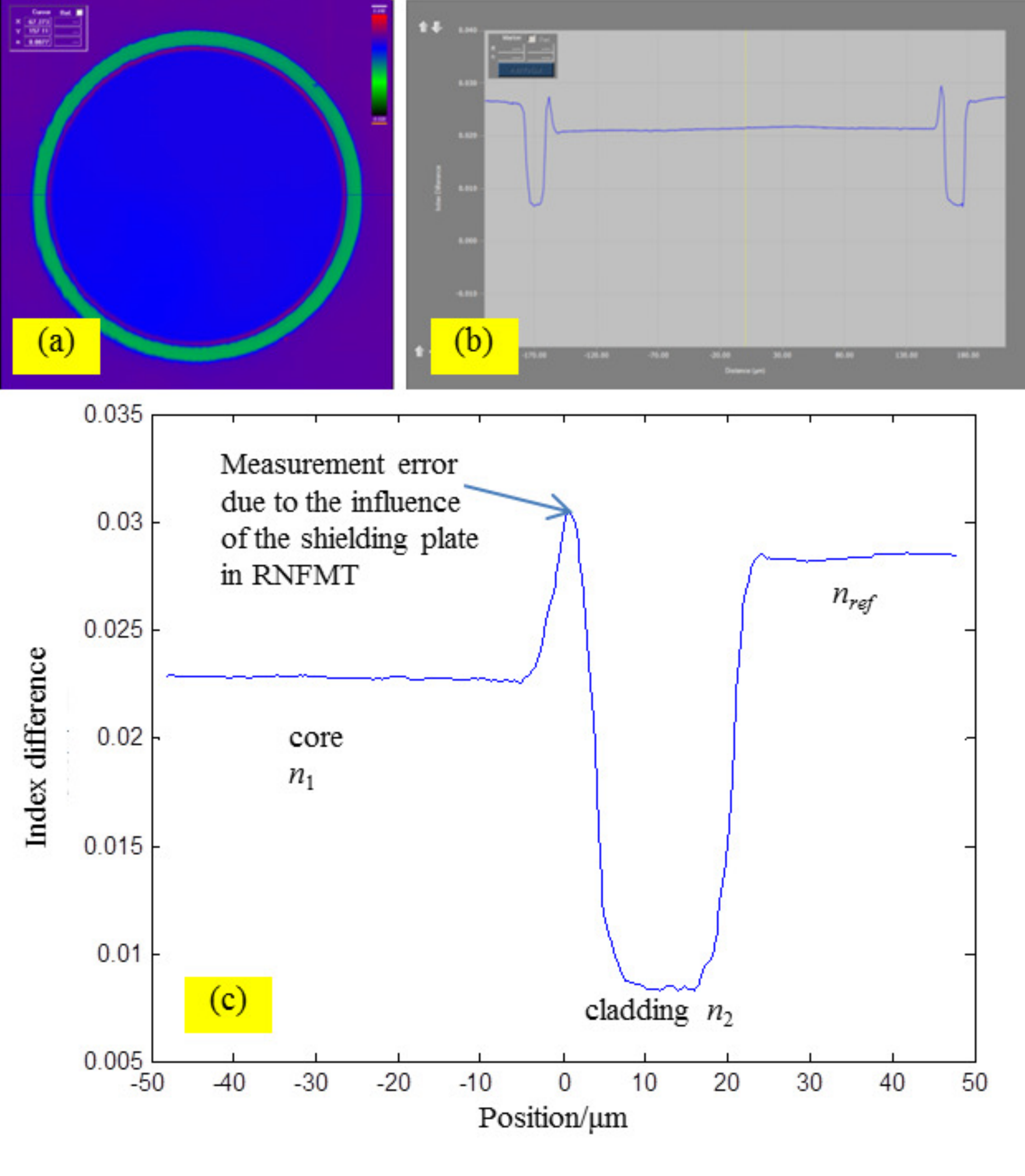}
      \caption{Refraction index distribution of a large core fibre of 320$\mu$m in LAMOST. (a): the contour map of the end face. (b): distribution cross the blue line in (a) which pass through the fibre centre. (c): the detailed distribution interception of the right part.}
         \label{fig:3}
   \end{figure}

We measured the index profile of the large core fibre in LAMOST by refracted near-field measurement technique (RNFMT), as shown in Fig.\ref{fig:3}. \emph{n$_{ref}$}  is the refractive index for reference.

\[{n_{ref}} = 1.4700 \Rightarrow \left\{ {\begin{array}{*{20}{c}}
{{n_1} = 1.4649 \pm 0.0008}\\
{{n_2} = 1.4495 \pm 0.0011}
\end{array}} \right.\]

According to RNFMT, a shielding plate should be placed behind the output end of the testing fiber to stop the guided modes and the leak modes, and only the refractive modes can be detected. The equivalent \emph{N.A.} of the plate will affect the measurement results which can not be too small or too large, and especially for a large core fiber, the undesired peak like in Fig.\ref{fig:3}(c) will occur around the boundary between the cladding and the core.\citep{Stewart1982Optical}

\begin{figure}
   \centering
   \includegraphics[width=\hsize]{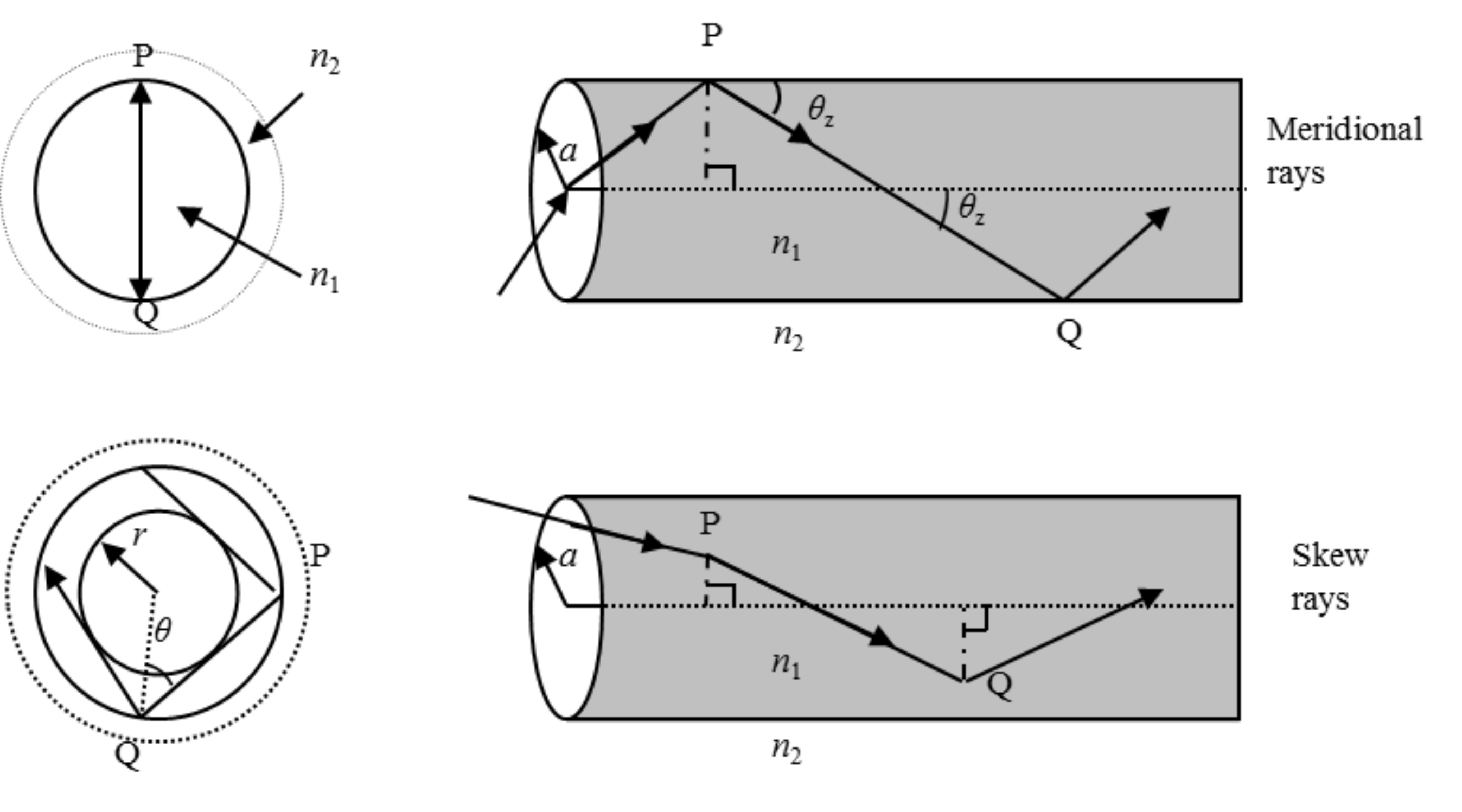}
      \caption{Ray-tracing analysis of meridional rays and skew rays in the cross section of a fibre. For meridional rays, the transverse component only exists in the meridional plane. For skew rays, the transverse component exists in an annular region near the boundary of the core and the cladding. The width of the ring is determined by the incident condition.}
         \label{fig:4}
   \end{figure}

The index and diameter of fibre core and cladding determine the \emph{N.A.} (\emph{N.A.}=0.21 in our experiment for fibres in LAMOST) and the acceptable light rays transmitting in fibres. Leaving out the absorption, here we only take the rays meet the condition of total internal reflection (TIR) into account. The rays can be characterized as meridional or skew rays in Fig.\ref{fig:4}. In ideal circumstance, no bending or imperfection exist in the fibre, the former only pass through in meridian plane containing the axis of the fibre with no transverse component, only to stay within the meridian plane into which they were launched. However, for skew rays, a helical path avoiding the axis establishes along the fibre because skew rays will strike the interface between the core and cladding at an angle with components both parallel and transverse to the meridian.

\begin{figure}
   \centering
   \includegraphics[width=\hsize]{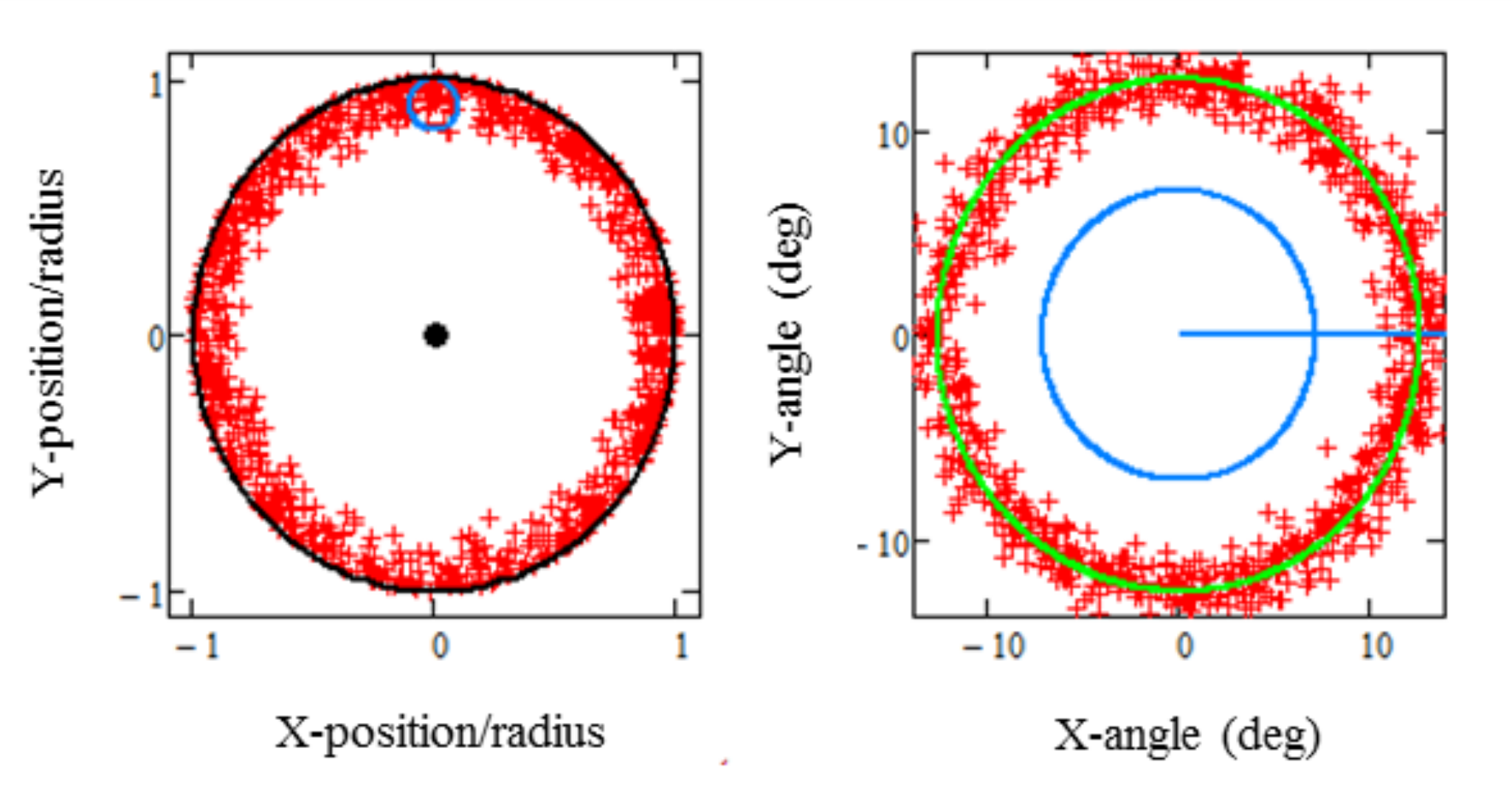}
      \caption{The near filed (left) and far field (right) distribution on the output end for eccentric incident condition at an angle of 12.7$^\circ$. The small blue circle in the left is the incident light spot. The green circle in the right represents the locus of expected data points. The large blue circle stands for the aperture of \emph{F}=4.5.\citep[see][figure 16]{Allington-Smith2012Simulation}}
         \label{fig:5}
   \end{figure}

In practical applications, both of the two types of rays transmit in fibres. Under the constraint of \emph{N.A.} and the TIR condition, rays in meridian plane with incident angle larger than the critical angle will partially go through into the cladding and emit the leaky modes, so the energy cannot be constrained in fibre core stably and dissipate along the fibre in a certain distance. But for skew rays, the incident angle limitation is not strictly constrained by \emph{N.A.} but by the TIR, that is, skew rays with a larger incident angle can also couple into the fibre but not for rays in meridian plane. In the meantime, noticing the path in the cross section, the light can only go through in a small annular region near the out circle and the larger the incident angle is, the thinner the ring will be to contain higher order modes. These phenomena can be explained by the relationship between the guided mode and geometrical optics ray tracing theory in Sec. 2.2 and has been simulated by Allington-Smith et al. \citep[see][figure 16]{Allington-Smith2012Simulation} as shown in Fig.\ref{fig:5}.

\subsection{Guided modes in geometrical optics}
As is shown in Fig.\ref{fig:5}, the output power distribution is related with input condition. In this section the combination of mode theory and geometrical optics is proposed to explain the phenomenon.

In the principle of optical fibre optics, there are two constraint conditions to emit stable guided modes: TIR condition and transverse resonance condition (TRC). The power distribution on the cross section can be considered as the superposition of lots of independent rays with the same incident angle. This kind of combination of guided modes can form standing wave. Both of the TIR condition and TRC constitute the necessary and sufficient condition of guided modes.

TIR condition is determined by the refraction index distribution and the TRC by the phase change during the transmission. According to the principle of standing wave, to meet the TRC, the phase difference (PD) should be of integer times of 2$\pi$, including two parts: 1) phase change from the back and forth movements of transverse component; 2) phase shift in the TIR.

   \begin{figure}
   \centering
   \includegraphics[width=\hsize]{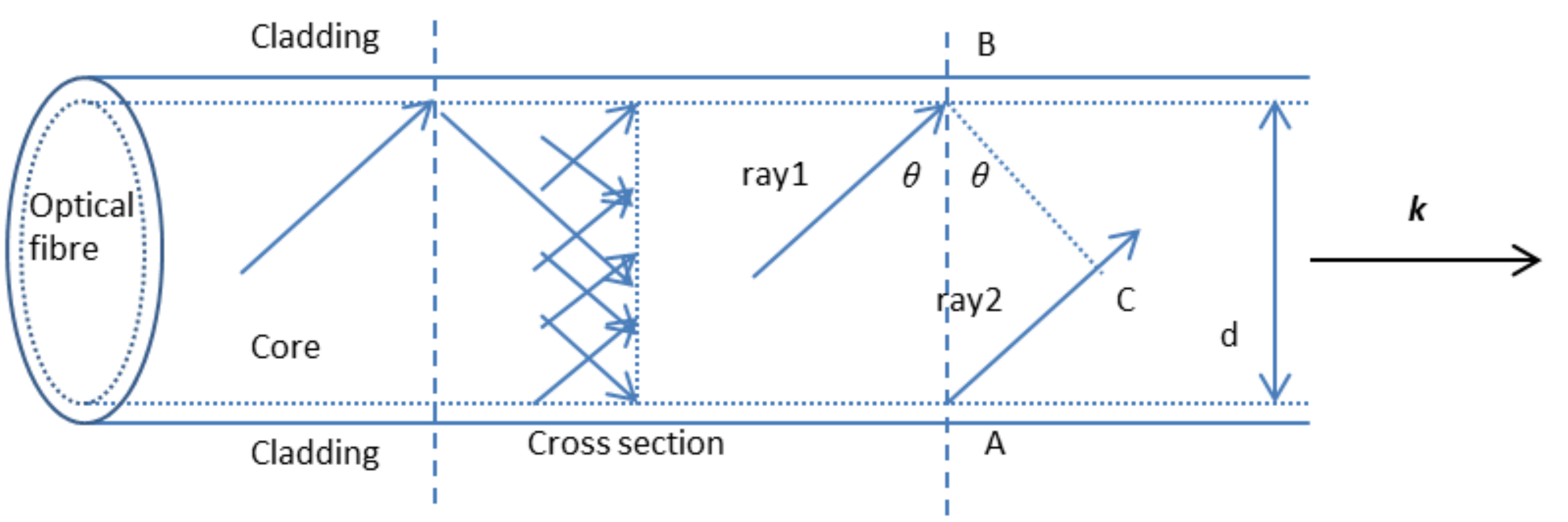}
      \caption{The schematic diagram of rays superposition to emit guided modes in a fibre. Standing wave forms in the cross section with same angles (consider the positive and negative angles as the same angles for the symmetry of fibres).}
         \label{fig:6}
   \end{figure}

In Fig.\ref{fig:6}, the parallel light rays of ray1 and ray2 belong to the same mode and the PD is generated by the optical path difference from A to B and back to A, where the BC stands for the wavefront which has the same wave phase. Then the PD can be acquired:
\begin{equation}\label{eq1}
{\delta }' = 2\pi  \cdot \frac{{AC}}{\lambda } = 2\pi  \cdot \frac{{d\cos \theta }}{\lambda } = 2\pi  \cdot \frac{{{n_1}d\cos \theta }}{{{\lambda _0}}}
\end{equation}
where the $\lambda $ and the ${\lambda _0}$  are the wavelength in dielectric medium and vacuum respectively, and ${n_1}$  is the refractive index of fibre core.

The module of light vector \emph{\textbf{k}} is wave number $k = \frac{{2\pi }}{\lambda }$, so the propagation constant standing for the component with respect to fibre axis can be written as $\beta  = k\sin \theta  = \frac{{2\pi }}{{{\lambda _0}}}{n_1}\sin \theta $, substituting in Eq.(\ref{eq1}), the total PD of transverse component can be written in the form:
\begin{equation}\label{eq2}
\begin{split}
& {\delta _1} = 2{\delta }' = 2dk\sqrt {1 - {{\sin }^2}\theta } \\
& \quad = 2d\sqrt {{k^2} - {{\left( {k\sin \theta } \right)}^2}}  = 2d\sqrt {{{\left( {{k_0}{n_1}} \right)}^2} - {\beta ^2}}
\end{split}
\end{equation}
where the ${k_0}$ is the wave number in the vacuum.

\begin{figure}
   \centering
   \includegraphics[width=\hsize]{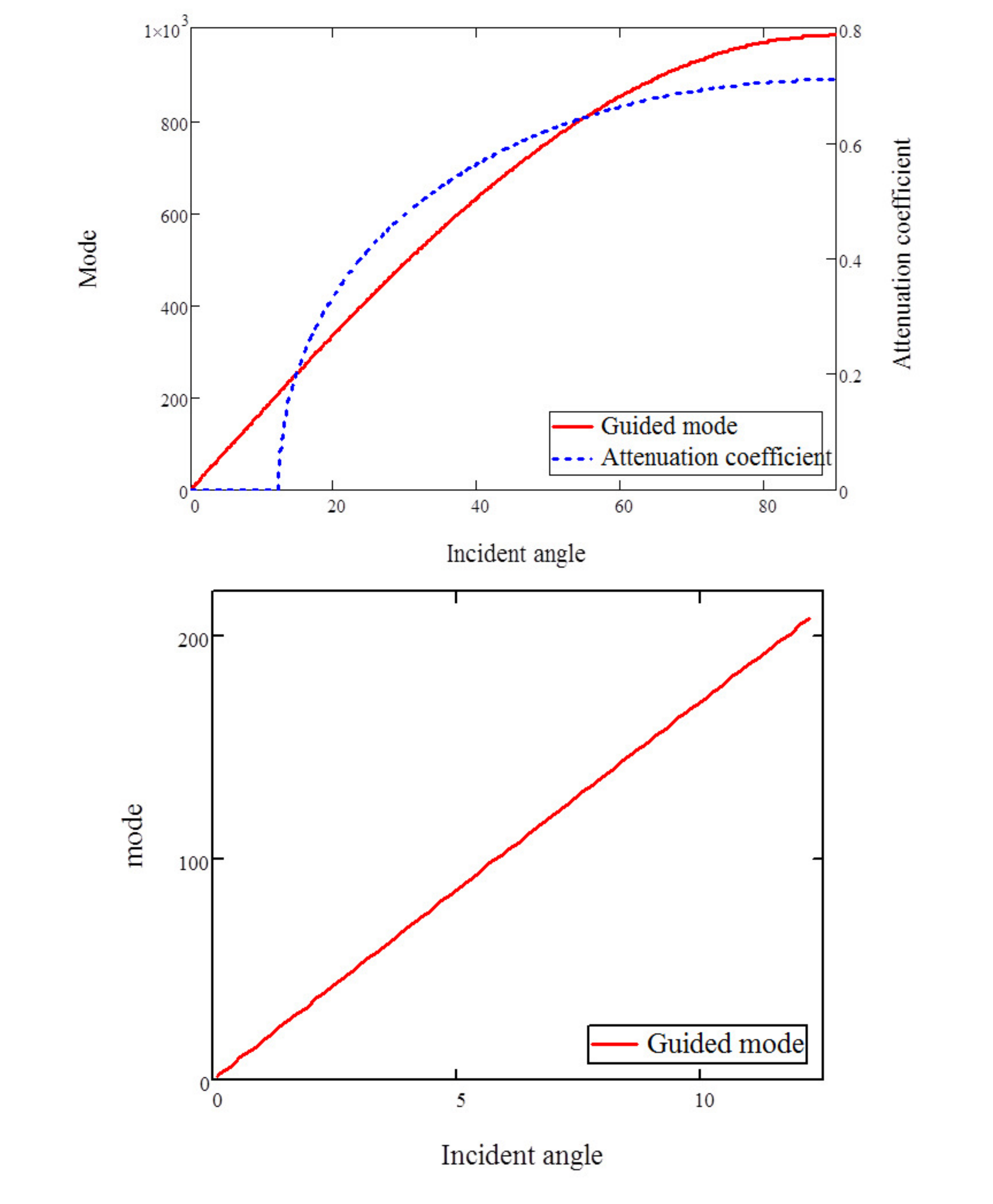}
      \caption{The relationship between guided modes and incident angle. According to the principle of TIR, the output and input angle will be the same for each guided mode, so in the output end only the modes with output angles meet the constraints of \emph{N.A.} can transmit. Rays of different angles will form different annular spot and the superposition of all the modes will lead to a ring spot or a Gaussian spot which depends on the input condition, including incident angle, type of input beam, and the fibre circumstance. When the incident angle is larger than 12.2$^\circ $, although it can emit some modes, the attenuation coefficient is increasing really fast, meaning that the power of the modes leak into the cladding and is quickly depleted in a certain length. So none stable guided modes will exist.}
         \label{fig:7}
   \end{figure}

When light reflect in the interface between the core and the cladding of different materials, the phase of reflecting light changes with respect to the incident light, and the light wave is electromagnetic wave, so the PDs of TE mode in Eq.(\ref{eq3}) and TM mode in Eq.(\ref{eq4}) have a small difference that generally can be ignored especially under the weak-guide condition.
\begin{equation}\label{eq3}
\begin{split}
& {\phi _{12}} = - 2\arctan \frac{{\sqrt {{n_1}{{\sin }^2}\theta  - {n_2}^2} }}{{{n_1}\cos \theta }} \\
& \qquad = - 2\arctan \frac{{\sqrt {{\beta ^2} - {k_0}^2{n_2}^2} }}{{\sqrt {{k_0}^2{n_1}^2 - {\beta ^2}} }}
\end{split}
\end{equation}
\begin{equation}\label{eq4}
\begin{split}
& {\phi _{12}}' =  - 2\arctan \frac{{{n_1}\sqrt {{n_1}{{\sin }^2}\theta  - {n_2}^2} }}{{{n_2}^2\cos \theta }}\\
& \qquad =  - 2\arctan \frac{{{n_1}^2\sqrt {{\beta ^2} - {k_0}^2{n_2}^2} }}{{{n_2}^2\sqrt {{k_0}^2{n_1}^2 - {\beta ^2}} }}
\end{split}
\end{equation}
To simplify the PD form, denote Eq.(\ref{eq3}) and Eq.(\ref{eq4}) as follow for convenience:
\begin{equation}\label{eq5}
\phi  = \phi \left( {{k_0},{n_1},{n_2},\beta } \right)
\end{equation}
Then the TRC demands the PD to satisfy the constraints in Eq.(\ref{eq6}):
\begin{equation}\label{eq6}
\begin{split}
& \delta  = {\delta _1} + \phi \\
& \quad = 2d\sqrt {{{\left( {{k_0}{n_1}} \right)}^2} - {\beta ^2}}  + \phi \left( {{k_0},{n_1},{n_2},\beta } \right) \\
& \quad = m \cdot 2\pi \begin{array}{*{20}{c}},& \qquad {m = 0,1,2, \cdot  \cdot  \cdot } \end{array}
\end{split}
\end{equation}
Also the TIR condition described as $\sin \theta  > {{{n_2}} \mathord{\left/ {\vphantom {{{n_2}} {{n_1}}}} \right. \kern-\nulldelimiterspace} {{n_1}}}$ should be considered according to the standing wave principle:
\begin{equation}\label{eq7}
\begin{split}
& \left\{ \begin{split}
& {k_0}{n_2} < \beta  \le {k_0}{n_1}\\
& 2d\sqrt {{{\left( {{k_0}{n_1}} \right)}^2} - {\beta ^2}}  + \phi \left( {{k_0},{n_1},{n_2},\beta } \right) = m \cdot 2\pi
\end{split} \right. ,\\
& \qquad \qquad m = 0,1,2, \cdot  \cdot  \cdot
\end{split}
\end{equation}
Fig.\ref{fig:7} shows the relationship between incident angle and the number of guided modes for \emph{N.A.} = 0.21 measured before in experiments, and the attenuation coefficient is in dashed blue.

The guided modes exist stably for the rays with incident angle smaller than 12.2$^\circ $, which is determined by \emph{N.A.} = 0.21. For other rays with larger angles, the attenuation coefficient increases very fast as exponential function, meaning that the power of higher order modes losses too quickly to keep the existence of stable guided modes.

\begin{figure}
   \centering
   \includegraphics[width=\hsize]{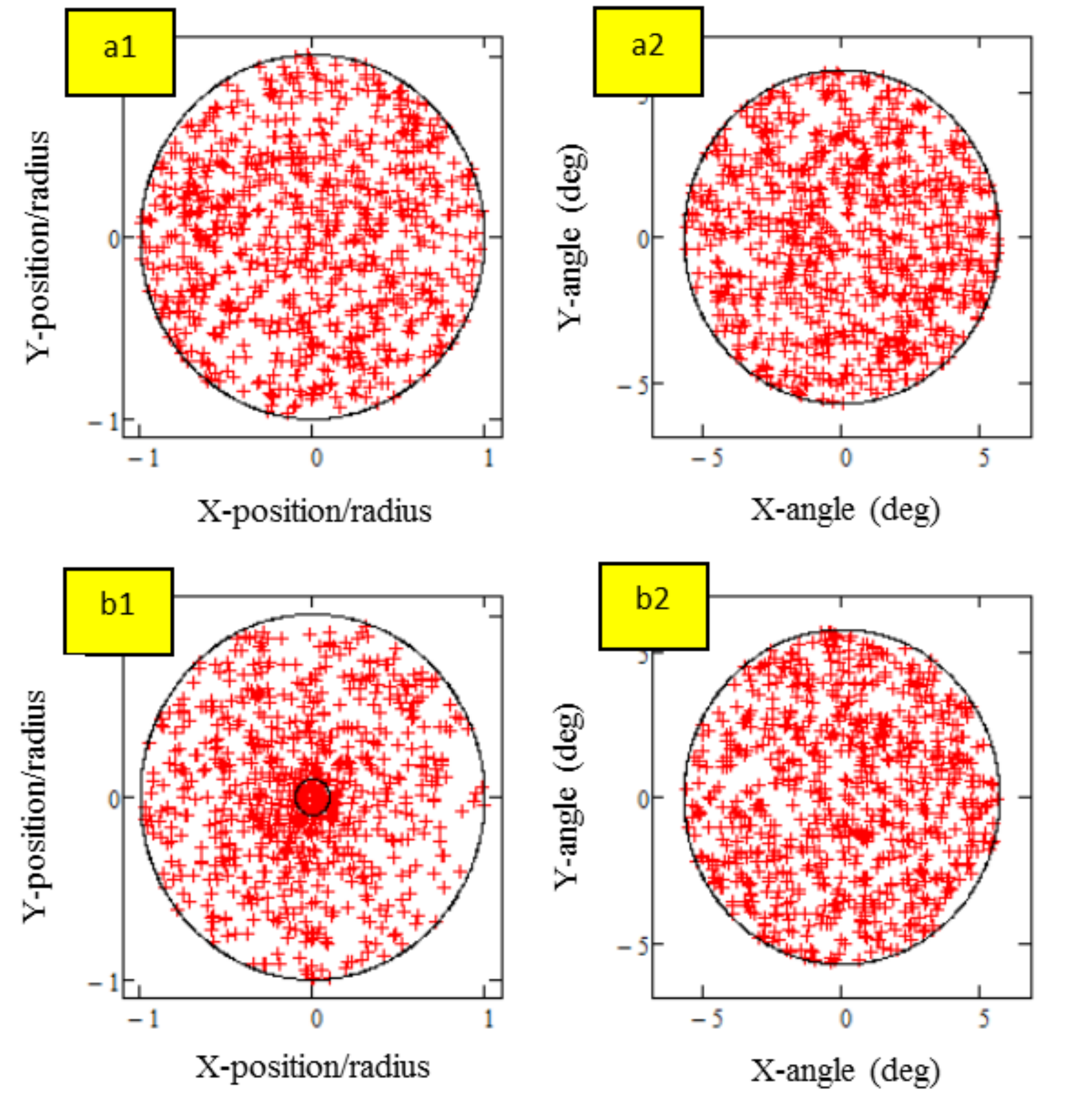}
      \caption{The simulated near field (a1 \& b1) and far field (a2 \& b2) of the rays superposition illuminated by a cone beam of \emph{F}$_{in}$ = 5.0. The beam covers the whole end face in (a1) \& (a2) and the near field and the far field are uniform. In (b1), the small black circle represents the centrally-located spot. The far field is also uniform but the near field becomes sharply peaked in the centre. \citep[see][figure 4]{Allington-Smith2012Simulation}}
         \label{fig:8}
   \end{figure}

For a cone beam situation, rays injecting into a fibre in the centre or covering the whole end face will generate a uniform spot in the far field because all the rays spread uniformly in meridian plane. But in the near field, the power distribution differs from each other. If the cone beam covers the whole end face, then the output end is uniform (Fig.\ref{fig:8}(a1)), otherwise the near field pattern is sharply peaked in the centre (Fig.\ref{fig:8}(b1)). Generally, the two cases can convert mutually during the transmission. Because the major difference between them is whether the cone beam covers the end face, if we consider several cross sections in different positions of the fibre as different input end faces (we consider it as the local input end face, denoted as LIEF) which will not influence the simulation, then some LIEFs are totally covered and some are illuminated by spots with different radii depending on the distance between the LIEF and the original input end face.

\begin{figure}
   \centering
   \includegraphics[width=\hsize]{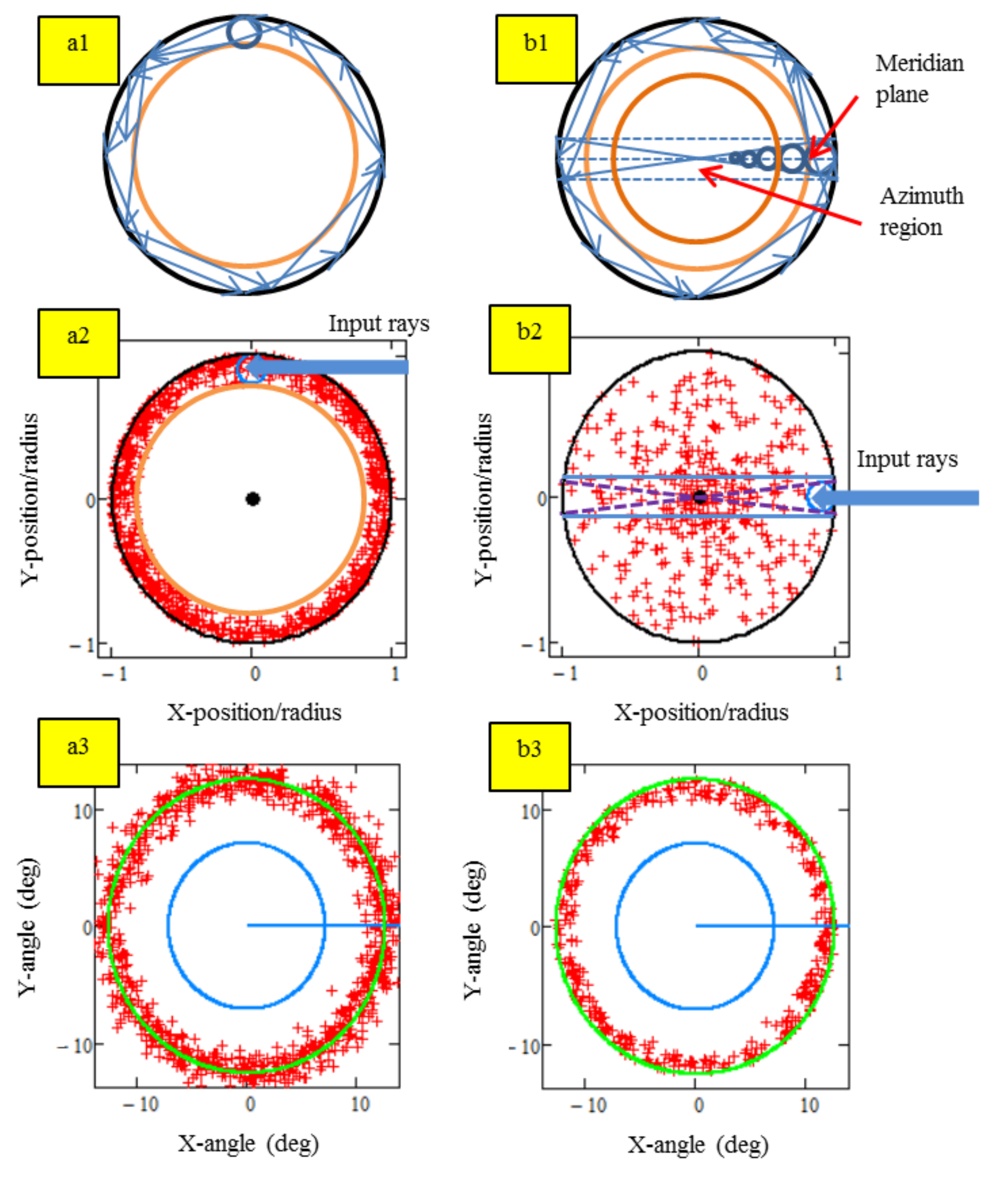}
      \caption{Simulation of the superposition of skew rays and meridional rays in near (a2, b2) and far (a3, b3) field which describe the guided modes distribution in the cross section illuminated by collimated beam at an angle of 12.7$^\circ$. The input beam is not an ideal pointolite, so we can treat it as the combination of a series of consequent points spread over the blue circle. But both the far fields appear to be ring spots.}
         \label{fig:9}
   \end{figure}

But in the practical application or the simulation, the input light spot has a certain area and is not an ideal point source, the output field can be separated into two kinds of situation. One is for total skew rays of eccentric beam in Fig.\ref{fig:9}a, and the other is for rays injecting in meridian plane with a light spot in Fig.\ref{fig:9}b. The light goes through in the fibre spirally for skew rays. When skew rays are projected onto the cross section, the rays only reflect and are limited in the annular region as shown in Fig.\ref{fig:9}(a2). But for the other situation, the path of rays is a little complicated. Generally the incident region can be considered as two parts of meridian plane and azimuth region. The input spot is treated as a series of consequent spots distributing on the meridian plane uniformly, so the output light spreads uniformly in every single meridian plane. But still, there is part of light incident on the azimuth region, and it is much more like the skew rays situation, and in every annular region the output light spreads uniformly too. Finally the superposition of the output field is demonstrated as a two dimensional Gaussian distribution as shown in Fig.\ref{fig:9}(b2). And the longitudinal components along with the direction of fibre axis have the same angles as the incident angle, so in the far field, both of the spots become rings with the output angle the same as input beam (the green circle represents the rays without scattering or FRD).

But in practical applications, multimode optical fibres employed in astronomical instruments more or less have some imperfections like the irregularities of the manufacture, microbending and stress caused by connectors or groove platforms. These imperfections will scatter the rays, which changes the incident angle reaching the interface between core and cladding or cause modes exchange between neighbour modes, then some rays couple into cladding in greater angles and FRD and light loss occurs.

\subsection{Power distribution model (PDM)}
A simple way to investigate the performance of optical fibres is to test the relationship between input and output light filed and this method can also measure the FRD and light loss with the output optical power \emph{P} being calculated.

\begin{figure}
   \centering
   \includegraphics[width=\hsize]{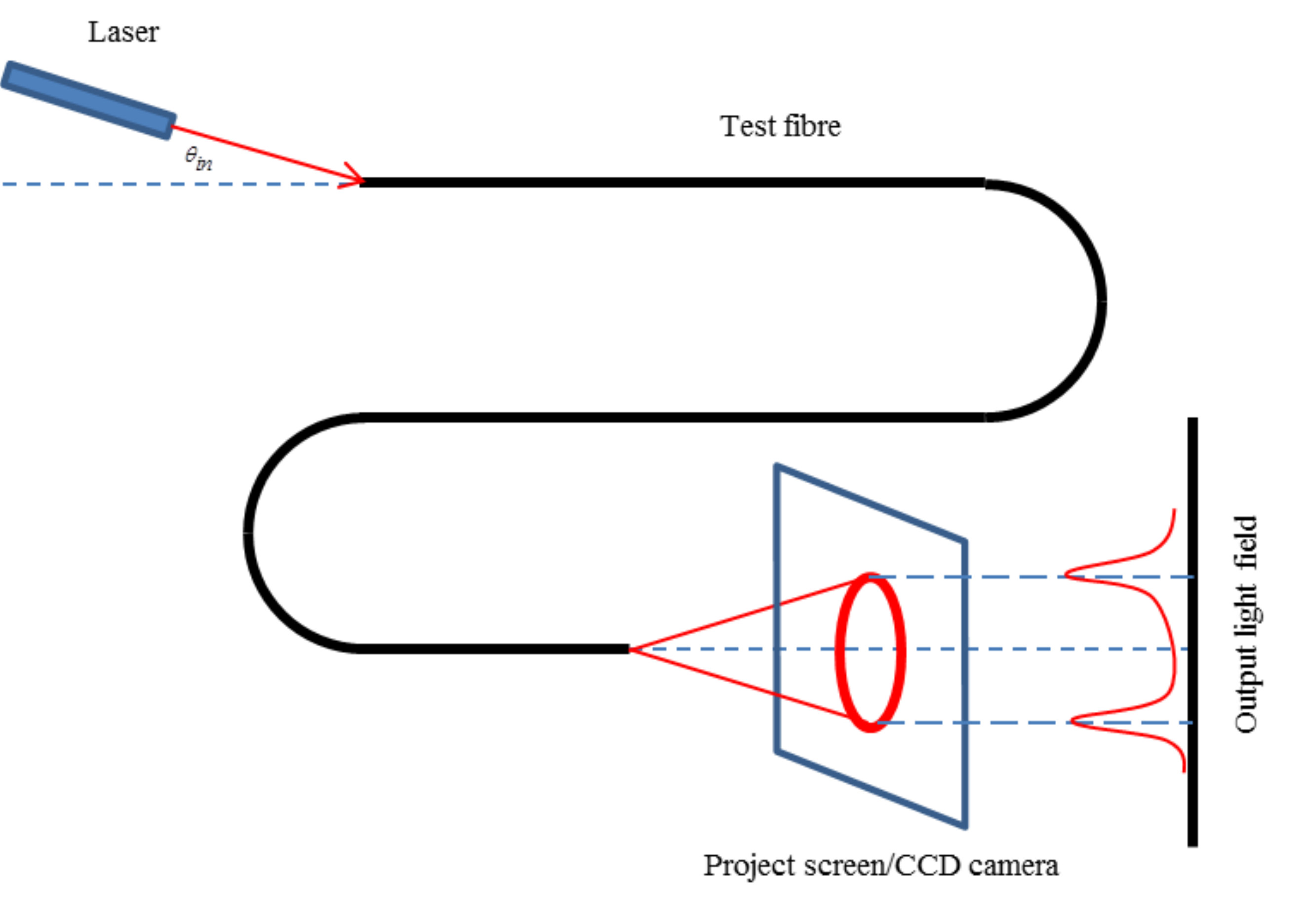}
      \caption{The schematic setup for the laser experiment. A collimated input beam is launched at an angle of   to inject into a test fibre. The far field is projected on to an observation screen or a CCD camera.}
         \label{fig:10}
   \end{figure}

\citet{Carrasco1994A} (hereafter CP94) and \citet{Poppett2010The} have shown that it is possible to use the model proposed by \citet{Gloge1972Derivation} with a single parameter \emph{D} to characterize its performance. The mode shows that the far field distribution image can represent the modal power distribution. Here we consider it as power distribution model (PDM).
\begin{equation}\label{eq8}
\begin{split}
& \frac{{\partial P\left( {{\theta _{out}},{\theta _{in}}} \right)}}{{\partial L}} = \\
& \qquad - A\theta _{in}^2P\left( {{\theta _{out}},{\theta _{in}}} \right) + \frac{D}{{{\theta _{in}}}}\frac{\partial }{{\partial {\theta _{in}}}}\left( {\theta \frac{{\partial P\left( {{\theta _{out}},{\theta _{in}}} \right)}}{{\partial {\theta _{in}}}}} \right)
\end{split}
\end{equation}
Where ${\theta _{out}}$ is the output angle, $A$ is an absorption coefficient and $D$ is a parameter that depends on the key parameter ${d_0}$ that characterizes microbending:
\begin{equation}\label{eq9}
D = {\left( {\frac{\lambda }{{2a{n_1}}}} \right)^2}{d_0}
\end{equation}
where $\lambda $ is the wavelength of light, $a$ is the core diameter and ${n_1}$ is the refraction index of the core. Both the value of $D$  and ${d_0}$  depend on a specific fibre and can be tested and acquired from experiments simply by testing the output annual spot excited by a collimated laser beam as shown in Fig.\ref{fig:10}. In the typical experimental setup, given the incident angle ${\theta _{in}}$, use a laser beam to couple into the fibre and a CCD camera to capture the output spot. Then the \emph{FHWM} can be acquired to estimate the parameter ${d_0}$.

Eq.(\ref{eq8}) has been solved by \citet{Gambling1975Mode} for the case of a collimated beam at an input angle of ${\theta _{in}}$:
\begin{equation}\label{eq10}
\begin{split}
& P\left( {{\theta _{out}},{\theta _{in}}} \right) =  \left[ {\frac{{\exp \left( {{{ - bL} \mathord{\left/ {\vphantom {{ - bL} 2}} \right. \kern-\nulldelimiterspace} 2}} \right)}}{{1 - \exp \left( { - bL} \right)}}} \right] \\
& \qquad \times \exp \left\{ { - \left( {\frac{{{\chi _{out}} + {\chi _{in}}}}{2}} \right)\left[ {\frac{{1 + \exp \left( { - bL} \right)}}{{1 - \exp \left( { - bL} \right)}}} \right]} \right\}\\
& \qquad \times {I_0}\left[ {\frac{{{{\left( {4{\chi _{out}}{\chi _{in}}} \right)}^{{1 \mathord{\left/
 {\vphantom {1 2}} \right.
 \kern-\nulldelimiterspace} 2}}}\exp \left( {{{ - bL} \mathord{\left/
 {\vphantom {{ - bL} 2}} \right.
 \kern-\nulldelimiterspace} 2}} \right)}}{{1 - \exp \left( { - bL} \right)}}} \right]
\end{split}
\end{equation}
where $\chi  = {\left( {{A \mathord{\left/
 {\vphantom {A D}} \right.
 \kern-\nulldelimiterspace} D}} \right)^{{1 \mathord{\left/
 {\vphantom {1 2}} \right.
 \kern-\nulldelimiterspace} 2}}}{\theta ^2}$, $b = 4{\left( {AD} \right)^{{1 \mathord{\left/
 {\vphantom {1 2}} \right.
 \kern-\nulldelimiterspace} 2}}}$, and ${I_0}$ is the modified Bessel function of zeroth order.

\begin{figure}
   \centering
   \includegraphics[width=\hsize]{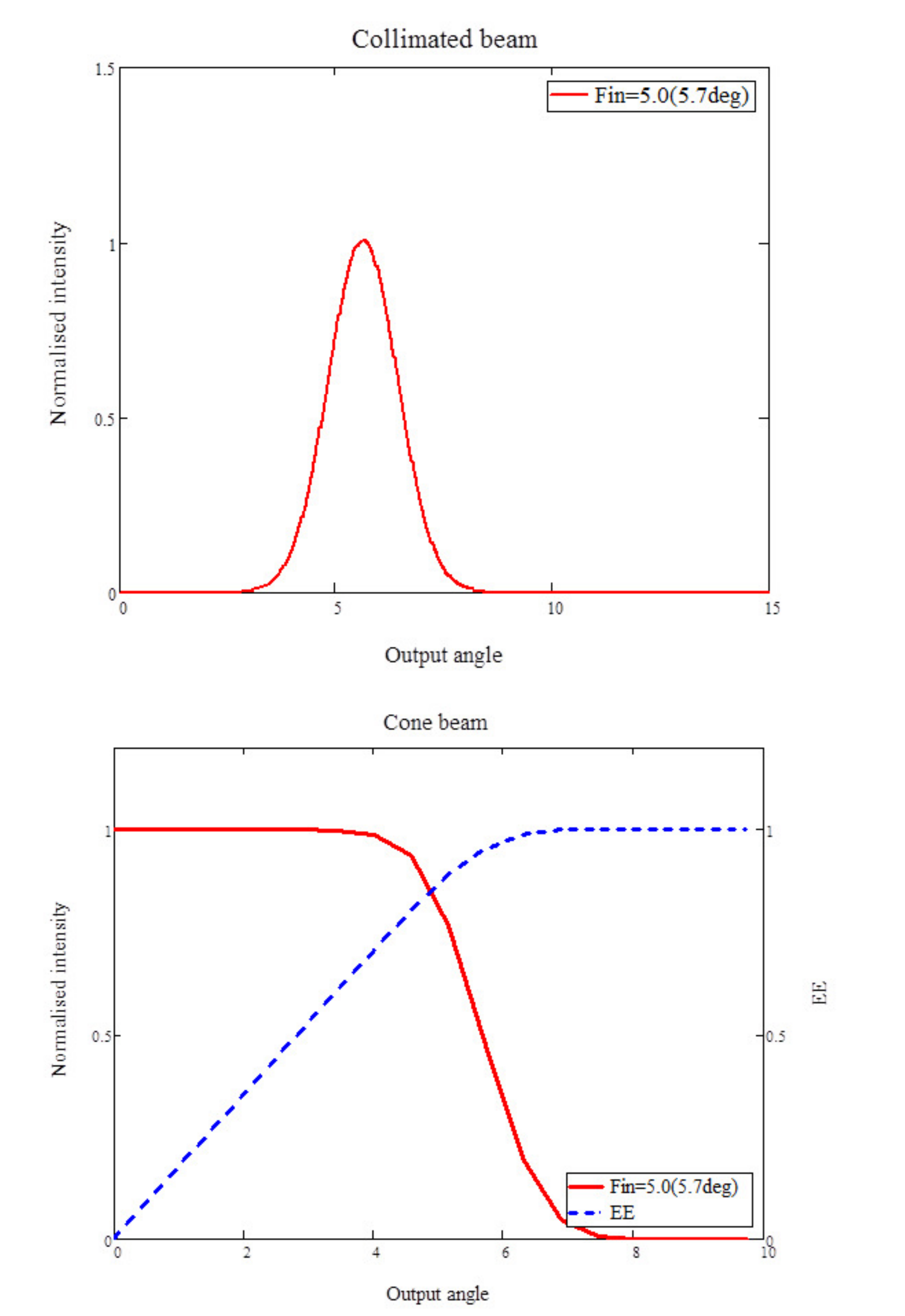}
      \caption{Simulation of normalised output filed for a fibre of 5m long with $D = 2.4 \times {10^{ - 5}}{m^{ - 1}}$ of a collimated beam at an angle of incidence ${\theta _{in}} = 5.7^\circ $ and a cone beam of \emph{F}$_{in}$ = 5.0 predicated by PDM. The dashed blue line is the ratio of EE.}
         \label{fig:11}
   \end{figure}

Eq.(\ref{eq10}) describes the output flux for the case of a collimated input beam, and for other situation, $G\left( {{\theta _{in}},\varphi ,{\theta _m}} \right)$ represents angular distribution of the input light, where ${\theta _{in}}$ is the incident angle with respect to the optical axis, $\varphi $ is the azimuthal angle and ${\theta _m}$ is the maximum angle against to the incident angle, so the output flux is given by:
\begin{equation}\label{eq11}
\begin{split}
& F\left( {{\theta _{out}},{\theta _{in}}} \right) = \\
& \qquad \int_0^{2\pi } {\int_0^\pi  {G\left( {{\theta _{in}},\varphi ,{\theta _m}} \right)P\left( {{\theta _{out}},{\theta _{in}}} \right)} } \sin {\theta _{in}}d{\theta _{in}}d\varphi
\end{split}
\end{equation}
There are particular cases that if the input light is symmetrical with respect to the fibre axis, the function \emph{G} will be only a function of ${\theta _m}$, and consider the case where the input beam is a uniform cone beam of the cone angle ${\theta _0}$ , then \emph{G} is a step function given by:
\begin{equation}\label{eq12}
G = \left\{ \begin{split}
1,\mathop {}\limits_{} \mathop {}\limits_{} \mathop {}\limits_{} \mathop {}\limits_{} \mathop {{\theta _{in}} < {\theta _0}}\limits_{} \\
0,\mathop {}\limits^{} \mathop {}\limits^{} \mathop {}\limits^{} \mathop {}\limits^{} \mathop {{\theta _{in}} > {\theta _0}}\limits^{}
\end{split} \right.
\end{equation}
Using these formulas to model different phenomena relevant to incident conditions can principally predict the output field and calculate the FRD somehow. And also this model provides a simple way to experimentally measure the parameter ${d_0}$ with the reverse thinking to acquire the output filed firstly and then obtain the \emph{FWHM}, and finally fitting the curve between incident angle and \emph{FWHM} to calculate the best estimated parameter ${d_0}$. Fig.\ref{fig:11} shows the simulated output field of two types of input light of collimated beam and a cone beam respectively. The dashed blue line represents the EE, which can be used to determine the output focal ratio and to calculate the throughput.

In this model, the influence factors for the case of a collimated beam include the fibre core diameter  ${a}$, the microbending parameter ${d_0}$, the wavelength of the injected light $\lambda $, and the length of the fibre $L$. Each of them would lead to different trends of FRD. For wavelength dependence, \citet{Murphy2008Focal} found a weak dependence on FRD but the results showed opposite trend against the predictions of the model, which shows that FRD will increase with the increasing wavelength. But nearly no evidence in experiments shows the length properties which longer fibres make FRD much worse predicted by the model, however, many tests were made at a relatively fast input focal ratio, for example \emph{F}$_{in}$ = 3.5 or much faster, where FRD is relatively inconspicuous compared to slower beams because the output beam approaches an asymptotic value in a fast input focal ratio. While in our lab, the length dependence shows another interesting phenomenon in Sec. 4.2 for input focal ratio at \emph{F}$_{in}$ = 5.0 the same as that in LAMOST and other two input conditions of \emph{F}$_{in}$ = 4.0 and \emph{F}$_{in}$ = 8.0.

\section{Experimental methods and FRD tests}
According to the model description in Sec. 2, the profile and the power distribution of output spot is related to the incident angle and the injecting position on the end-face of the fibre. To optimize the incidence conditions, mainly to reduce the angle misalignment error and to locate the incident point in the centre as much as possible, active optics and adaptive optics are implied in large telescopes to eliminate the gravity and thermal effects and compensate distortion mainly caused by atmosphere turbulence or seeing. So efficiently coupling the light into a fibre is of great importance.

Many groups have reported their experiments on FRD properties, but the results differ from one to another. Even worse, sometimes the results acquired from the same experiment for fibres from the same manufacture and the same type but in different time conflict against each other for the bad reliability and repeatability, most of which were attributed to the poor stability of illumination system and various ways that how they deal with the fibre preparation including fibre end preparation, the mounting method, distribution trace of fibre and so on. In this section we proposed a novel method to contribute to improve the feasibility, reliability and repeatability by introducing \emph{f-intercept} into EDM.

\subsection{Illumination system setup}
Either laser or white light can be used as light source, but the output spots surfer serious laser speckle in laser illumination system, which will affect the profile and is difficult to calculate the diameter of the spots. But laser is convenient for angle control and the waist is much smaller than white light.

\begin{figure*}
   \centering
   \includegraphics[width=\hsize]{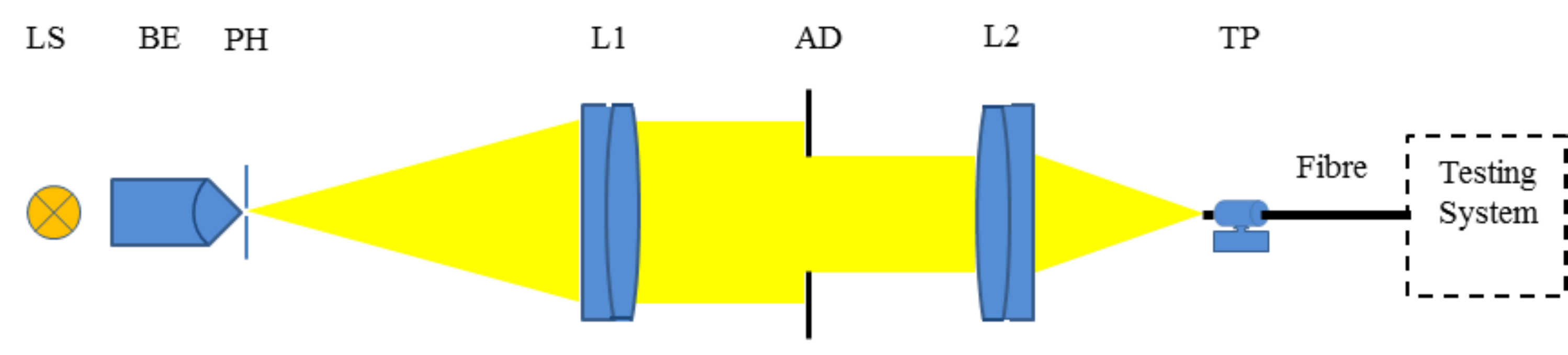}
      \caption{The schematic diagram of illumination system. LS: light source (laser or white light); BE: beam expander. PH: pinhole of 30$\mu $m$\sim $500$\mu$m (replaceable); L1\&L2: apochromatic lens; AD: adjustable diaphragm to control the input focal ratio; TP: transfer platform combined with six-axis translation stage (precision less than 1$\mu$m).}
         \label{fig:12}
   \end{figure*}

First, we should determine the incident condition to satisfy the test system. Fig.\ref{fig:12} shows the experimental illumination system.

\begin{figure}
   \centering
   \includegraphics[width=\hsize]{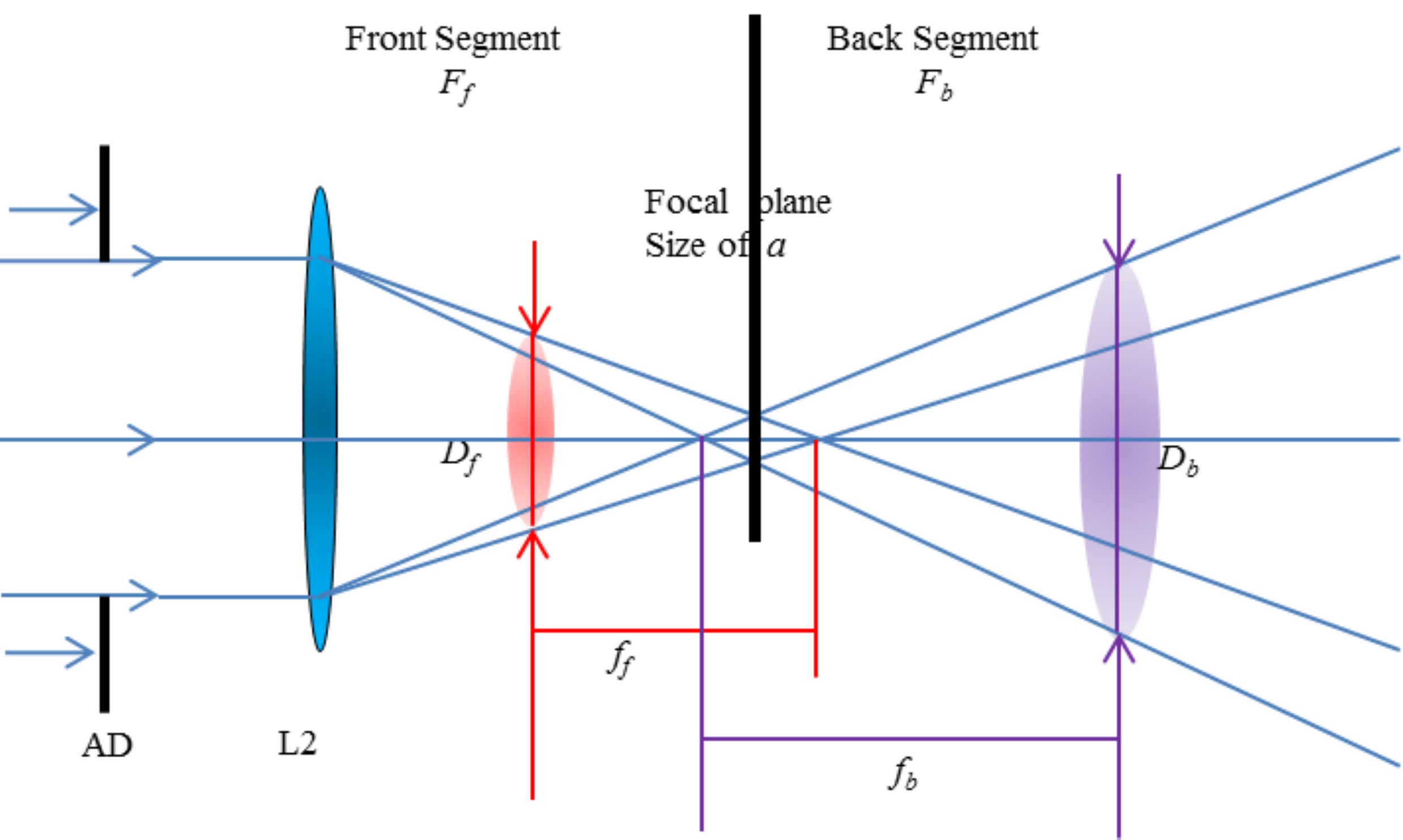}
      \caption{Light field distribution through the imaging system. The light spot is not an ideal pointolite on the focal point and the spot is getting larger when light goes across the focal plane.}
         \label{fig:13}
   \end{figure}

As for the limitation of the experiment devices, when the light goes through the imaging system, there exists a spot of a limited size on the focal plane because of the intrinsic scale of light source, scattering and aberration, and even for the laser system it does too. In the symmetric optical system, the aperture of the imaging spot on the focal plane is approximately the same size as the entrance pinhole as in Fig.\ref{fig:13}.

According to the definition of focal ratio in the fibre, given the focal length of the lens \emph{f} and the aperture size \emph{D} and the size of the light source spot \emph{a} in imaging plane, using triangle geometrical relationship, the output focal ratio can be written as:
\begin{equation}\label{eq13}
\begin{split}
& {F_f} = \frac{{{f_f}}}{{{D_f}}} = \frac{f}{{D - a}}\\
& {F_b} = \frac{{{f_b}}}{{{D_b}}} = \frac{f}{{D + a}}
\end{split}
\end{equation}
So we get the relationship of
 \[{F_f} > {F_b} .\]
The focal ratio in front of the focal point is larger than that of the back segment, which means the distribution of the energy becomes much diffuser along the light goes through the focal point. Leaving out the scattering and aberration, the difference of focal ratio between $F_f$ and $F_b$ can be up to 5\% for a 0.5mm entrance pinhole.

The input light is set to meet the condition as on LAMOST, and the input focal ratio is \emph{F}$_{in}$ = 5.0. The focal length of lens is \emph{f} = 150mm and the aperture of adjustable diaphragm \emph{D} = 30mm. The diameter of the aperture is controlled by the opening and closing of several fan-shaped blades and the accuracy results in an input focal ratio error of about 1\%. The output focal ratio was derived from EMD. An appropriate EE ratio is important to evaluate the performance of FRD \citep{Wang2013The}. We choose EE95 to determine the diameter of spots because at this ratio of EE the results best agreed the setup of illumination system.

\subsection{Energy distribution method and the \emph{f-intercept} (EDMF)}
\begin{figure}
   \centering
   \includegraphics[width=\hsize]{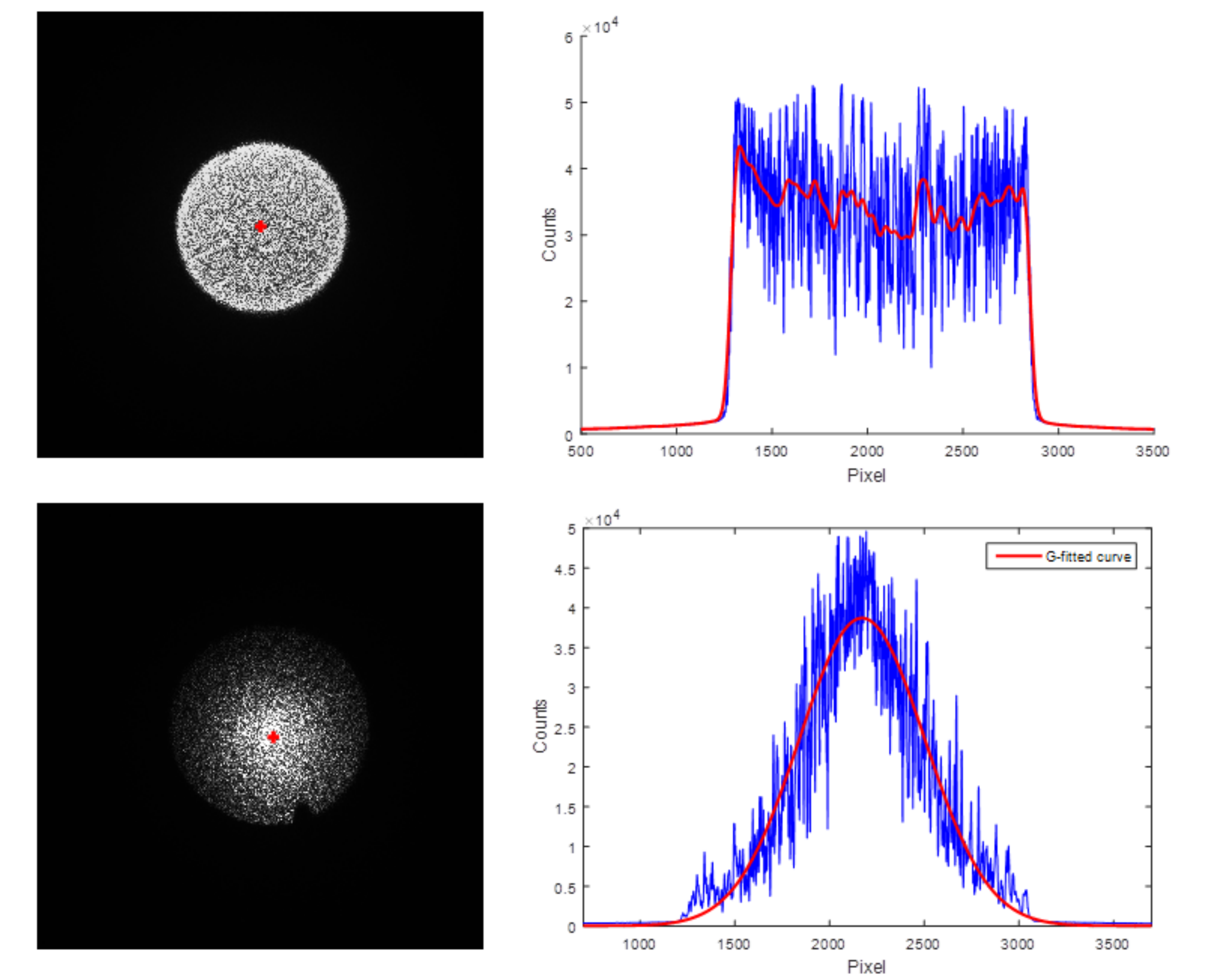}
      \caption{Laser beam as the light source. The speckles make serious fluctuation on the profile.}
         \label{fig:14}
   \end{figure}

\begin{figure}
   \centering
   \includegraphics[width=\hsize]{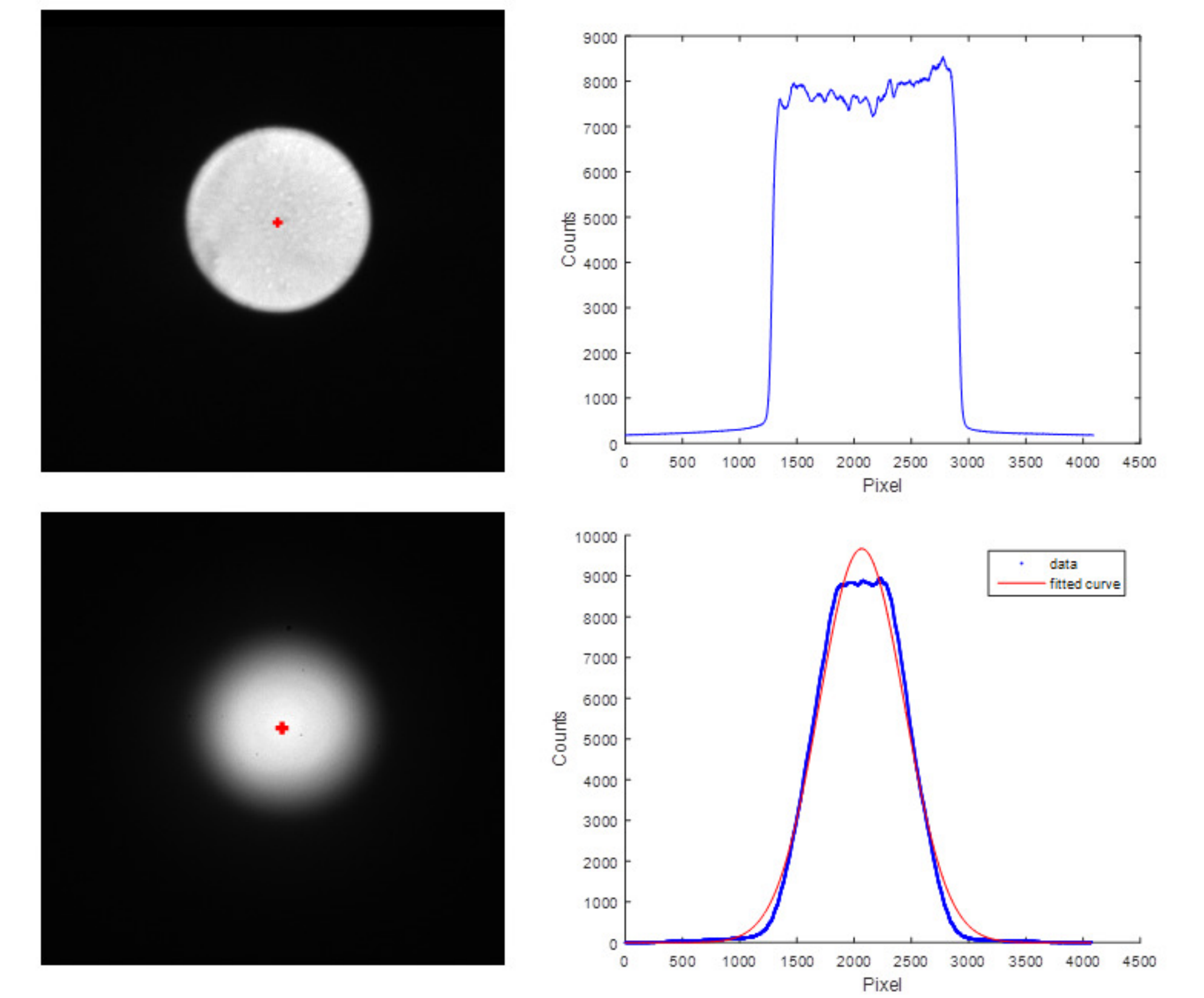}
      \caption{White light (LED) as the light source. The profile is much smoother and the power distribution is uniform in the front segment in the upper image.}
         \label{fig:15}
   \end{figure}

In the experiment, a set of consequent spots were captured by CCD to conduct linear regression to fit the linear relationship between the focal distances and the diameters of output spots to calculate the focal ratio. Fig.\ref{fig:14} and Fig.\ref{fig:15} show the output spots illuminated by laser and white light (emitted by a LED) respectively.

Serious speckle occurs in Fig.\ref{fig:14} and the profile of power distribution has serious fluctuations. When illuminated by LED, the speckle has been well repressed. The two dimensional fitting curves of both spots captured in front segment show an approximate rectangle distribution, which means the beam in front segment are likely to be an uniform cone beam. But in the back segment, the light field tends to be a Gaussian beam, meaning that the power of light has been redistributed when go across the focal plane.

\begin{figure}
   \centering
   \includegraphics[width=\hsize]{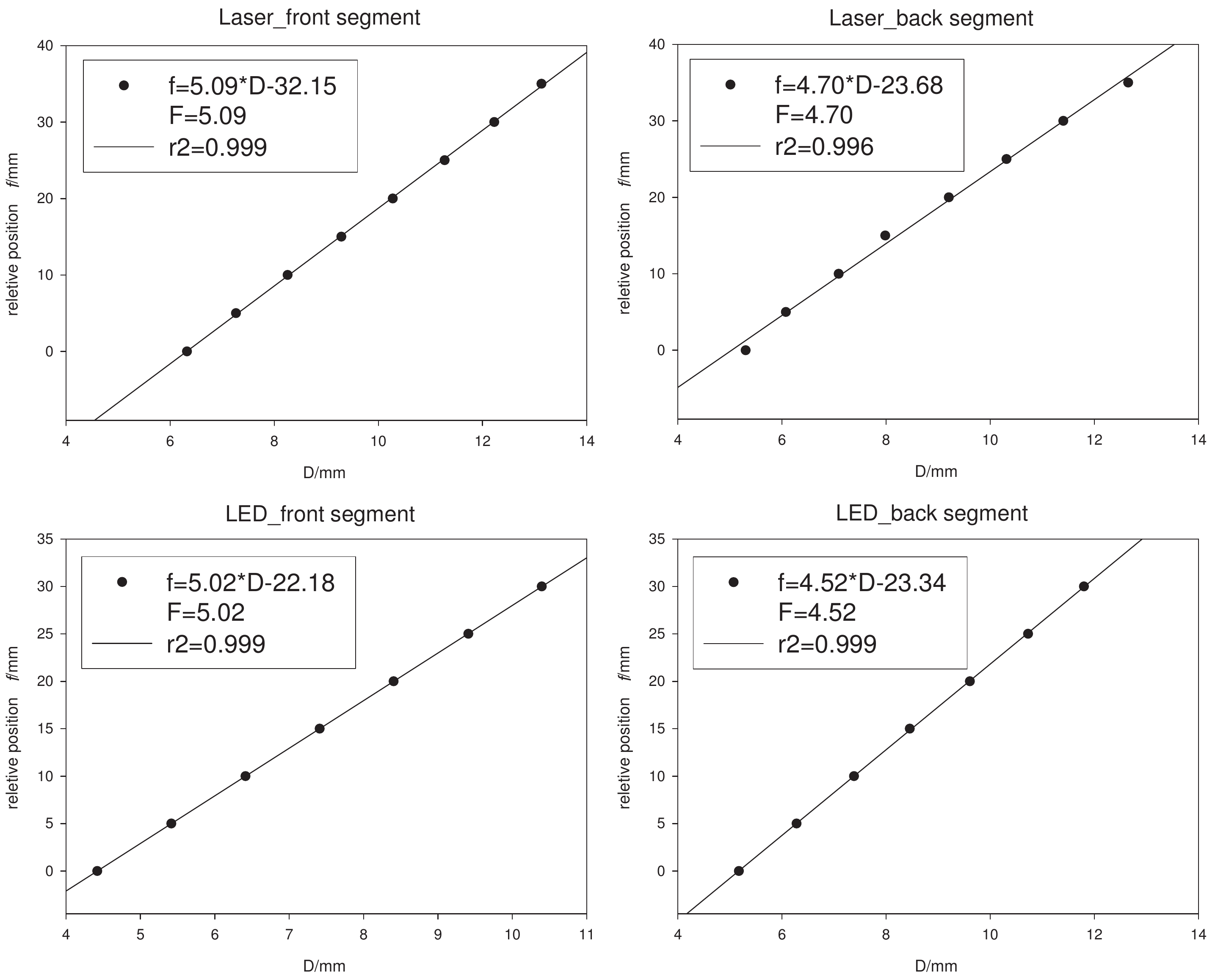}
      \caption{The slope of fitting curve represents the focal ratio according to EDM. \emph{r}$^2$ is the coefficient of determination (0 < $r^2$ < 1, the larger the better), an indicator to evaluate the linearity between the original data and the fitting curve.}
         \label{fig:16}
   \end{figure}

Energy distribution method \citep{Murphy2008Focal,Xue2013LAMOST} is fit for our experiment to process the CCD data for its convenience that there is no need to measure the precise distance of the initial position where CCD captures the first output spot. And the increment of relative position represents the shift in distance with respect to the initial position where is set to be zero point in the fitting curve. Fig.\ref{fig:16} are the fitting curves drew according to EDM, from which we can acquire the slope of the fitting curve standing for focal ratio and can also get the \emph{f-intercept} containing the information of original position of the focal point. The results are assembled in Table \ref{tab:1}. IP stands for the initial position of the transfer platform and FP is the actual position of focal plane calculated according to the \emph{f-intercept} and IP.

\begin{table*}
\caption{Parameters calculated according to EDM for laser and LED illumination system}             % title of Table
\begin{threeparttable}[b]
\label{tab:1}      % is used to refer this table in the text
\centering                          % used for centering table
\begin{tabular}{c c c c c c c}        % centered columns (4 columns)
\hline                 % inserts double horizontal lines
Light source & & Focal ratio & \emph{r}$^2$ & \emph{f-intercept}/mm & \tnote{1} IP/mm & \tnote{2} FP/mm \\    % table heading
\hline
%   \multirow{14}{*}{Laser} & \multirow{6}{*}{\tabincell{c}{Front \\ segment}} & 5.09 & 0.999 & -32.15 & 70.0 & 102.15 \\
   \multirow{14}{*}{Laser} & \multirow{6}{*}{Front segment} & 5.09 & 0.999 & -32.15 & 70.0 & 102.15 \\
   & & 5.11 & 0.999 & -32.19 & 70.0 & 102.19 \\
   & & 5.08 & 0.999 & -32.18 & 70.0 & 102.18 \\
   & & 5.13 & 0.999 & -34.13 & 68.0 & 102.13 \\
   & & 5.05 & 0.999 & -42.27 & 60.0 & 102.27 \\
   & & 5.10 & 0.999 & -44.35 & 58.0 & 102.35 \\
   & Average & 5.09$\pm$0.02 & & & & 102.20 \\
\cline{2-7}
%   & \multirow{6}{*}{\tabincell{c}{Back \\ segment}} & 4.70 & 0.996 & -23.68 & 126.0 & 102.32 \\
   & \multirow{6}{*}{Back segment} & 4.70 & 0.996 & -23.68 & 126.0 & 102.32 \\
   & & 4.65 & 0.995 & -23.55 & 126.0 & 102.45 \\
   & & 4.68 & 0.995 & -23.64 & 126.0 & 102.36 \\
   & & 4.55 & 0.993 & -27.53 & 130.0 & 102.47 \\
   & & 4.66 & 0.996 & -32.64 & 135.0 & 102.36 \\
   & & 4.71 & 0.995 & -37.61 & 140.0 & 102.39 \\
   & Average & 4.66$\pm$0.05 & & & & 102.41 \\
\hline
%   \multirow{14}{*}{LED} & \multirow{6}{*}{\tabincell{c}{Front \\ segment}} & 5.02 & 0.999 & -22.18 & 80.0 & 102.18 \\
   \multirow{14}{*}{LED} & \multirow{6}{*}{Front segment} & 5.02 & 0.999 & -22.18 & 80.0 & 102.18 \\
   & & 5.00 & 0.998 & -22.00 & 80.0 & 102.00 \\
   & & 5.00 & 0.997 & -22.03 & 80.0 & 102.03 \\
   & & 4.97 & 0.999 & -31.86 & 70.0 & 101.86 \\
   & & 5.07 & 0.998 & -32.35 & 70.0 & 102.35 \\
   & & 4.97 & 0.998 & -31.97 & 70.0 & 101.97 \\
   & Average & 5.01$\pm$0.03 & & & & 102.06 \\
\cline{2-7}
%   & \multirow{6}{*}{\tabincell{c}{Back \\ segment}} & 4.52 & 0.999 & -23.34 & 126.0 & 102.66 \\
   & \multirow{6}{*}{Back segment} & 4.52 & 0.999 & -23.34 & 126.0 & 102.66 \\
   & & 4.58 & 0.983 & -23.17 & 126.0 & 102.83 \\
   & & \tnote{*} 4.78 & \tnote{*} 0.981 & \tnote{*} -24.72 & \tnote{*} 126.0 & \tnote{*} 101.38 \\
   & & 4.48 & 0.992 & -32.52 & 135.0 & 102.48 \\
   & & 4.46 & 0.981 & -37.81 & 140.0 & 102.19 \\
   & & 4.42 & 0.986 & -42.50 & 145.0 & 102.50 \\
   & Average & 4.50$\pm$0.05 & & & & 102.53 \\
\hline                                   %inserts single line
\end{tabular}
\begin{tablenotes}
  \item Notes:
  \item[1] IP: The initial position of the transfer platform.
  \item[2] FP: The actual position of focal plane according to the \emph{f-intercept} and IP.
  \item[*] '*': Data with '*' might be seriously influenced by background or ambient light to be an outlier.
\end{tablenotes}
\end{threeparttable}
\end{table*}

   \begin{figure}
   \centering
   \includegraphics[width=\hsize]{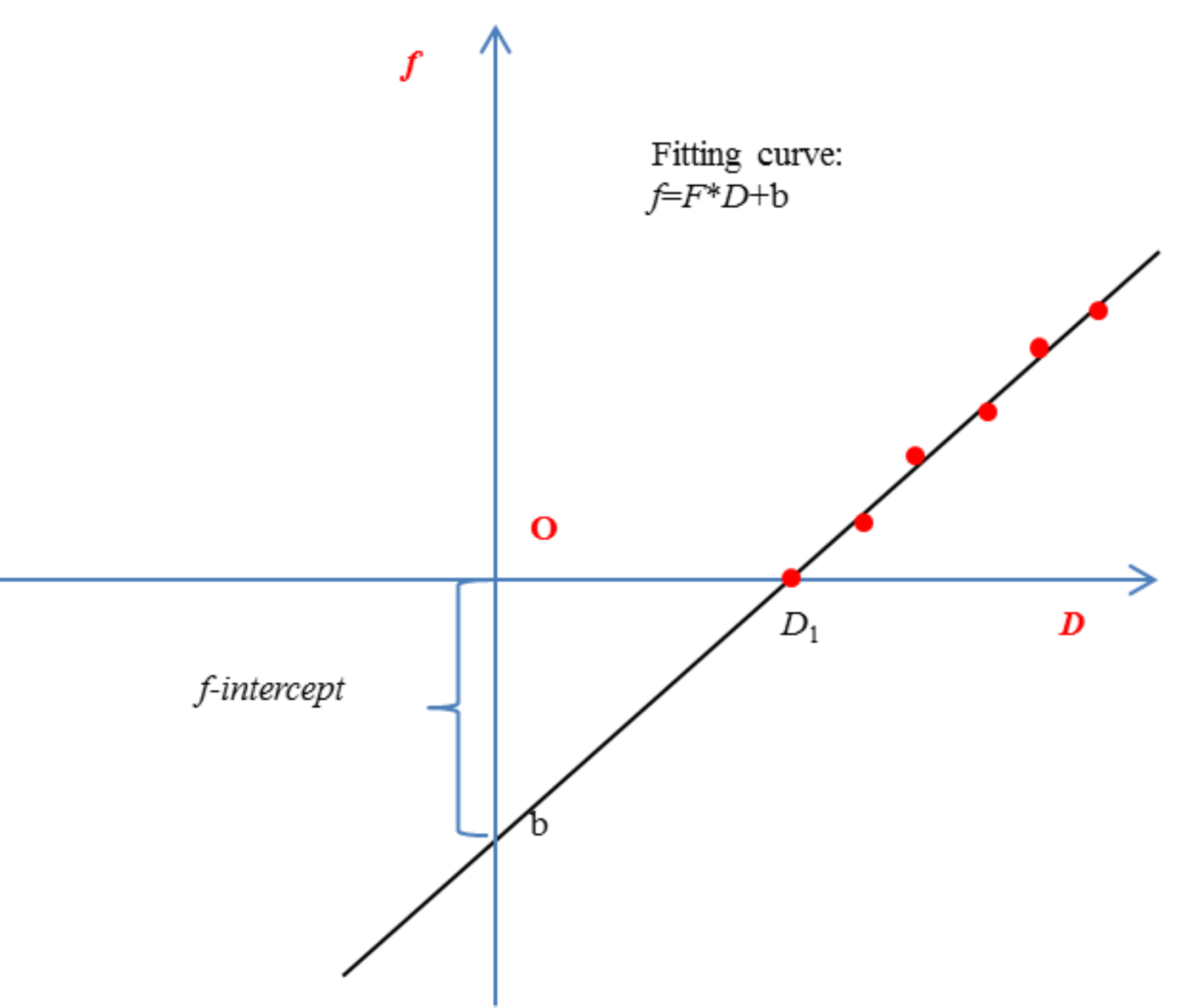}
      \caption{Fitting process of EDM. The convenience of EDM is that there is no need to measure the actual distance between the CCD camera and the focal point. But it brings the uncertainties to the measured focal ratio since we are unaware of the accurate position where is the focal point of light source system. According to the regression equation of fitting curve, we can acquire the \emph{f-intercept} to reveal the information of focal point.}
         \label{fig:17}
   \end{figure}

   \begin{figure}
   \centering
   \includegraphics[width=\hsize]{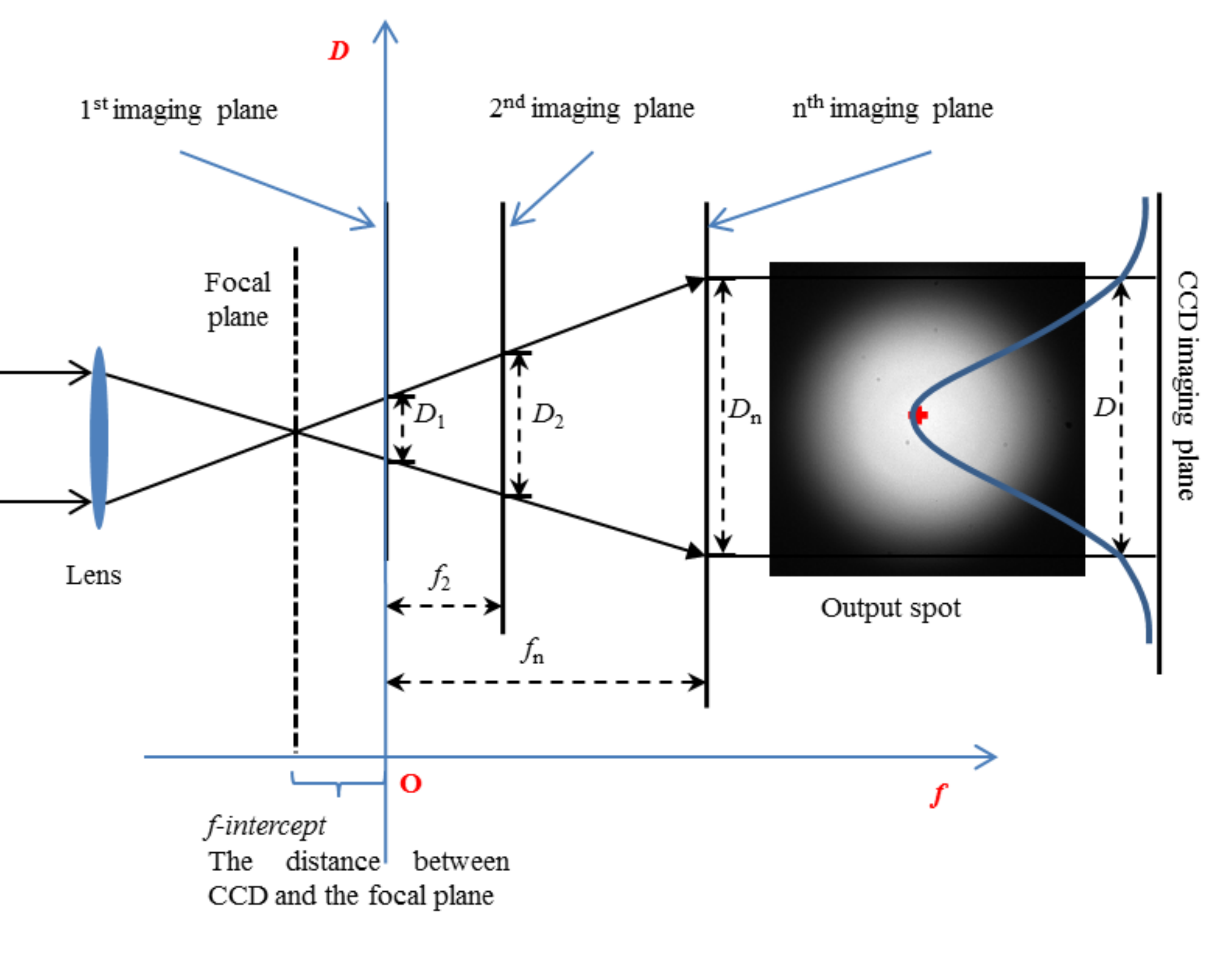}
      \caption{The process of capturing the spots. Each time the distance between the CCD camera and the initial position was measured and the initial position was set to zero point. The focal ratio is calculated with the difference of relative position and the spot's diameter. Theoretically, the \emph{f-intercept} should be all the same to determine the actual focal point.}
         \label{fig:18}
   \end{figure}

Fig.\ref{fig:17} and Fig.\ref{fig:18} show the process of EDM and the physical meaning of \emph{f-intercept}. A series of output spots were captured in different positions and the slope of fitting curve can indicate the focal ratio and the fitting constant \emph{b} is the \emph{f-intercept} to estimate the real position of focal point or focal plane. In Fig.\ref{fig:18}, the distance \emph{f} starts from \emph{f}$_1$ = 0, and then we only need to measure the increment of relative distance \emph{f}$_2$ from the next position to the initial position and so forth. So the actual position of the focal plane will be a negative value and has a shift in distance of \emph{f-intercept} compared to the first measurement position. With this relationship we can conform the focal point easily to guarantee the incident condition unchanged when couple the beam into fibre always at the same place to improve the reliability and repeatability.

From the results we can see that the focal ratio in the front segment of illumination system is larger and more acceptable and agrees to the pre-set value better compared to that in the back segment. Considering the changes of focal ratio, size of light source, aberration, redistribution and many other factors, the power distribution is more diffuser in the back segment and input focal ratios will be different when the light couples into fibre in different places, especially a faster focal ratio occurs in the back segment. In both types of illumination systems (laser and LED), the focal ratio is more stable in the front segment for a smaller relative error. The focal ratio is closer to the pre-set value using LED as the light source, and a probable reason is that the speckle in laser illumination system leads to an unstable bias in calculating the diameters of spots. The fall of power distribution between the nearest pixels can be very huge in the speckle with serious fluctuation, so the sum of encircled energy would suddenly jump to a unstable high level or just stay the same without any change when the variable diameter in the program reaches to the next pixel in the process of EDM, thus making the diameter estimation to be biased. But even for the laser condition, still the stability is better in the front segment. As for LED, there is no worry about the speckle, so the light field is very smooth and the growth of encircled energy is steady, and the possibility of jump in the diameters of spots is significantly reduced. But there are also some other factors we should take into consideration. In LED illumination system, the power of light is relatively low, so the system is more sensitive to the ambient light and the noise, especially in the back segment where the power distribution is weaker and diffuser, enhancing the sensitivity. The data with the symbol '*' in Table \ref{tab:1} for instance might be influenced by background and noise to be an outlier.

We have acquired the \emph{f-intercept} and the initial position of transfer platform in Table \ref{tab:1}, and the estimated actual position of focal plane is located in the range of 102.20$\sim$102.41mm for laser smaller than that of 102.06$\sim$102.53mm for LED, which also means the properties of point source of laser is better than LED, so the laser is usually used as a point source to conveniently control the incident position. Then we can adjust the transfer platform to fix at the estimated position of focal point to mount the fibre which will make sure the input focal ratio the same as pre-set value and in this way to guarantee the consistency of incident condition.

\subsection{Fibre tests and FRD properties}
\begin{figure}
   \centering
   \includegraphics[width=\hsize]{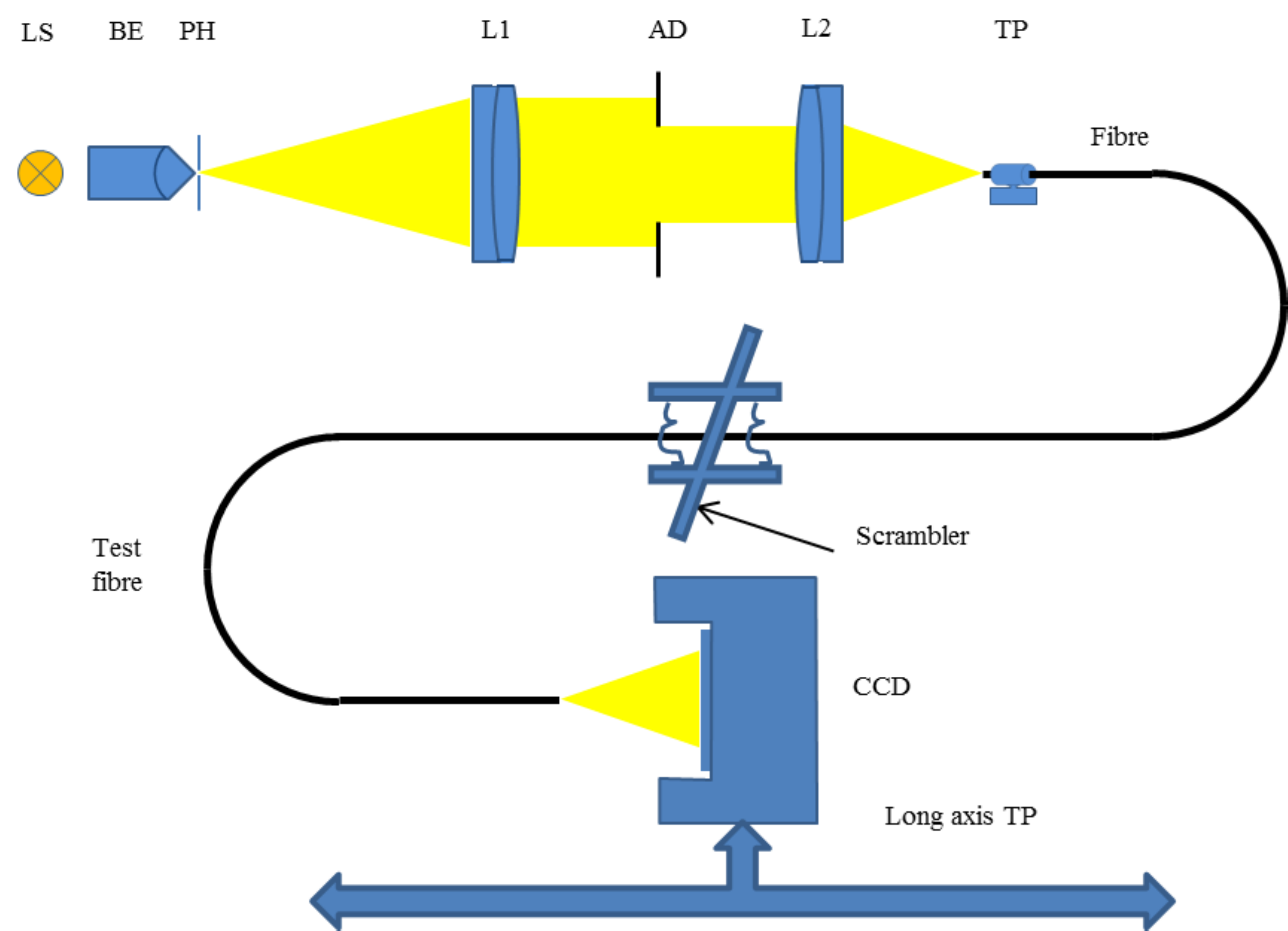}
      \caption{The schematic diagram of the fibre testing setup. LS: light source (laser or white light); BE: beam expander. PH: pinhole of 200$\mu$m; L1\&L2: apochromatic lens; AD: adjustable diaphragm to control the input focal ratio; TP: transfer platform combined with six-axis translation stage (precision less than 1$\mu$m). The light source can be laser or LED and the scrambler is to repress speckle to smooth the output spots for laser illumination system.}
         \label{fig:19}
   \end{figure}

We've tested the focal ratio in the front segment and back segment of illumination system. And the \emph{f-intercept} can determine the focal point where the fibre should be mounted to couple light into. Similarly, the focal-plane plate in the telescope is like the fibre end, and to regulate the position of each fibre end on the plate is not only to aim the celestial body but also to find the best incident point. And this experiment will testify the optimized method for fibres to efficiently improve the FRD performance and the throughput. The experiment system is shown in Fig.\ref{fig:19}.

\begin{figure}
   \centering
   \includegraphics[width=\hsize]{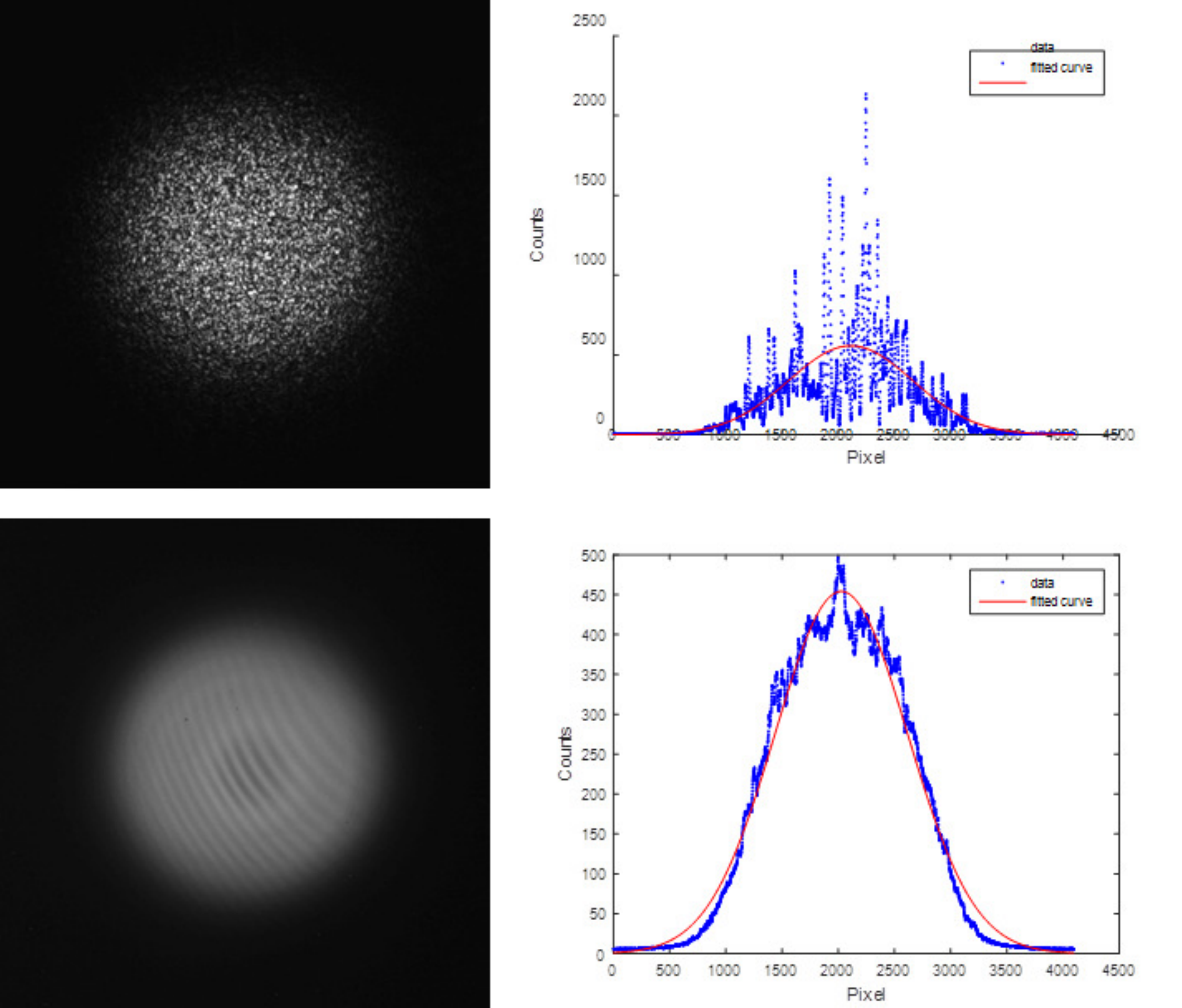}
      \caption{Output spots captured by CCD. Serious speckle exists without scrambling in the upper image and a smooth spot is better after scrambling for determination of the diameter. The fringe pattern in the lower image is probably caused by the detector window which is also noticed by \citet{Reynolds2014A}.}
         \label{fig:20}
   \end{figure}

Tests were carried out on a type of 320$\mu$m core fibre which is currently used in LAMOST, and another type of common used fibre of 200$\mu$m core. The parameters of fibres are in Table \ref{tab:2}. Since the laser speckle in the fibre has great influence on the profile and the energy distribution of the output spot, a vibrating mode scrambler is introduced into the system to repress the speckle and smooth the spot. In this way the diameter of the spot can be calculated more precisely \citep{Sun2014Mode}. Fig.\ref{fig:20} shows the power distribution of output spots for laser system. The grains are obvious and the fluctuation of energy varies greatly without scrambling. A well scrambled spot has good degree of fitting to a Gaussian spot.

\begin{table}
\caption{Parameters of testing fibres}             % title of Table
\label{tab:2}      % is used to refer this table in the text
\centering                          % used for centering table
\begin{tabular}{c c c c}        % centered columns (4 columns)
\hline                 % inserts double horizontal lines
Core/$\mu$m & \emph{N.A.} & Limited \emph{F}$_{in}$ & Length/m \\    % table heading
\hline
   320 & 0.22$\pm$0.02 & $\ge$2.45 & $\sim$4.0 \\
   200 & 0.22$\pm$0.02 & $\ge$2.45 & $\sim$4.0 \\
\hline                                   %inserts single line
\end{tabular}
\end{table}

\begin{figure}
   \centering
   \includegraphics[width=\hsize]{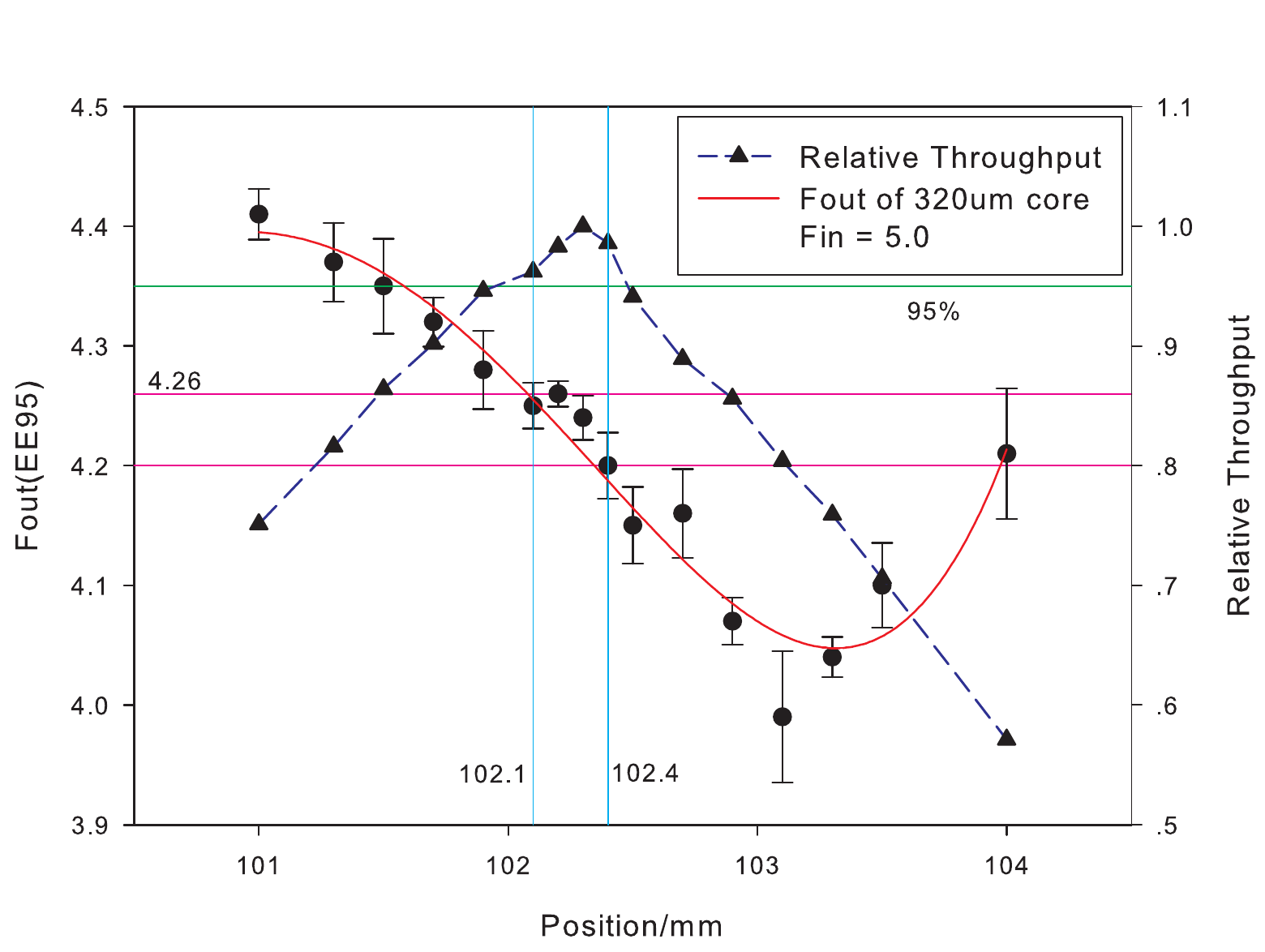}
      \caption{Output focal ratio and relative throughput vs position. The positions between 102.1mm and 102.3mm are promising choices for fibre test with the focal ratio difference of 0.02 and the relative throughput up to 96\%.}
         \label{fig:21}
   \end{figure}

\begin{figure}
   \centering
   \includegraphics[width=\hsize]{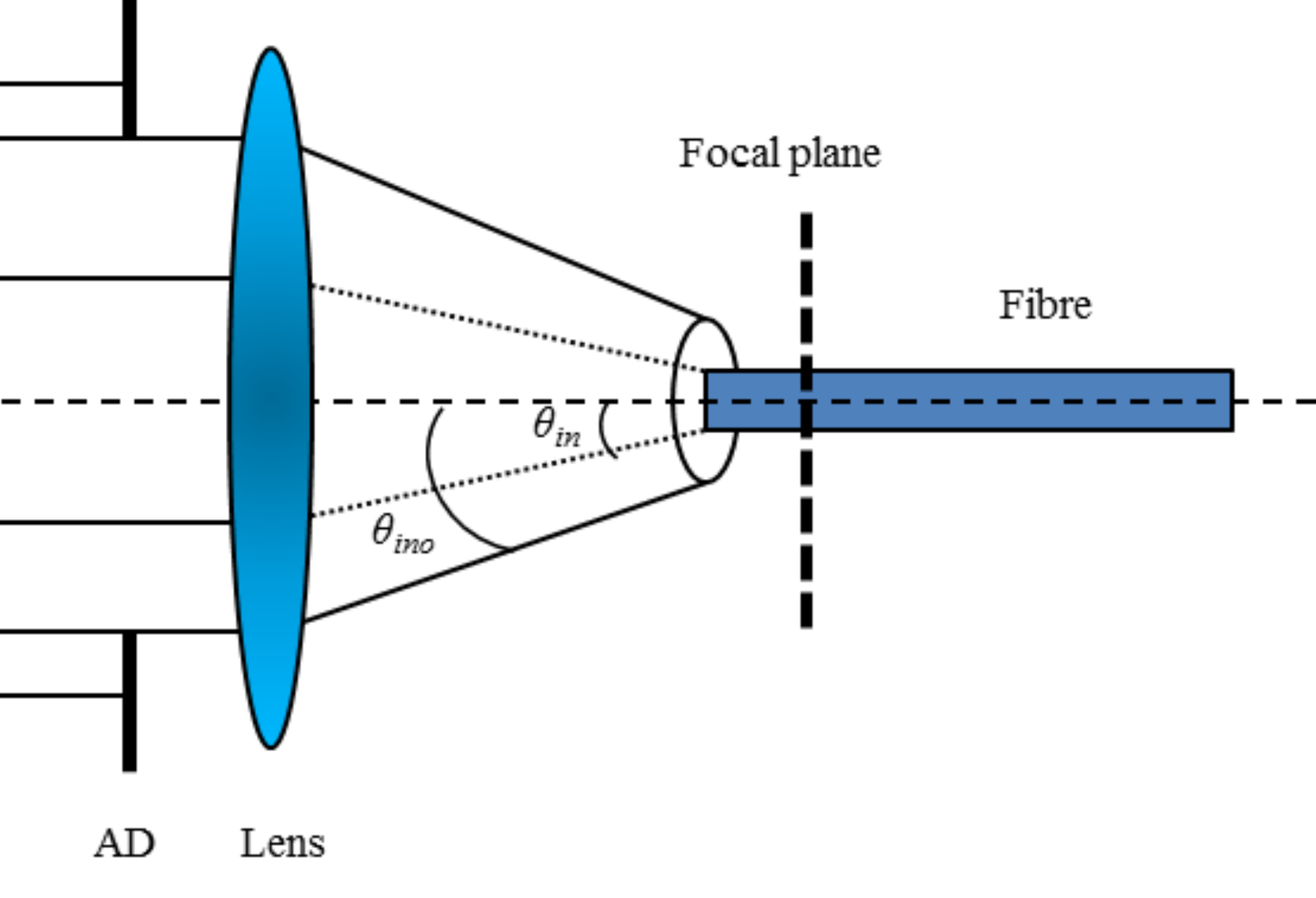}
      \caption{The fibre end is out of focus in the front segment far deviate from the focal plane. Under this condition, the actual incident angle $\theta _{in}$ is smaller than the pre-set value of $\theta _{ino}$, so the actual input focal ratio becomes larger. But in the meantime, the energy is less injected into the fibre.}
         \label{fig:22}
   \end{figure}

To avoid the subsidiary stress effect from the fibre fixture, the fibre tube is chosen instead of the squeeze type jig to mount the fibre, which has been tested previously to prove no evident influence on FRD. The FRD performance was tested in both laser and LED illumination systems with input focal ratio of \emph{F}$_{in}$ = 5.0. Different places were chosen in the range of 101.0$\sim$104.0mm which overlap the whole range of distances estimated by \emph{f-intercept}. From the results shown in Fig.\ref{fig:21} and Table \ref{tab:3}, the closer the fibre is to the focal point, the better the FRD performance is and the variability of output focal ratio is smaller and the degree of linearity of fitting curve is better. In the relative far distance away from the focal point in both left and right sides, the variabilities of output focal ratio become unstable though the output focal ratio is larger in left side close to the lens which seems to be much better FRD performance, while in fact the relative throughput is lower. These phenomena may be caused by the different power distribution and the actual input focal ratio of inner and outer of the focal point region. When the fibre was mounted out of focus in the back segment, only part of light can couple into fibre. As the input power distribution is not uniform, power of different modes would change along with the rays going cross each other, the incident angles spread so different in a large range that more of higher modes would be stimulated according to the principle of guided mode, so the output light field is diffuser and serious FRD occurs. If the fibre was mounted on the left side in the front segment close to the lens, the output focal ratio increases because the actual input focal ratio is larger as shown in Fig.\ref{fig:22}, and the total energy coupling into the fibre decreases at the same time. To determine the best position is very important for fibre testing and also has great influence on the throughput and observing efficiency in telescopes.

\begin{table*}
\caption{Output focal ratio in different positions (partially illustrated)}             % title of Table
\label{tab:3}      % is used to refer this table in the text
\centering                          % used for centering table
\begin{tabular}{c c c c c c c c c c c}        % centered columns (4 columns)
\hline                 % inserts double horizontal lines
   \multirow{2}{*}{Position} & \multicolumn{2}{c}{102.1mm} & \multicolumn{2}{c}{102.2mm} & \multicolumn{2}{c}{102.3mm} & \multicolumn{2}{c}{102.4mm} & \multicolumn{2}{c}{102.5mm} \\    % table heading
\cline{2-11}
   & \emph{F$_{out}$} & r$^2$ & \emph{F$_{out}$} & r$^2$ & \emph{F$_{out}$} & r$^2$ & \emph{F$_{out}$} & r$^2$ & \emph{F$_{out}$} & r$^2$ \\
\hline
%   \multirow{6}{*}{\tabincell{c}{Output \\ focal \\ ratio}} & 4.27 & 0.998 & 4.26 & 0.999 & 4.22 & 0.999 & 4.22 & 0.996 & 4.13 & 0.988 \\
    \multirow{6}{*}{Output focal ratio} & 4.27 & 0.998 & 4.26 & 0.999 & 4.22 & 0.999 & 4.22 & 0.996 & 4.13 & 0.988 \\
    & 4.25 & 0.999 & 4.25 & 0.998 & 4.22 & 0.999 & 4.23 & 0.998 & 4.13 & 0.998 \\
    & 4.22 & 0.992 & 4.25 & 0.999 & 4.24 & 0.994 & 4.18 & 0.999 & 4.18 & 0.997 \\
    & 4.27 & 0.991 & 4.27 & 0.992 & 4.24 & 0.998 & 4.17 & 0.995 & 4.21 & 0.999 \\
    & 4.26 & 0.996 & 4.26 & 0.999 & 4.26 & 0.993 & 4.24 & 0.996 & 4.15 & 0.996 \\
    & 4.23 & 0.999 & 4.28 & 0.996 & 4.27 & 0.996 & 4.18 & 0.999 & 4.12 & 0.997 \\
    Average & \multicolumn{2}{c}{4.25$\pm$0.02} & \multicolumn{2}{c}{4.26$\pm$0.01} & \multicolumn{2}{c}{4.24$\pm$0.02} & \multicolumn{2}{c}{4.20$\pm$0.03} & \multicolumn{2}{c}{4.15$\pm$0.03} \\
\hline                                   %inserts single line
\end{tabular}
\end{table*}

The stability of laser test in a period of two hours indicated a good reliability with the power error of 3.2\%. The output energy of fibre was valued by the integrated grey value of all the encircled pixels of EE95 and was normalized to compare with each other as the relative throughput. The output focal ratio decreases from 4.41 down to 3.99 along with the fibre position goes across the focal point from front segment to back segment. The relative throughput reaches to the peak at the position of 102.2mm. The results show that the positions between 102.1mm and 102.3mm are promising choices for fibre test with the focal ratio difference of 0.02. Making this comprehensive consideration, the appropriate position for FRD properties test is chosen at 102.2$\pm$0.01mm and the relative throughput is more than 96\%. Under this condition, experiments on fibres of 320$\mu$m core and 200$\mu$m core were implemented with LED illumination system.

   \begin{figure}
   \centering
   \includegraphics[width=\hsize]{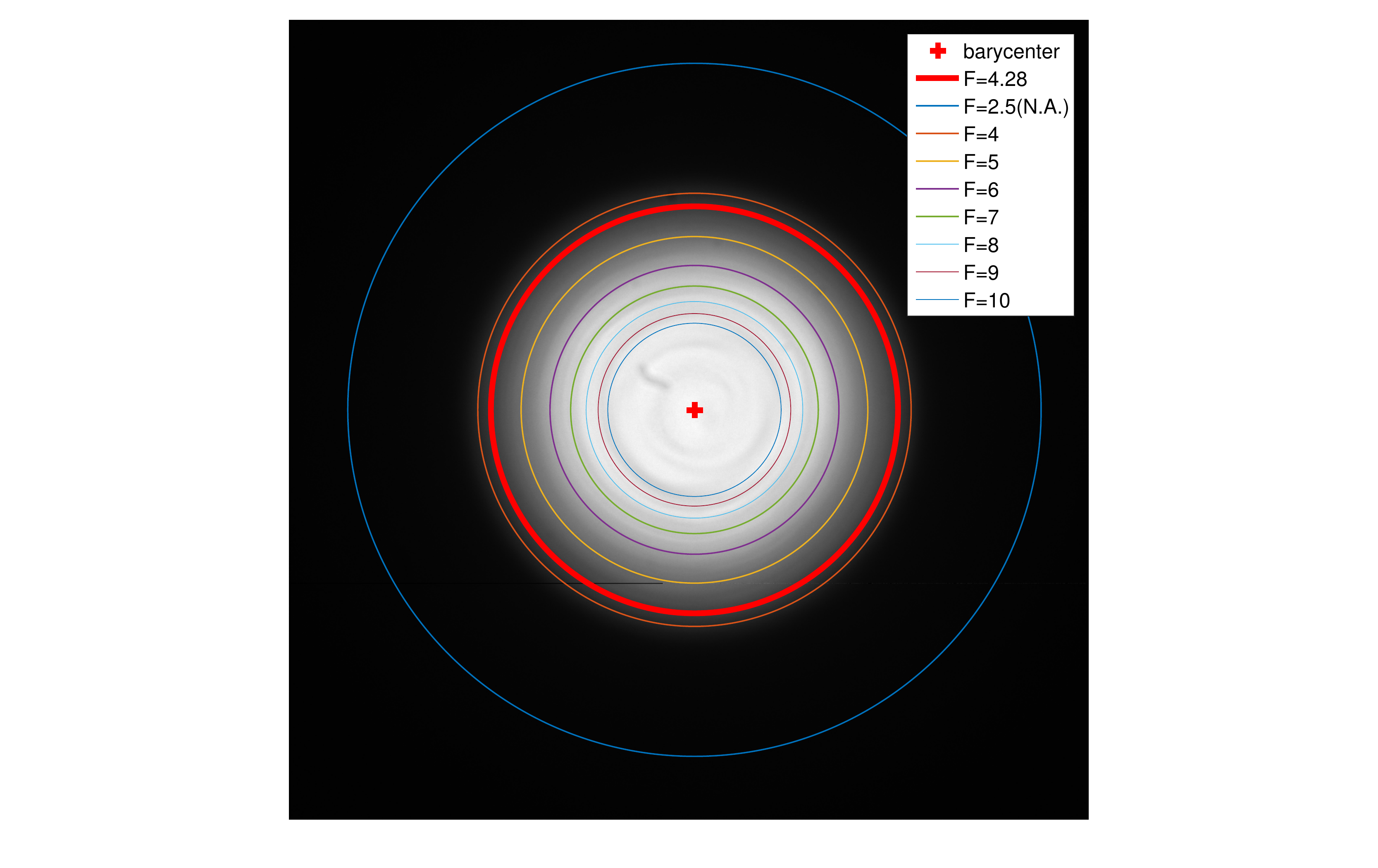}
      \caption{Mathematic used to obtain the EE ratio of the spot from the fibre end. Encircled energy by several rings in the image of output spot is integrated for different aperture defined by focal ratio. The thick red circle is the measured focal ratio of 4.28 under EE95. The outer blue circle covers the whole spot to collected all the energy at \emph{F} = 2.5 close to \emph{N.A.} = 0.20.}
         \label{fig:23}
   \end{figure}

The curve in Fig.\ref{fig:23} is used to define the efficiency over a range of focal ratio at the exit of the fibre, where each focal ratio value contains the summation of partial energy emergent from the fibre. The limited focal ratio that can propagate in the tested fibre is approximately \emph{F} = 2.45 considering the \emph{N.A.} of this fibre to be \emph{N.A.} = 0.22$\pm$0.02. Therefore, we have defined \emph{F} = 2.5 to be the outer limit (region of integration) of the external annulus within which all of the light from the test fibre will be collected which is enough for energy integration and can reduce the influence of the noise of ambient light and dark current.

   \begin{figure}
   \centering
   \includegraphics[width=\hsize]{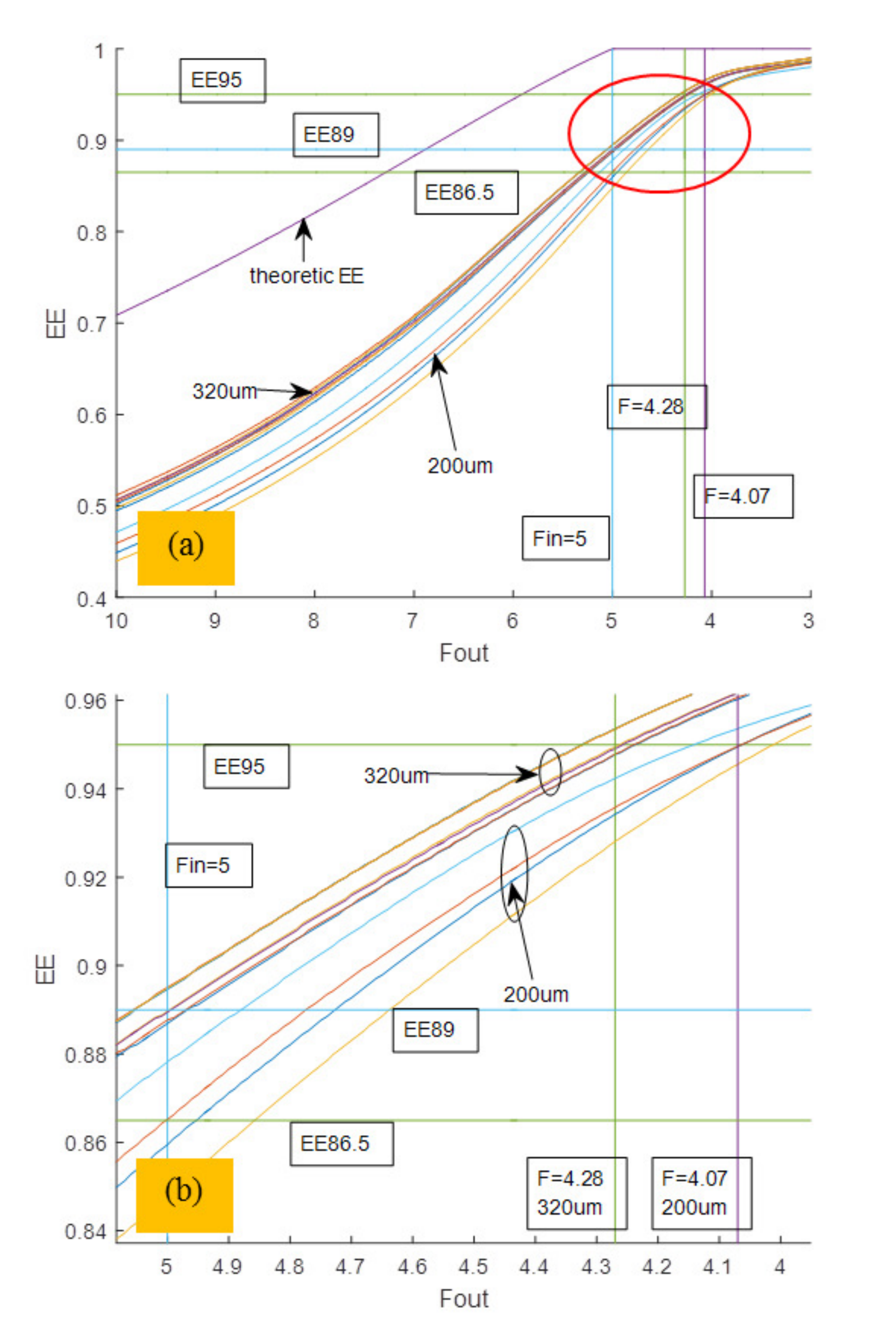}
      \caption{Curves show the relationship between EE ratio and output focal ratio for fibres of 320$\mu$m core and 200$\mu$m core. The red circled region in (a) is shown magnified as in (b).}
         \label{fig:24}
   \end{figure}

The red thick line in Fig.\ref{fig:23} is the diameter of EE95, and the output focal ratio is \emph{F}$_{out}$ = 4.28. For the aperture of \emph{F} = 3, most of the light has been collected with the throughput around EE = 97.8\% for the fibre of 320$\mu$m core in Fig.\ref{fig:24}. The diameter of the aperture decreases with the increasing focal ratio and the differences of the aperture size become smaller and smaller. When the focal ratio is set to \emph{F} > 7, the aperture only changes a little but the throughput decreases from 68\% (\emph{F}$_{out}$ = 7) down to 51\% (\emph{F}$_{out}$ = 10), so the FRD can significantly affect the energy usage efficiency. This experiment was carried out with LED illumination and the output focal ratio is \emph{F}$_{out}$ = 4.28 similar to that in the laser illumination system and the relative error is very small in multiple measurements, indicating that the reliability and repeatability are satisfying. According to the theoretic EE curve, the throughput should be EE = 100\% at \emph{F}$_{out}$ = 5.0 the same as input focal ratio, but for the existence of FRD, the throughput is much smaller (EE89 for 320$\mu$m, EE86.5 for 200$\mu$m). We also notice that the EE ratio can affect the relative error of measurements in output focal ratio. In lower EE ratio region, the relative errors between multiple measurements become larger than that in higher EE ratio region and both of fibres performance good stability when the EE ratio reaches to 95\%. So choosing EE95 as the criteria is fit for investigating the FRD properties in our system. So introducing \emph{f-intercept} into EDM is a promising method to improve the reliability and precision of experiments and this method can be applied to adjust the position of the focal-plane plate in the telescope.

\section{Modification of PDM (MPDM) and length properties of FRD}
In this chapter we proposed a modified method MPDM to optimize the model due to an additional dimension of parameter \emph{f} (focal distance: the distance between the fibre end and the CCD) was measured and introduced into PDM model. By combining the experimental data and the MPDM, we are able to acquire the parameter ${d_0}$, with which the length properties can be predicted and compared with experimental results.

\subsection{MPDM}
The output light field is mainly determined by one parameter ${d_0}$ in original PDM model. The parameter ${d_0}$ in Eq.(\ref{eq2}) can be considered as microbending factor, while in our experiments, it is better to treat it as the eigenvalue to describe the power of scattering or the defects of fibre of which the microbending is certainly included and considered. Anyhow, the parameter ${d_0}$ can impact the transmission performance.

\begin{figure}
   \centering
   \includegraphics[width=\hsize]{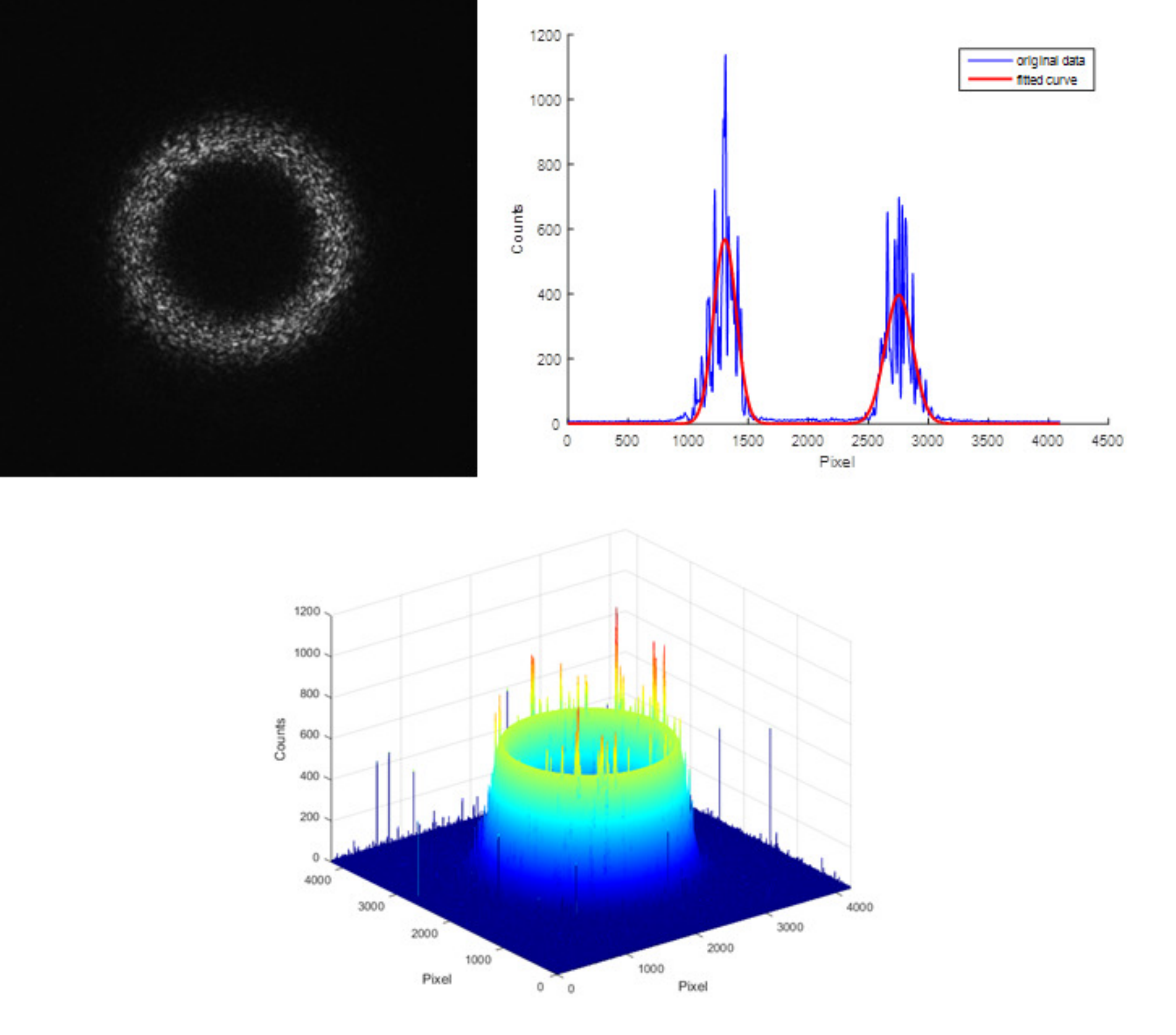}
      \caption{The 2-D fitting curve and 3-D fitting surface of laser illumination system. The 2-D profile is asymmetry in one direction because of the serious speckle. Under this condition, the calculation of \emph{FWHM} will be affected and biased if we only consider a single direction. Even the scrambler cannot make it uniform in every direction but it can smooth the large peaks and deep valleys. A viable way is to fit the ring in a three dimensional space to reduce the influence caused by speckle to eliminate the random errors and the selection effect.}
         \label{fig:25}
   \end{figure}

The collimated laser beam technique in Fig.\ref{fig:9} in Sec. 2.3 to measure parameter ${d_0}$ applied in our experiment for a long fibre of 25m is presented in CP94 and Fig.\ref{fig:25} shows the output ring, the two dimensional curve across the barycentre and the three dimensional fitting surface which can eliminate the biased error caused by asymmetry.

   \begin{figure}
   \centering
   \includegraphics[width=\hsize]{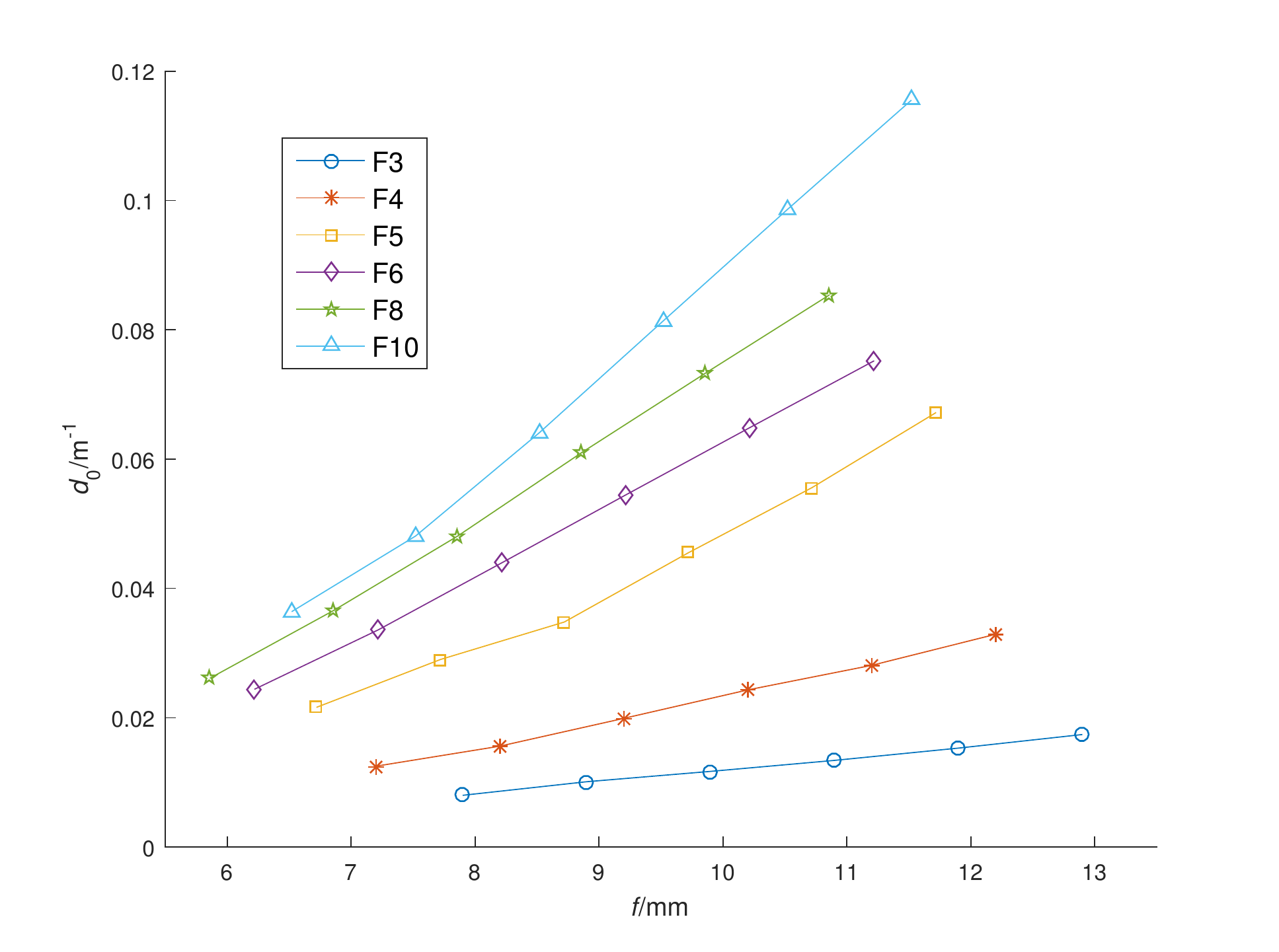}
      \caption{Parameter ${d_0}$ calculated by original PDM model. The results cannot give a stable value and differ with each other. However the increasing values indicate the positive correlation with measurement positions.}
         \label{fig:26}
   \end{figure}

The data in Fig.\ref{fig:26} is derived from the original PDM model. Adopting original PDM to calculate the parameter ${d_0}$ cannot get a stable value and it differs with each other in different measurement positions. The original PDM model in Eq.(\ref{eq8}) and Eq.(\ref{eq9}) merely relates the output angle to parameter ${d_0}$ but it does not explain how the position of output field, i.e. the focal distance \emph{f} between the output field to fibre end, will affect the measurements. Therefore we introduce a new factor, the focal distance \emph{f}, into PDM to modify the model to be more appropriate for similar measurement techniques like ours.

   \begin{figure}
   \centering
   \includegraphics[width=\hsize]{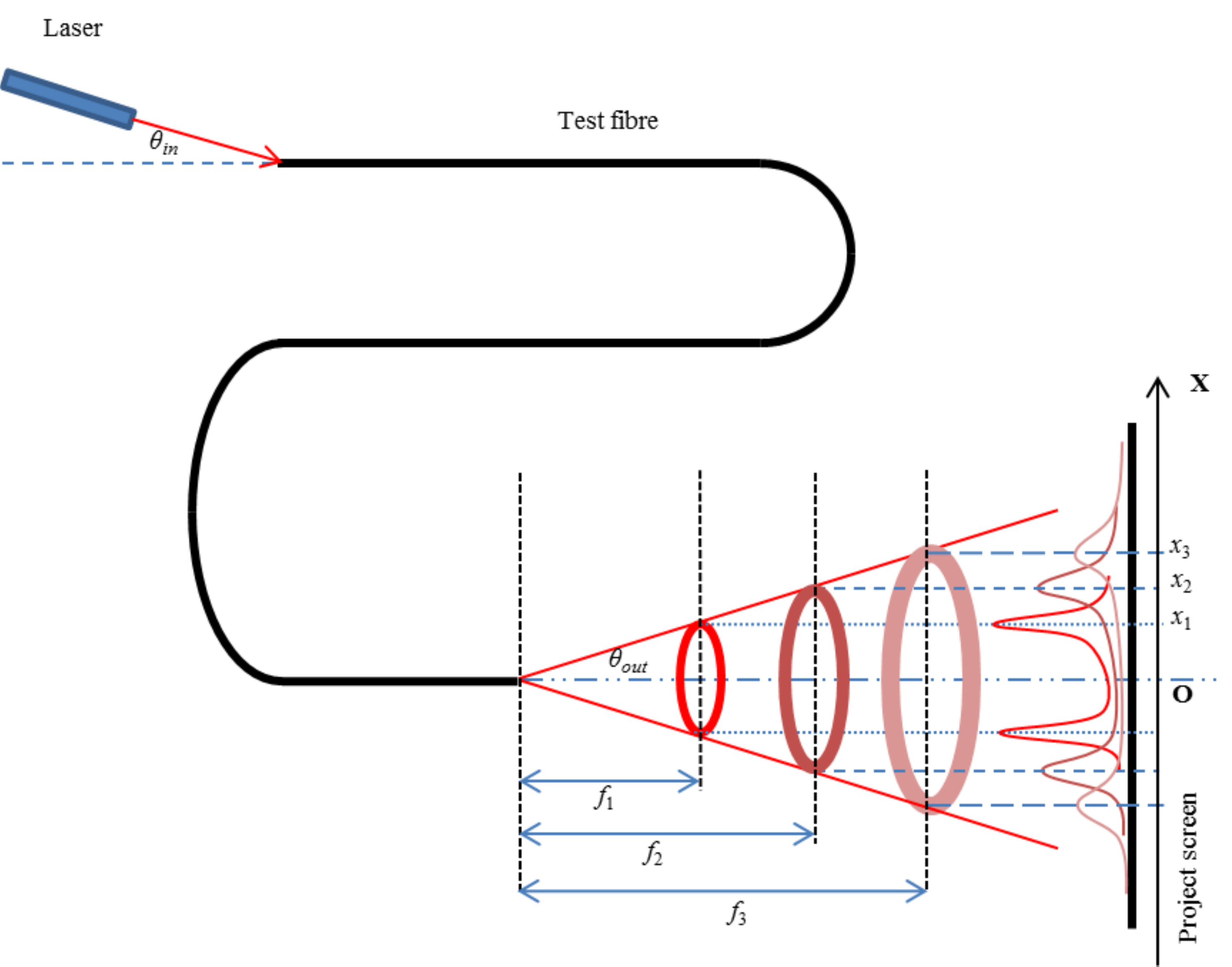}
      \caption{The schematic diagram of measuring process. The collimated laser beam is injected into the fibre at an angle ${\theta _{in}}$ and the output light field is a ring. A CCD camera is to record a series of output spots with different \emph{FWHM}s in different positions.}
         \label{fig:27}
   \end{figure}

\begin{figure}
   \centering
   \includegraphics[width=\hsize]{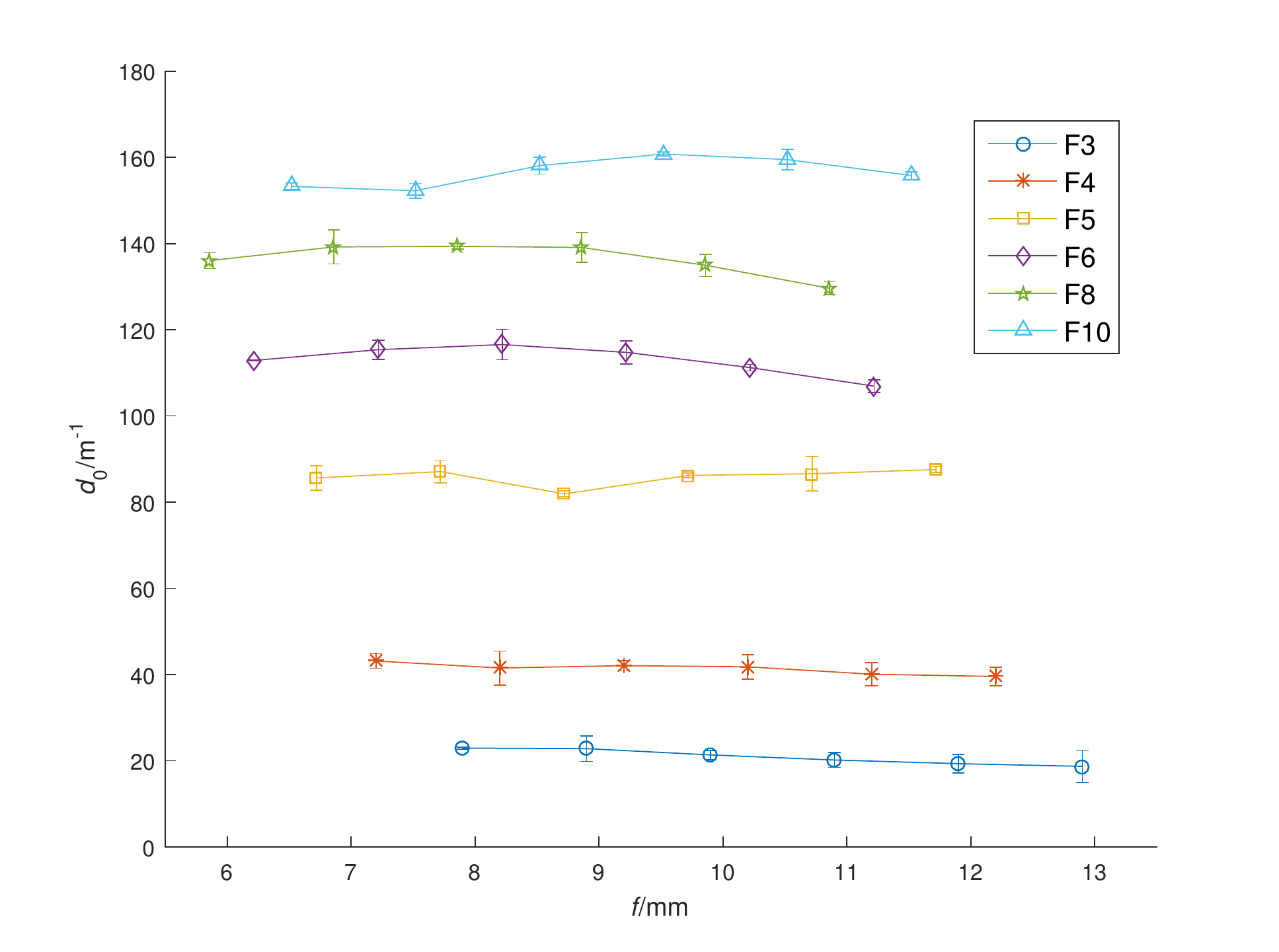}
      \caption{The distribution of parameter ${d_0}$ calculated by MPDM in different input focal ratio. MPDM eliminates the uncertainties of distance and the parameter $d_0$ remains stable for each input focal ratio, but still the result is different from the results of CP94 which shows no evident distance dependence.}
         \label{fig:28}
   \end{figure}

The principle and measuring process is shown in Fig.\ref{fig:27}. A series of output spots are imaged in different positions on CCD, so the focal distance \emph{f} should be taken into consideration. Generally, the order of magnitude of parameter $D$ is about $10^{-4}$ and absorption coefficient $A < 1$, making the approximation, so we can take $bL \ll 1$ for short fibres unless the fibre reaches hundreds or thousands meters and therefore $\exp \left( { - bL} \right) \approx 1 - bL$, then the Eq.(\ref{eq10}) can be written as
\begin{equation}\label{eq14}
P\left( {{\theta _{out}},{\theta _{in}}} \right) \approx \frac{1}{{bL}}\exp \left( { - \frac{{{\theta _{out}}^2 + {\theta _{in}}^2}}{{4DL}}} \right){I_0}\left( {\frac{{{\theta _{out}}{\theta _{in}}}}{{2DL}}} \right)
\end{equation}
The output angle is limited by \emph{N.A.}, then we can consider small angle approximation in Fig.\ref{fig:27} and the output angle is given by
\begin{equation}\label{eq15}
{\theta _{out}} \approx \tan {\theta _{out}} = \frac{{{x_i}}}{{{f_i}}}
\end{equation}
Substituting for ${\theta _{out}}$ from Eq.(\ref{eq15}), we get
\begin{equation}\label{eq16}
\begin{split}
& P\left( {{x_i},{f_i},{\theta _{in}}} \right) \approx \\
& \qquad \frac{1}{{bL}}\exp \left( { - \frac{{{x_i}^2 + {{\left( {{\theta _{in}}{f_i}} \right)}^2}}}{{4DL{f_i}^2}}} \right){I_0}\left( {\frac{{{x_i}{\theta _{in}}}}{{2DL{f_i}}}} \right)
\end{split}
\end{equation}
The output spot will be a Gaussian centred spot or a ring spot depending on the incident angle ${\theta _{in}}$, and the Gaussian cross section and width is
\begin{equation}\label{eq17}
{\sigma _i} = {\left( {2DL} \right)^{{1 \mathord{\left/
 {\vphantom {1 2}} \right.
 \kern-\nulldelimiterspace} 2}}}{f_i}
\end{equation}
Further, Eq.(\ref{eq17}) can be rewritten in the form
\begin{equation}\label{eq18}
{\sigma _i}^2 = 2 \cdot {\left( {\frac{\lambda }{{2a{n_1}}}} \right)^2}L{d_0}{f_i}^2 \propto {d_0}{f_i}^2
\end{equation}
Eq.(\ref{eq16}) and Eq.(\ref{eq18}) describes the output distribution and the Gaussian profile in different places, with which we can recalculate the parameter ${d_0}$ shown in Table \ref{tab:4} and Fig.\ref{fig:28}.

\begin{table*}
\caption{Parameter \emph{d}$_0$(m$^{-1}$) from MPDM in different positions}             % title of Table
\begin{threeparttable}[b]
\label{tab:4}      % is used to refer this table in the text
\centering                          % used for centering table
\begin{tabular}{c c c c c c c c c}        % centered columns (4 columns)
\hline                 % inserts double horizontal lines
   & Position 1 & Position 2 & Position 3 & Position 4 & Position 5 & Position 6 & Average & \emph{FWHM}($^\circ $)\\
\hline
   \emph{F}$_{in}$ = 3 & 22.9 & 22.8 & 21.4 & 20.2 & 19.3 & 18.7 & 20.9$\pm$1.6 & 1.18$\pm$0.05 \\
   \emph{F}$_{in}$ = 4 & 43.1 & 41.5 & 42.1 & 41.8 & 40.1 & 39.6 & 41.4$\pm$1.2 & 1.66$\pm$0.03 \\
   \emph{F}$_{in}$ = 5 & 85.6 & 87.1 & 81.9 & 86.2 & 86.6 & 87.6 & 85.9$\pm$1.9 & 2.39$\pm$0.03 \\
   \emph{F}$_{in}$ = 6 & 112.9 & 115.4 & 116.6 & 114.8 & 111.2 & 106.9 & 113.0$\pm$3.2 & 2.74$\pm$0.04 \\
   \emph{F}$_{in}$ = 8 & 136.0 & 139.2 & 139.4 & 139.1 & 135.0 & 129.6 & 136.4$\pm$3.5 & 3.01$\pm$0.04 \\
   \emph{F}$_{in}$ = 10 & 153.3 & 152.2 & 158.1 & 160.8 & 159.5 & 155.8 & 156.6$\pm$3.1 & 3.22$\pm$0.03 \\
\hline                                   %inserts single line
\end{tabular}
\begin{tablenotes}
  \item Notes:
  \item Position *: The different positions where the output spots were captured by CCD.
  \item REMINDER: Even though the sequence number of Position * in the table is the same in each column, i.e. Position 1, the actual position in the experiment is not all the same for different input focal ratio, and the detailed information of positions can be seen in Fig.\ref{fig:26} and Fig.\ref{fig:28}.
\end{tablenotes}
\end{threeparttable}
\end{table*}

\begin{figure}
   \centering
   \includegraphics[width=\hsize]{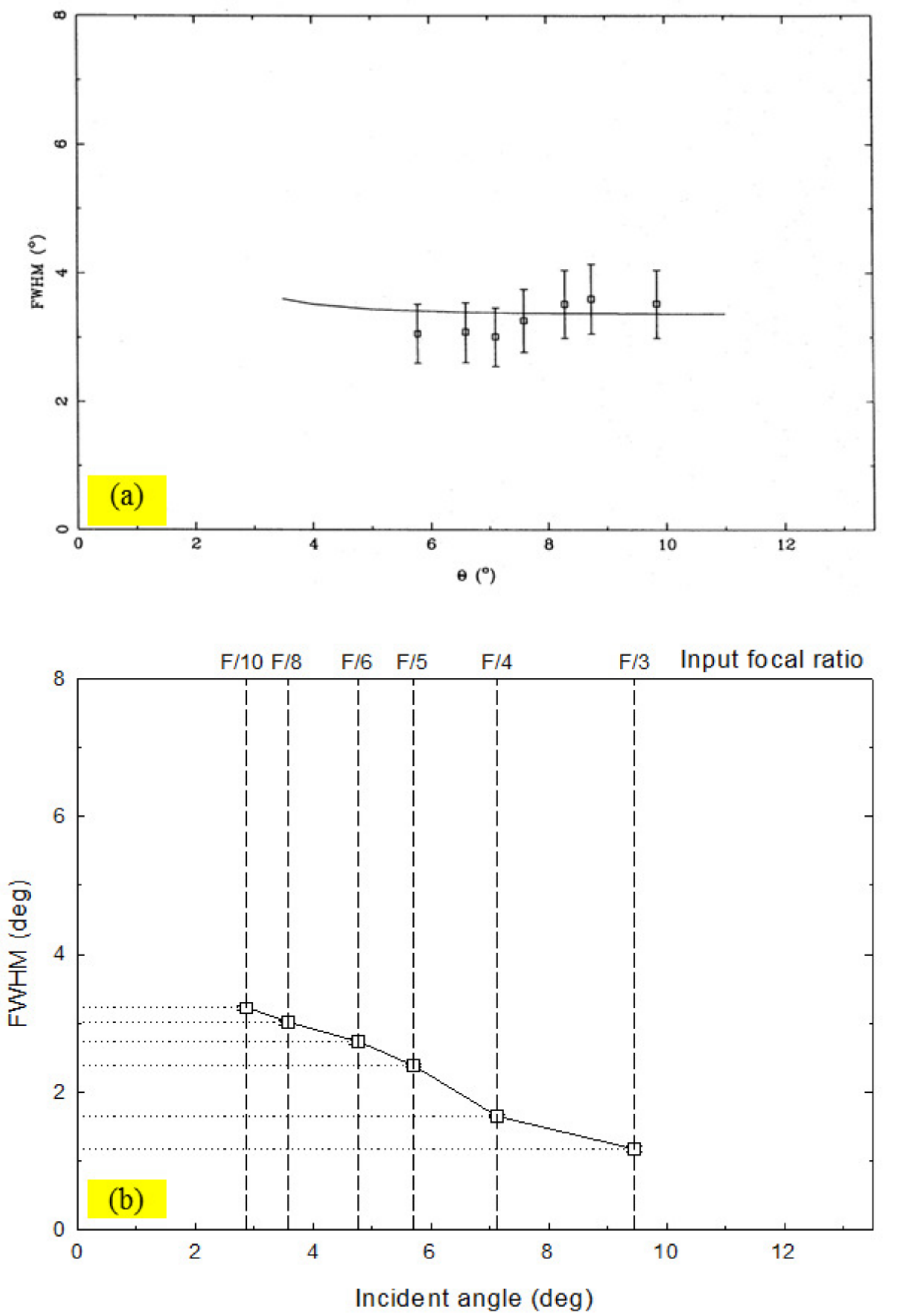}
      \caption{\emph{FWHM} versus input angle. (a) The results in CP94 show clearly that the \emph{FWHM} remains more or less constant within a range of angles of incidence from 5.7$^\circ $ to 10$^\circ $ corresponding to input focal ratio from \emph{F}/2.8 to \emph{F}/5.0. \citep[see][figure 5]{Carrasco1994A}. (b) The results in our work show a different trend that the \emph{FWHM} decreases with incident angle from 2.8$^\circ $ to 9.5$^\circ $ or, in other words, increases with input focal ratio from \emph{F}/3.0 to \emph{F}/10.0.}
         \label{fig:29}
   \end{figure}

In MPDM the uncertainties of different positions have been eliminated. But the results suggest that the parameter ${d_0}$ has a dependence on input focal ratio, which is different from the results of CP94 shown in Fig.\ref{fig:29}. In their experiment, the \emph{FWHM} of output ring does not change with the input angle, which means the parameter ${d_0}$ is independent on input focal ratio. In the ideal situation, Eq.(\ref{eq16}) and Eq.(\ref{eq18}) indicate that the parameter ${d_0}$ and input focal ratio are indeed independent, which is agreed with the results of CP94 but inconsistent with ours. How does it happen that it differs from ours? This may involve the problem about the relationship between input and output focal ratio. Many groups has reported that the FRD performance gets worse in slower input focal ratio \citep{Marcel2006Application,Avila2006Photometrical,Murphy2008Focal,Bryant2014Focal,Pazder2014The} and the parameter ${d_0}$ is a eigenvalue to character the transmission properties, so in slower input focal ratio situation, \emph{F}$_{in}$ = 10 for example, the degree of FRD is worse, which means the parameter ${d_0}$ should be larger and it should be smaller on the contrary. In this view, the parameter ${d_0}$ is not just a simple factor that it might be a combination of several independent unresolved parameters that determine the transmission properties. In our experiment system, we can acquire the parameter ${d_0}$ in different input condition, so the relationship among the parameter ${d_0}$, focal distance \emph{f} and input focal ratio is indirectly resolved.

\begin{figure}
   \centering
   \includegraphics[width=\hsize]{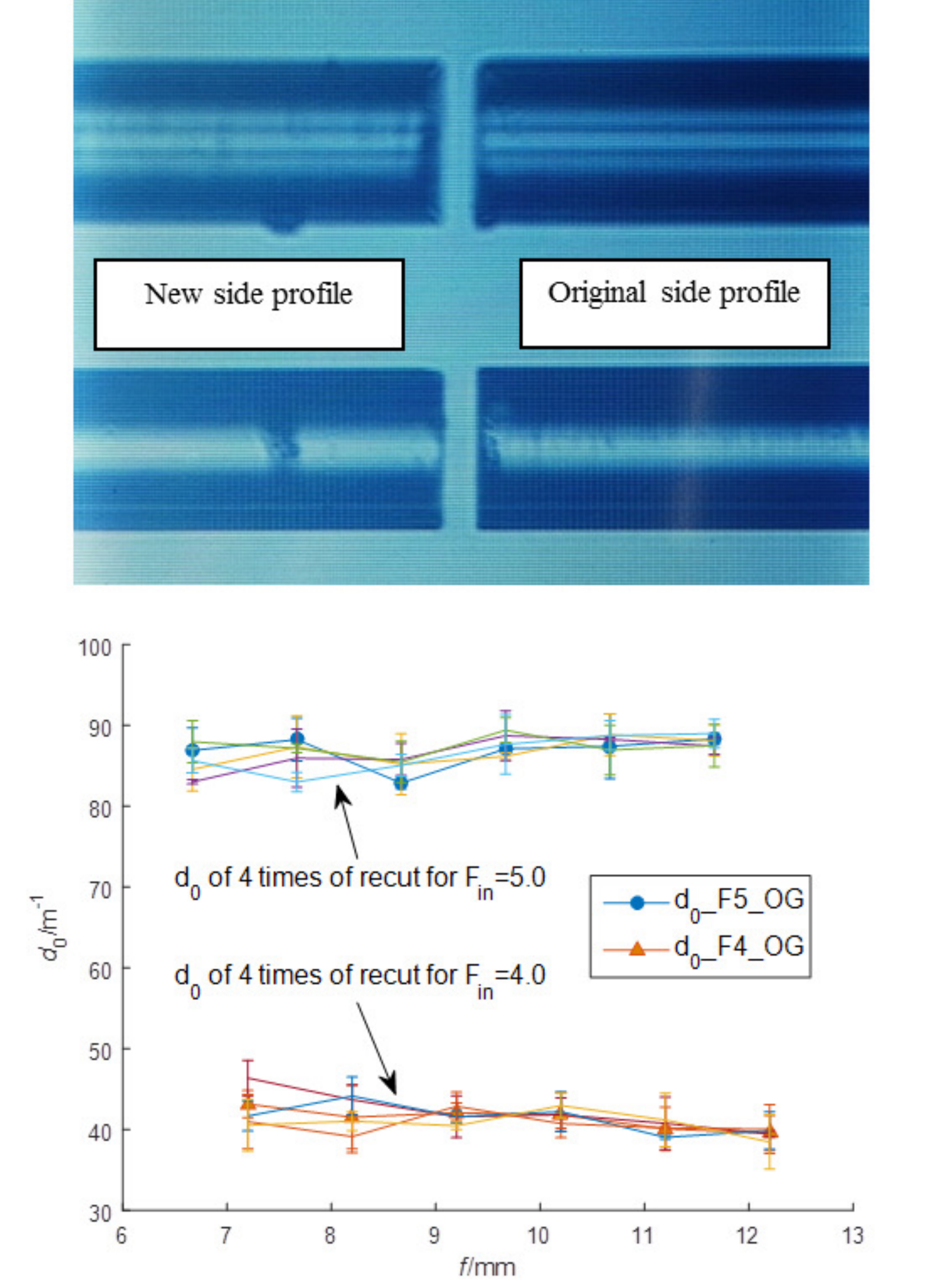}
      \caption{The upper image is the fibre side profile before and after recut to ensure that they are similar. The lines with symbols in the lower one show the parameter ${d_0}$ of the results of original end face. 85.9$\pm$1.9 for \emph{F}$_{in}$ = 5.0 and 41.4$\pm$1.2 for \emph{F}$_{in}$ = 4.0. The others are newly measured of new end faces. The new results remain 85.8$\pm$1.7 for \emph{F}$_{in}$ = 5.0 and 41.3$\pm$1.8 for \emph{F}$_{in}$ = 4.0. The results suggest that a well prepared fibre end has negligible influence on the parameter ${d_0}$.}
         \label{fig:30}
   \end{figure}

The experiment blow is to investigate the stability that the fibre is recut and replaced by a fibre of 200$\mu$m core to measure the parameter ${d_0}$ again. \citet{Poppett2010A} and \citet{Allington-Smith2013End} discussed the end effects on FRD of different end termination techniques and serious roughness or imperfections of the end face would make FRD more serious. So a well prepared end face is necessary for applications in the fibre system. The fiber end is cleaved by a fibre cleaver (product model: FK11) and the end face is very smooth without serious grains which usually exist in a polished end face. But we should notice the imperfections on the edge of the fibre near the cut incision and whether the surface normal is parallel to z-direction (along the fiber). So the fibre end face is eye-checked in a microscope to reject those with terrible defects near the incision. At the same time, the surface normal is inspected via fiber fusion splicer which can check the angle of the fiber end in both x- and y-direction. Only the fiber end with slope angle less than 0.5$^\circ $ is acceptable. Each time the fibre end was recut and checked to ensure it was similar to original one as shown in Fig.\ref{fig:30}. In the stability test, the input focal ratio is \emph{F}$_{in}$ = 4.0 and 5.0.

   \begin{figure}
   \centering
   \includegraphics[width=\hsize]{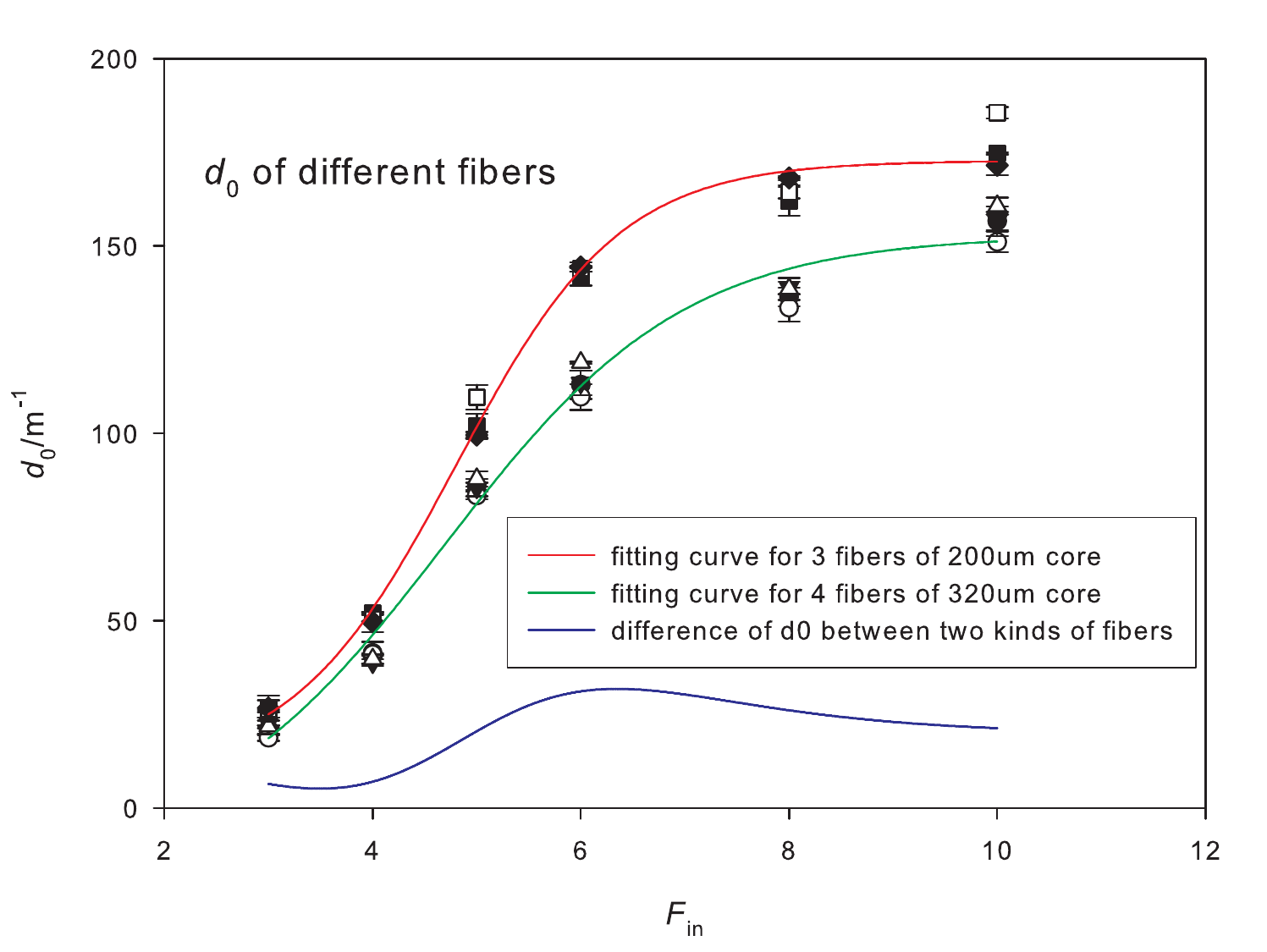}
      \caption{Stability test for laser illumination system with two kinds of fibres of different cores. The red and green lines are the fitting curves for several tests. Generally, the value of parameter ${d_0}$ of 200$\mu$m core is larger and the trend is similar. The relative difference by subtracting each other is shown in blue line. The difference is getting smaller in both very slow and fast input focal ratio conditions.}
         \label{fig:31}
   \end{figure}

The measured value of recut fibre remains around 85.8$\pm$1.7 for \emph{F}$_{in}$ = 5.0 and 41.4$\pm$1.2 for \emph{F}$_{in}$ = 4.0 similar to the results of 85.9$\pm$1.9 and 41.3$\pm$1.8 in the before. So the stability of recut is good for preparing the fibre end carefully and a well recut end face can reduce the difference in parameter ${d_0}$. When the fibre is replaced with a fibre of 200$\mu$m core used in Sec. 3.3, the parameter ${d_0}$ becomes larger as shown in Fig.\ref{fig:31}, so the FRD is more serious agreeing with the results in Sec. 3.3. Both of the two fitting curves show a similar trend that the parameter ${d_0}$ increases fast with relatively fast input focal ratio when \emph{F}$_{in}$ < 7.0 but the growths remain gradually in slower input focal ratio when \emph{F}$_{in}$ > 7.0. It also can be seen from the blue line of difference of two types of fibres that the relative variation increases at first to the peak around \emph{F}$_{in}$ = 7.0 and then decreases with the increasing input focal ratio. These phenomena are reasonable, because in the very fast input focal ratio condition approaching the limitation of \emph{N.A.}, the FRD is not so evident that the parameter ${d_0}$ will be nearly the same. Similarly, when the input focal ratio is very slow, larger than 8.0 for example, the relative change of input condition will be very small just like the situation in Fig.\ref{fig:23} that the size of input spot nearly remains unchanged, and if the input focal ratio continues to increase larger than 10.0, at this time the input light can be treated as a collimated beam incident vertically into the fibre, so the parameter ${d_0}$ will remain steady for each fibre. At the same time, the parameter ${d_0}$ is more about other factors than input focal ratio, like the arrangement of fibre paths, bending, stress and so on.

\begin{figure}
   \centering
   \includegraphics[width=\hsize]{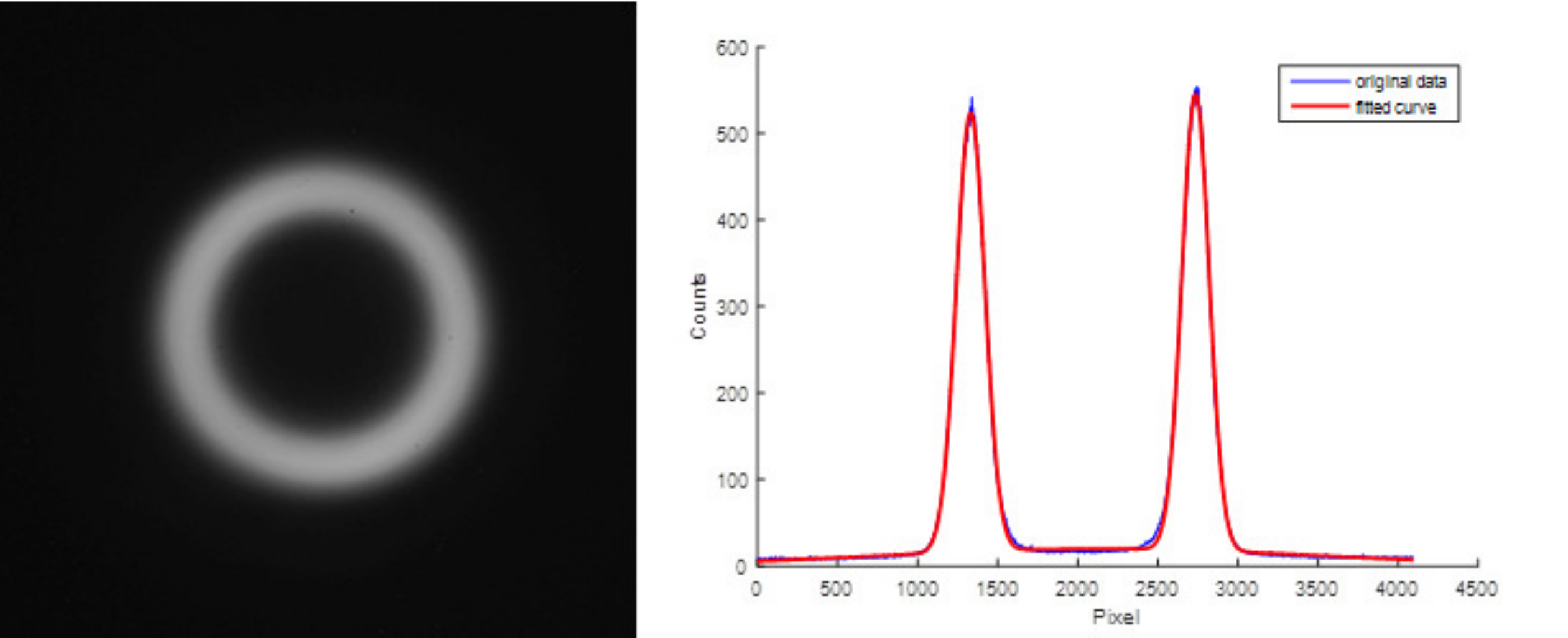}
      \caption{Output spots and 2-D fitting curve of LED illumination system. The uniformity and symmetry are much better than laser condition.}
         \label{fig:32}
   \end{figure}

\begin{figure}
   \centering
   \includegraphics[width=\hsize]{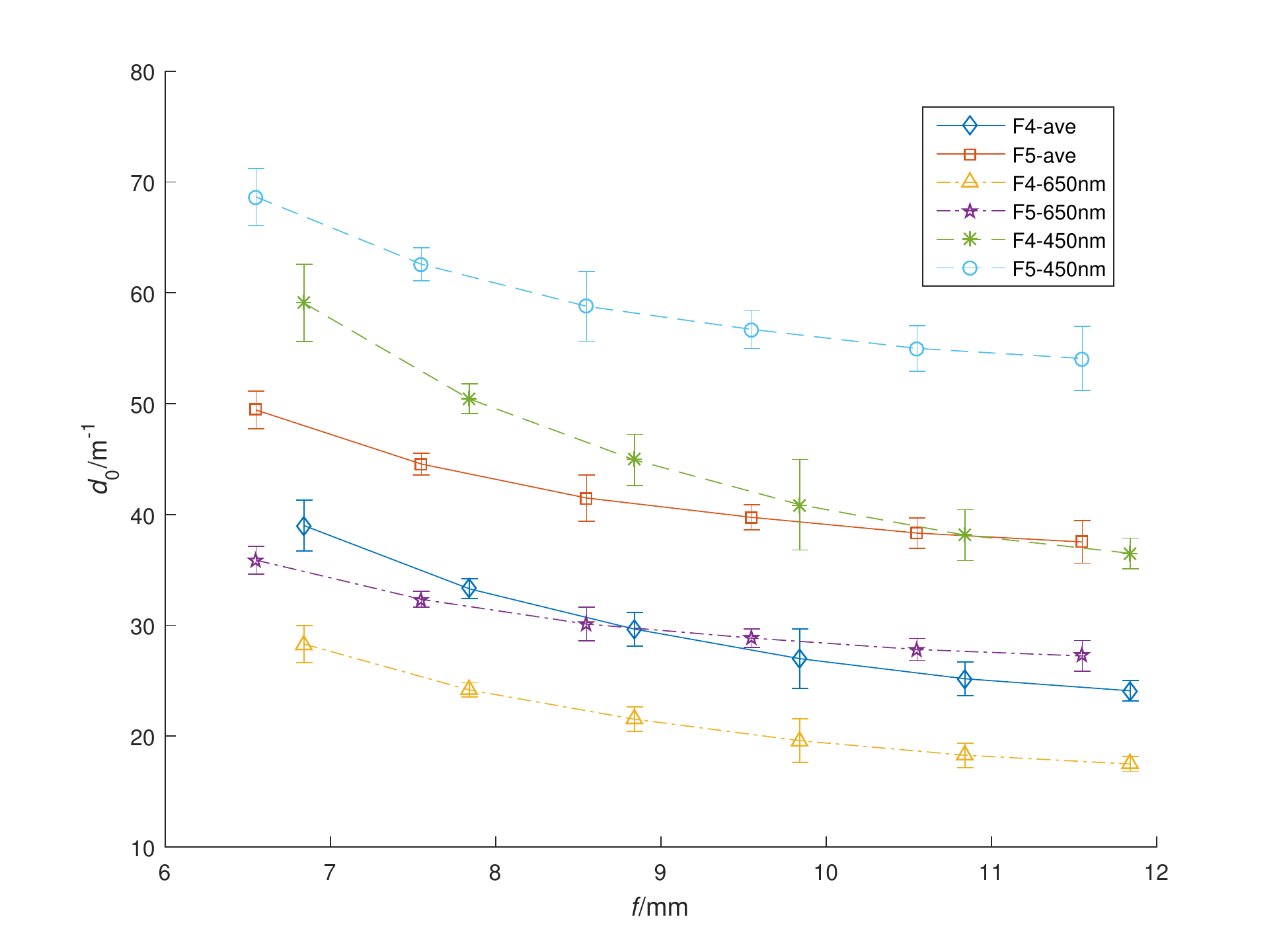}
      \caption{Parameter ${d_0}$ measured in LED illumination system by weighting technique. The weight for each given wavelength is derived from the SED.}
         \label{fig:33}
   \end{figure}

Using laser as light source, the output spot suffers speckle and it will result in errors in measurement of parameter ${d_0}$. If replaced with white light, the speckle can be perfectly eliminated in Fig.\ref{fig:32} but the problem of determination of wavelength occurs since the white light is combining a broad band of wavelengths. Regardless of the coherence of monochromatic light, assuming that the contribution of each wavelength to parameter ${d_0}$ is equivalent to the spectral energy distribution (SED), the parameter ${d_0}$ can be derived by weighting technique as is shown in Fig.\ref{fig:33}.

The parameter ${d_0}$ is monotone decreasing with the distance and the measured value by both monochrome technique and weighting technique is smaller than that in laser illumination system. The decreasing property indicates that the divergence speed is slower than that of the increasing distance, which means the growth of the \emph{FWHM} of the output ring is relatively small compared with the increasing speed of input focal ratio. The smaller parameter ${d_0}$ shows that the FRD performance is better in LED illumination system. Both kinds of the phenomena point to a fact that polychromatic light makes the power distribution smoother and more concentrated near the ring because of the elimination of speckle and coherence. But to measure the parameter ${d_0}$, laser illumination system is better compared to the unsteady value with decreasing properties in LED case.

Even though the output spot will suffer the speckle in laser illumination system, the monochromatic light with definite wavelength is appropriate for the calculation of parameter . White light can avoid the speckle but the wavelength problem should be faced with and the results in Fig.\ref{fig:33} yet cannot give out a steady result. The method MPDM by introducing focal distance \emph{f} into original model and re-establishing the output light field provides a new way for testing in multi-position measurement experiment system.

\subsection{FRD dependence on optical fibre length}
\begin{figure}
   \centering
   \includegraphics[width=\hsize]{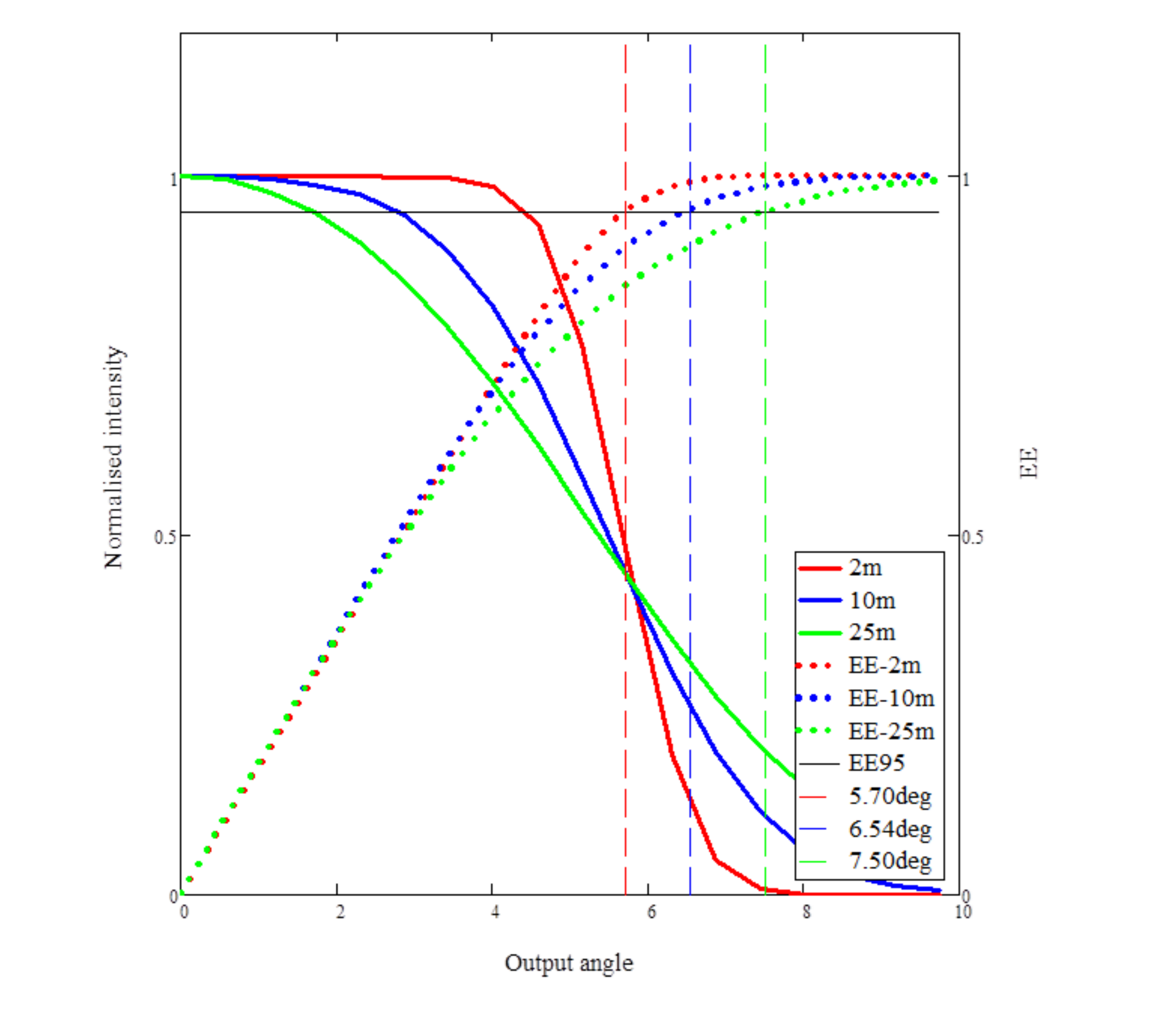}
      \caption{Predicted power distribution of output spots with different lengths of fibre by MPDM. The input light is a cone beam of \emph{F}$_{in}$ = 5.0. The half of opening angles of the output spots under EE95 are 5.70$^\circ$, 6.54$^\circ$ and 7.50$^\circ$ corresponding for 2m, 10m and 25m respectively. The output spot from a longer fibre is much diffuser and the diameter increases very fast.}
         \label{fig:34}
   \end{figure}

As is shown in Eq.(\ref{eq9}) and Eq.(\ref{eq16}), the parameter \emph{D} depends on the fibre core diameter $a$, the parameter ${d_0}$ and the wavelength of the injected light $\lambda $. The output power distribution in Eq.(\ref{eq10}) also depends on the length of the fibre $L$. To present how fibre length affects the FRD, we use MPDM model to predict the output power distribution by changing the input parameters, especially the fibre length, as shown in Fig.\ref{fig:34}, and the input focal ratio is \emph{F}$_{in}$ = 5.0 (incident angle ${\theta _{in}}$ = 5.7$^\circ$).

Apparently, the output power distribution changes obviously with different fibre lengths. The output light field becomes diffuser from a longer fibre. Given a fixed EE = 95\%, the diameter of output spot is larger and FRD performance is worse of a longer fibre. While the results in \citet{Poppett2010The} and \citet{Avila2006Photometrical} showed no evident length dependence and \citet{Brunner2010APOGEE} found that the output focal ratio of a short fibre of 10m was sensitive to the alignment changes of input position on the fibre end but much less sensitive for a long fibre of 40m.

Another assumption is that if we consider the laser beam as an extremely fine ray of a cone beam, and then we can directly acquire the relationship between ${\theta _{in}}$ and \emph{FWHM} to the input and output focal ratios according to the approximate relationships:
\begin{equation}\label{eq19}
{F_{in}} = \frac{1}{{2\tan {\theta _{in}}}}
\end{equation}
\begin{equation}\label{eq20}
{F_{out}} = \frac{1}{{2\tan \left( {{\theta _{in}} + {{FWHM} \mathord{\left/
 {\vphantom {{FWHM} 2}} \right.
 \kern-\nulldelimiterspace} 2}} \right)}}
\end{equation}
Eq.(\ref{eq19}) and Eq.(\ref{eq20}) can also describe the FRD properties because FRD can broaden the output ring in both inner and outer directions, which makes the ring diffused and widens the \emph{FWHM} \citep{Murphy2008Focal,Murphy2013The,Mattias2009Air}. Here we consider it as \emph{FWHM} model.

\begin{figure}
   \centering
   \includegraphics[width=\hsize]{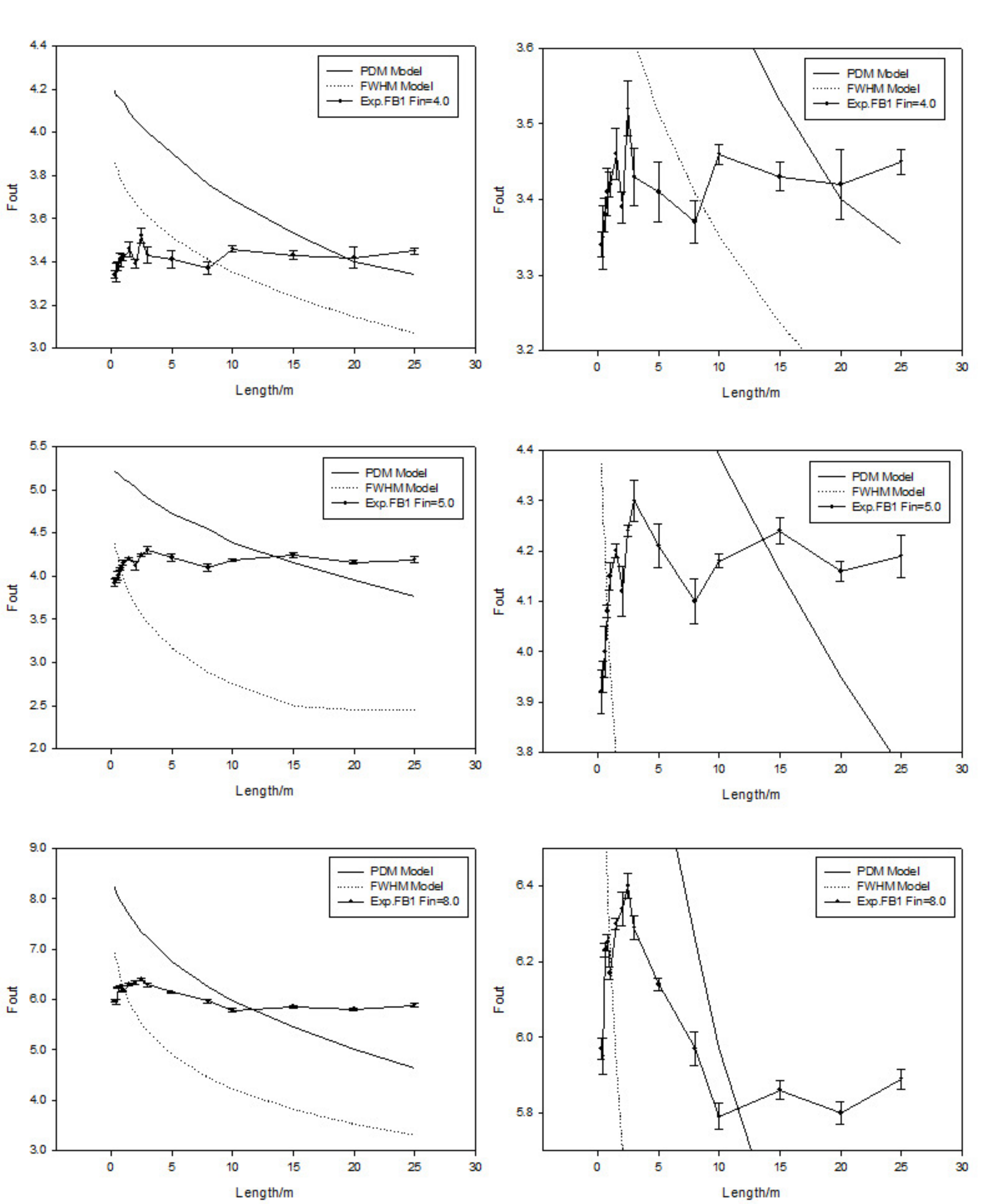}
      \caption{Output focal ratio of fibre 1 vs lengths.}
         \label{fig:35}
   \end{figure}

\begin{figure}
   \centering
   \includegraphics[width=\hsize]{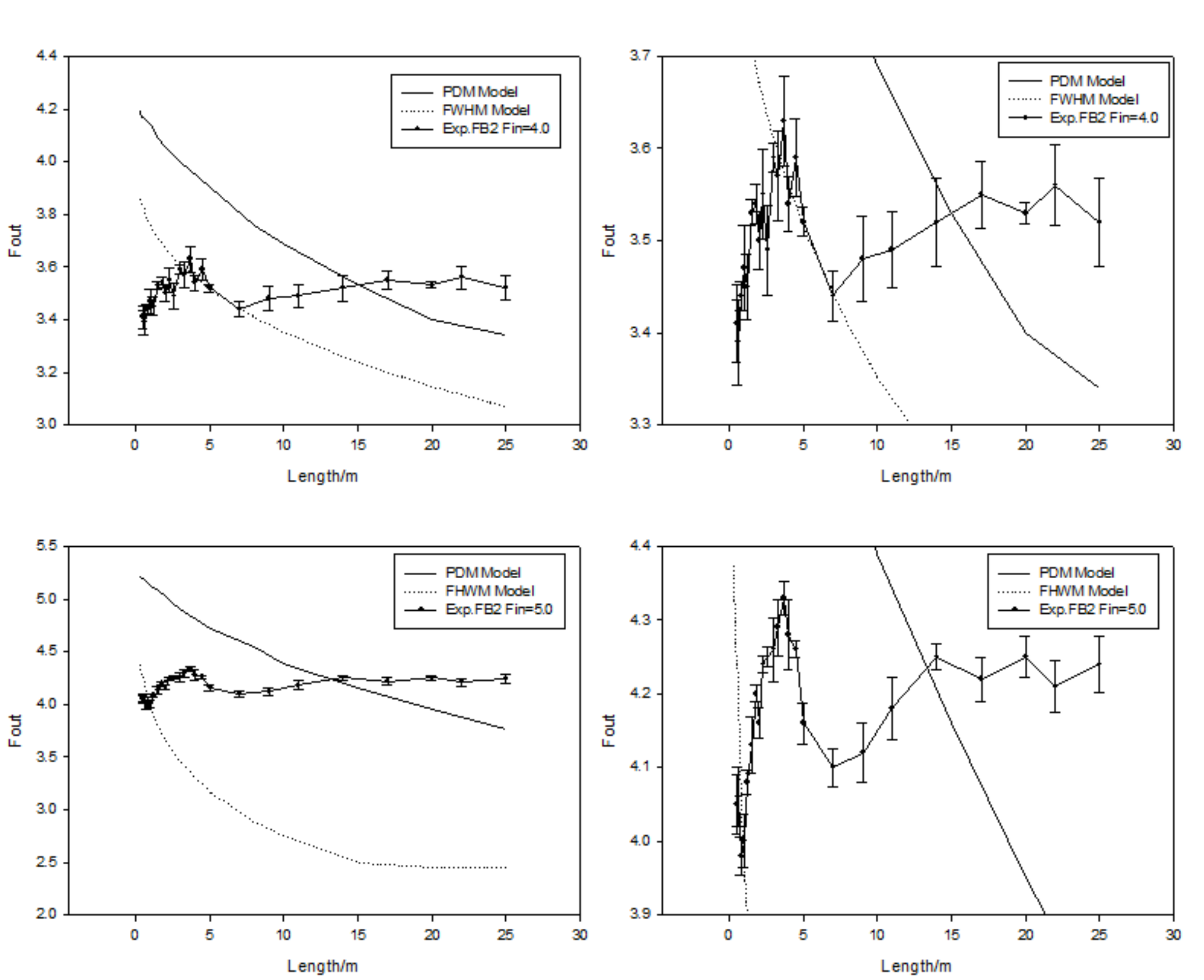}
      \caption{Output focal ratio of fibre 2 vs lengths.}
         \label{fig:36}
   \end{figure}

\begin{figure}
   \centering
   \includegraphics[width=\hsize]{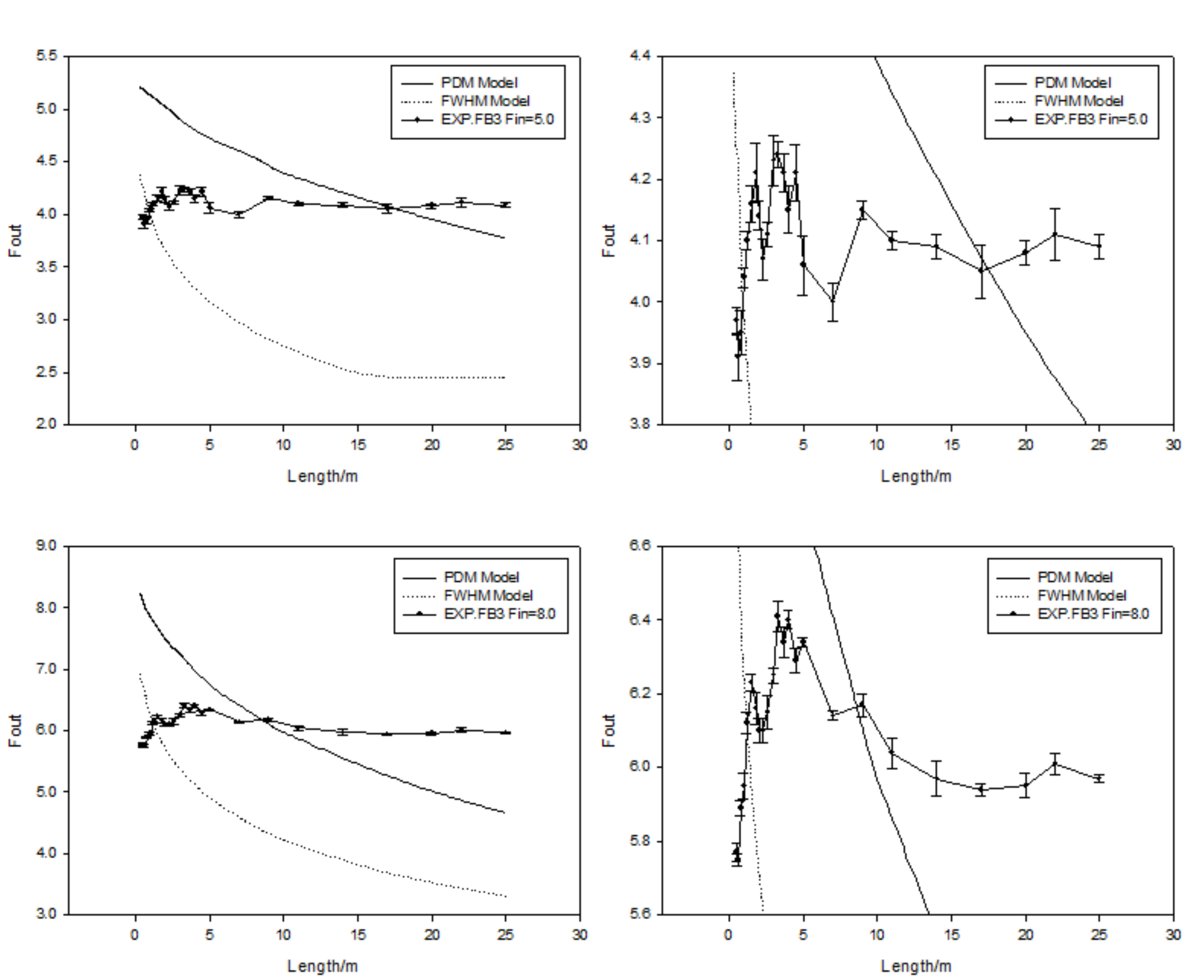}
      \caption{Output focal ratio of fibre 3 vs lengths.}
         \label{fig:37}
   \end{figure}

The two models of PDM and \emph{FWHM} are implemented to simulate the relationship between FRD and fibre length and the results are compared with experiments in Fig.\ref{fig:35}, Fig.\ref{fig:36} and Fig.\ref{fig:37}. The curve is shown magnified on the right. The necessary parameter ${d_0}$ is measured in Sec. 4.1 by MPDM corresponding to different input focal ratio. In the experiments and the simulations, the input focal ratio is set to \emph{F}$_{in}$ = 4.0, 5.0, 8.0, the value of parameter ${d_0}$ = 41.4, 85.9, 136.4 is shown in Table \ref{tab:4} and all the output focal ratio is calculated under the condition of EE95.

Both of the MPDM and \emph{FWHM} model are sensitive to length. The output focal ratio decreases fast with increasing length and the FRD is much more serious in \emph{FWHM} model. Again the prediction results of two models show that the performance of FRD is sensitive to input focal ratio and agree with the results that slow input focal ratio leads more serious FRD.

\begin{figure*}
   \centering
   \includegraphics[width=\hsize]{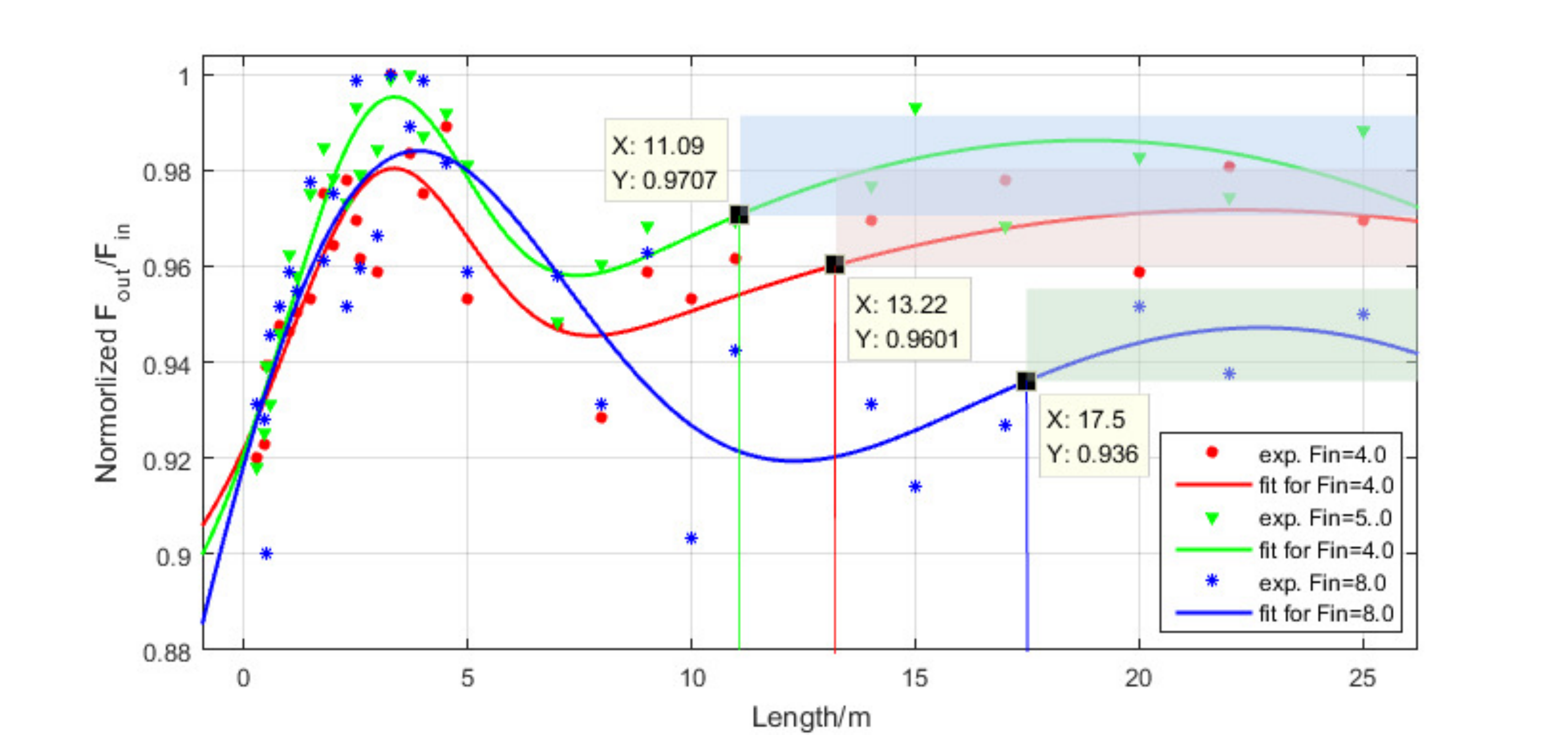}
      \caption{Normorlized \emph{F}$_{out}$/\emph{F}$_{in}$ vs fibre length. The peak occurs at the length of 3$\sim$4m. To improve the FRD performance, different fibre lengths should be chosen for different input conditions that the shaded areas indicate the stable \emph{F}$_{out}$ zone of the relative variation within 2\% that 13.22m for \emph{F}$_{in}$ = 4.0, 11.09m for \emph{F}$_{in}$ = 5.0 and 17.5m for \emph{F}$_{in}$ = 8.0. Generally, fibres longer than 15m are of good stability with output focal ratio.}
         \label{fig:38}
   \end{figure*}

But the experimental results do not follow the trend and show no evident decreases compared to the two models. Even in short fibres conditions, the FRD is not necessarily better than other cases. While an interesting phenomenon shows that the output focal ratio fluctuates in short fibres when fibre length is less than 5m, and reaches the maximum in the range of length between 3m to 4m, but remains steady in longer fibres especially when the fibre is longer than 15m, where the output focal ratio rises up a little sometimes and the curve becomes to be an approximate straight line parallel to the horizontal axis. And the promising length choice of fibre for different input focal ratio is given in Fig.\ref{fig:38}. The possible reason is the energy exchange or the modes converting. The modes in fibres converting between higher and lower modes and the absorption both make the output power distribution different in varies fibre lengths. Then the output spots become either larger or smaller, making the output focal ratio changes with different fibre lengths. At the same time, higher modes which are very sensitive to the fibre bending including macrobending and microbending can be conducted through the fibre before their dissipation in a relatively short fibre, so the fluctuation of output focal ratio occurs in a short fibre for the difficulty to control bending conditions especially the microbending defects which is mainly determined by the fabrication technology and the manufacture. But for longer fibres, most of higher modes fade away during the transmission and just lower modes can transform with small output angle, so the output focal ratio remains stable and sometimes increases. A probable method is to measure the transmitting efficiency to identify whether the energy loss is mainly caused by the dissipation of higher modes to optimize the length properties in FRD.

\begin{figure}
   \centering
   \includegraphics[width=\hsize]{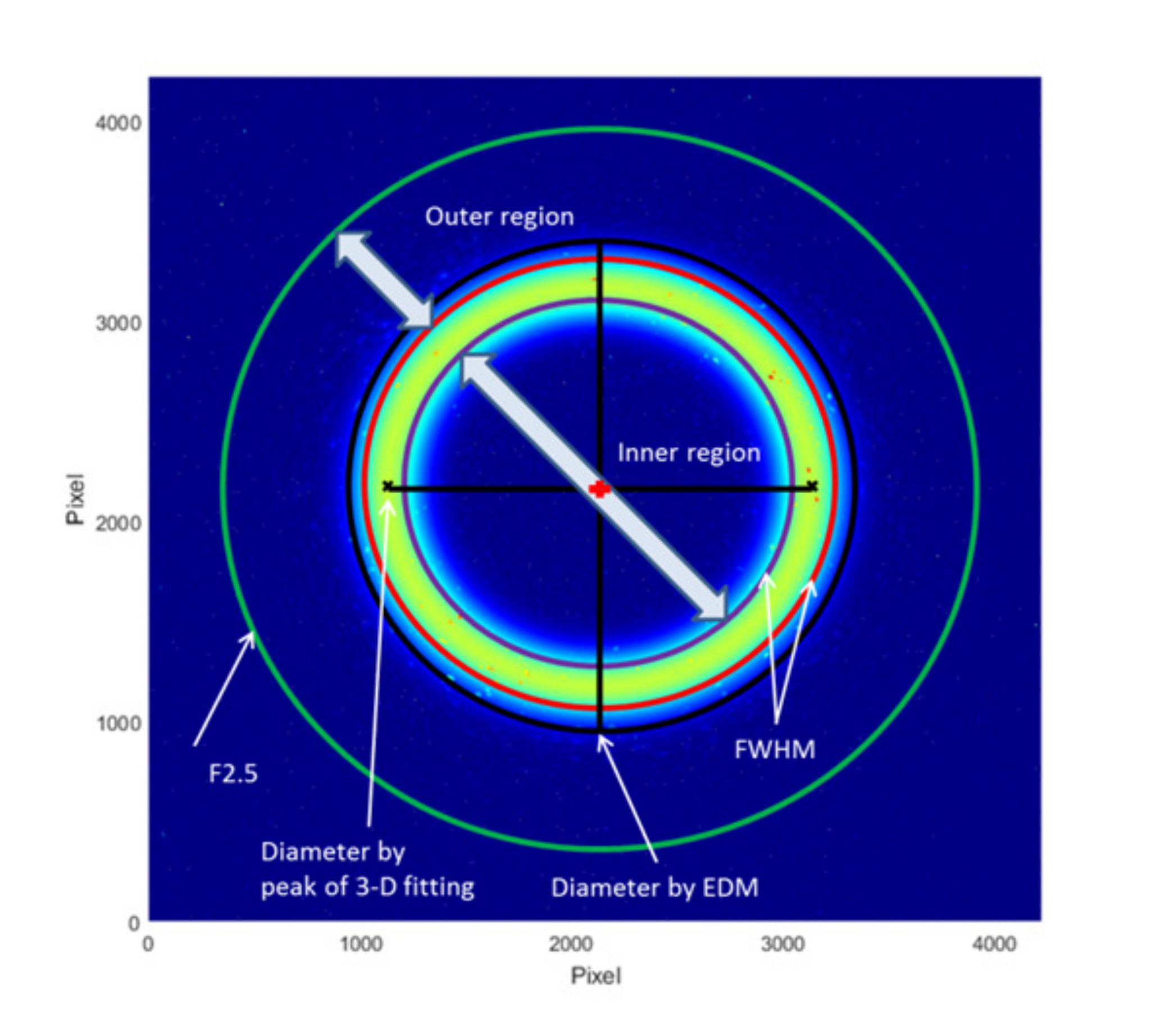}
      \caption{The schematic diagram of a 3-D fitting surface to measure the diameter and intensity. \emph{F} = 2.5 represents the aperture limited by \emph{N.A.} = 0.22.}
         \label{fig:39}
   \end{figure}

\begin{figure}
   \centering
   \includegraphics[width=\hsize]{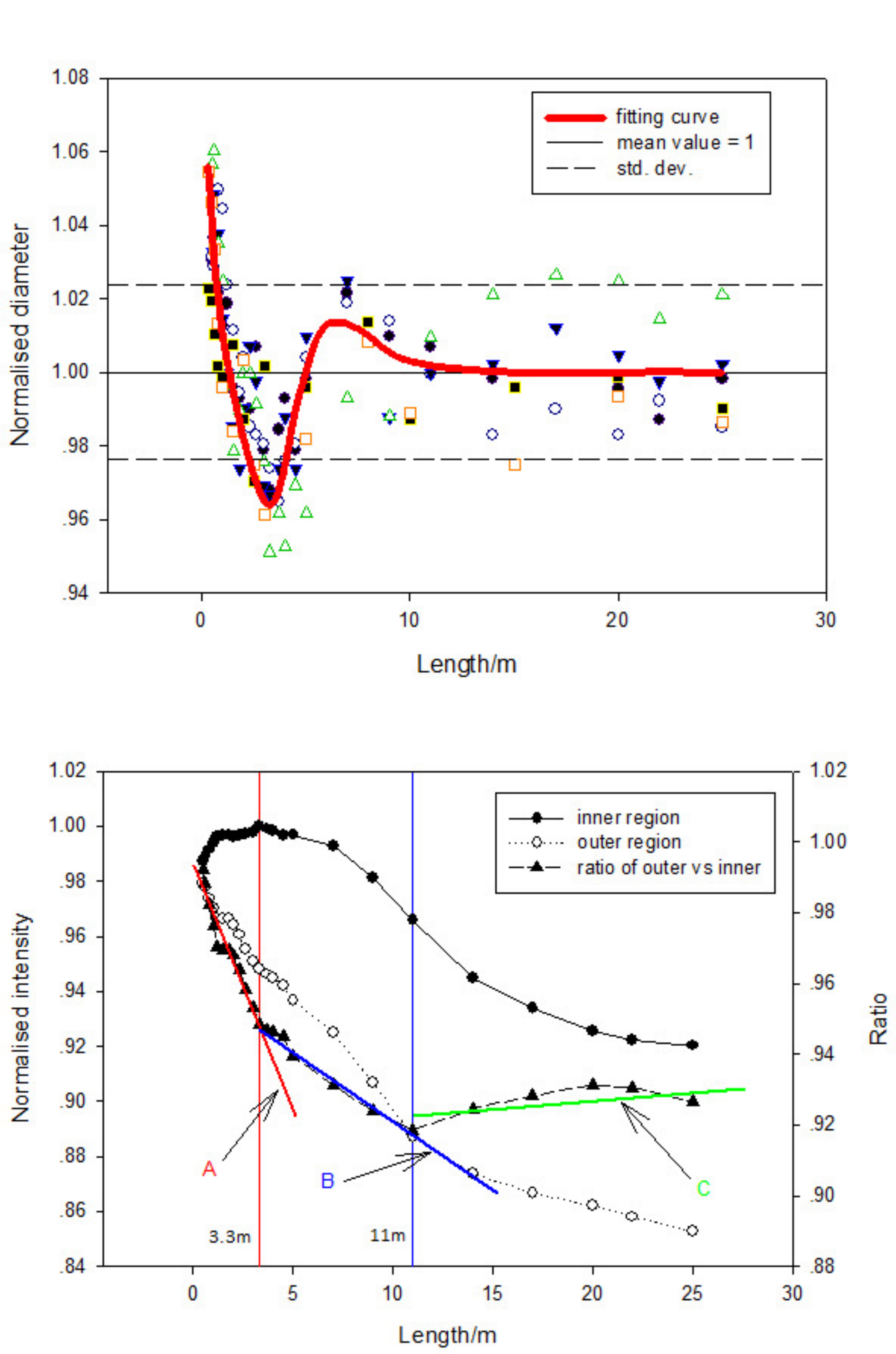}
      \caption{Normorlized diameter and intensity vs fibre length for input focal ratio \emph{F}$_{in}$ = 5.0. According to the definition of focal ratio in a fibre, the trends of diameters measured by peaks of the 3-D fitting surface are in consistent with the results acquired by EDM, but there is a little difference that the range of variation in diameters from peaks is most less than 4.5\% and smaller than that from EDM which is up to around 6\% in Fig.\ref{fig:38}. The intensity distribution may reveal a possible reason to explain the phenomena that the different depletion rates lead to three staged ratios of increasing slopes as in fitting lines A, B \& C. So both the energy converting and the shifts of the peaks contribute to the changes in the diameters.}
         \label{fig:40}
   \end{figure}

According to the guided mode theory in Sec. 2.2, higher modes are easy to couple into the cladding and dissipate very fast with large attenuation coefficient. To reveal the power converting process during the modes converting to study the FRD properties on length, we integrate the energy in inner and outer part of the output spot separated by \emph{FWHM} to compare the energy changes of different modes in Fig.\ref{fig:39}. In the meantime, we also determined diameters from the peaks of the fitting surface to compare the trends in sizes of output spots in the two methods. Fig.\ref{fig:40} shows the trends of diameters and intensity.

Two inflection points in Fig.\ref{fig:38} and Fig.\ref{fig:40} of fibre length are in consistent with each other: 3.3m and 11m. The two inflection points divide the fibre length into three ranges by the slope of fitting curves as A, B \& C shown in Fig.\ref{fig:40}. In A stage, the slope of ratio is negative and it indicate that the energy of the outer region losses faster than that of the inner part. Under this condition, the diameters will become smaller by using EDM, so the output focal ratios rise up with increasing length. Noticing that the power of the inner region increases in the beginning phase in A stage shows that the modes converting process transmits more energy from higher modes to lower modes though it is bidirectional. In B stage, the slope is still negative but smaller which means the power variation of both inner and outer region is nearly synchronous. Considering the dissipation, when EDM is used to determine the diameter, more energy is needed to reach the threshold of EE95, so the diameter starts to increase and output focal ratio becomes smaller. In the last C stage, the slope is positive and opposite to A stage, so generally the output focal ratio is smaller. And in a relatively long fibre, the intensity of both inner and outer region decreases slowly because extreme high order modes have been depleted and lower modes trend to be stable and the slope is close to zero. So the intensity of both parts remains stable to each other and the diameter and output focal ratio keep nearly unchanged in the last C stage.

At the same time, the range of variation induced by drifts of \emph{FWHM} is less than 4.5\% and cannot make up the total difference up to about 6\% of output focal ratio in different lengths. This small difference in diameters between EDM and \emph{FWHM} shows that both the energy converting and the shifts of the peaks contribute to the changes of output focal ratio. And this can also be a certain side of proof of energy exchanges caused by modes converting.

Except the length property, some other factors also should be considered in the experiments or in the practical applications like stress and bending effects. In the experiments, the short fibre is easy to control to be well prepared in a straight line, but it is too difficult for a long fibre to arrange it in a line but a loop to eliminate stress and bending effects which would affect the FRD performance. Anyhow, we were unable to guarantee the consistency of stress and bending environment but we tried to minimize the difference, and the results still show that the FRD performance of long fibres is relatively stable and not very sensitive to the changes of fibre length. So a good choice of appropriate fibre length is important to improve the FRD performance in a telescope to conquer the changing testing environment.

\section{Conclusions}
We deduced guided modes in geometric optics and combined the PDM model to simulate the transmission properties. The two theories together can describe the output light field distribution under different incident conditions and explain the influences of different parameters on FRD.

FRD is a challenging topic in astronomy for several reasons. First, the consistency of different measurement methods hinders both the communication of requirements and the comparison of different groups. Second, FRD measurement is related to so many factors that it is easily contaminated by external factors. Last, but not the least, the criteria of measuring FRD is not always the same in different systems. So we proposed a new method EDMF by combining EDM and \emph{f-intercept} to acquire the actual position of the focal point of the illumination system, therefore, the input condition can be well controlled to guarantee the consistency of the experiment environment and improve the reliability and comparability of the experiment results. The output focal ratio changes less than 2\% and the relative throughput can be up to 95\% with this method and it is with good performance in both laser and LED illumination systems.

At the same time, in the practical applications, the spherical aberration caused by the main mirror in a telescope will also cause the undesirable errors or diffusion of spots on the focal plane, so EDMF can be used to determine the best position of focal-plane plate.

In the original PDM model, FRD performance relates to the length of fibres, and specifically, the output focal ratio monotonically decreases with increasing fibre length. To simulate the FRD performance, the parameter   should be measured in advance. But the original PDM model has difficulty in the application of multi-position measurements in our experiments, so MPDM was proposed to resolve the relationship between focal distance and parameter ${d_0}$. The experimental results did not show the decreasing property of output focal ratio but an interesting trend of length dependence is that the output focal ratio of the two fibres increases at first and decreases later and finally remains steady in long fibres compared with the short fibres, where the value varies a lot, so to ensure a better FRD performance, a fibre longer than 15m is a promising choice.

In summary, we have modified EMD and PDM to be available for multi-position measurement experiment system. With EDMF the illumination system and the input condition can be easily tested and controlled. MPDM provides a new method to measure the parameter ${d_0}$ for simulations and predictions. With the method and the model, the dependence of FRD performance and throughput on length can be predicted and tested to optimize the length choice for the design of the spectroscopy telescope and in the meantime allow different work groups in the field to compare results much more conveniently and reliably.

\section*{Acknowledgements}

The authors would like to show our deepest gratitude to our colleague who have provided us with valuable guidance. Special thanks to Prof. Cui X. Q. (NIAOT) and Prof. Granham J. Murray (Durham University) for productive discussions. Without their enlightening instruction, impressive kindness and patience, this paper could not have been completed.

Guoshoujing Telescope (the Large Sky Area Multi-Object Fiber Spectroscopic Telescope, LAMOST) is a National Major Scientific Project built by the Chinese Academy of Sciences. Funding for the project has been provided by the National Development and Reform Commission. LAMOST is operated and managed by the National Astronomical Observatories, Chinese Academy of Sciences.

%%%%%%%%%%%%%%%%%%%%%%%%%%%%%%%%%%%%%%%%%%%%%%%%%%



\bibliographystyle{mnras}
\bibliography{MPDM} % if your bibtex file is called example.bib

%\begin{thebibliography}{99}
%\end{thebibliography}

%%%%%%%%%%%%%%%%%%%%%%%%%%%%%%%%%%%%%%%%%%%%%%%%%%

% Don't change these lines
\bsp	% typesetting comment
\label{lastpage}
\end{document}